\documentclass[fleqn,usenatbib]{mnras}

\usepackage{newtxtext,newtxmath}
\usepackage[T1]{fontenc}

\DeclareRobustCommand{\VAN}[3]{#2}
\let\VANthebibliography\thebibliography
\def\thebibliography{\DeclareRobustCommand{\VAN}[3]{##3}\VANthebibliography}

\usepackage{graphicx}	
\usepackage{amsmath}	
\usepackage{gensymb}
\usepackage{xspace}
\usepackage{caption, subcaption}
\usepackage{longtable, colortbl}
\usepackage{supertabular,booktabs}
\usepackage{lscape}
\usepackage{float}
\usepackage{natbib}
\usepackage{times}
\usepackage{graphics,epsfig}
\usepackage{comment}
\usepackage{orcidlink}

\newcommand{\lcdm}{$\rm \Lambda{\textrm{CDM}}$\xspace}               
\newcommand{\ee}[1]{\ensuremath{\times\ 10^{#1}}}                    
\newcommand{\eten}[1]{\ensuremath{10^{#1}}}                          
\newcommand{\kms}{\textrm{km~s}\ensuremath{^{-1}}}	                 
\newcommand{\msun}{\ensuremath{\rm M_{\odot}}}			             
\newcommand{\mstar}{\ensuremath{M_{\ast}}}			                 
\newcommand{\mspsqpc}{\ensuremath{\rm M_{\odot}\ pc^{-2}}}           
\newcommand{\HI}{H{\sc i}}                                           
\newcommand{\SHI}{\ensuremath{S_{\rm {\mathrm{H\textsc{i}}}}}}       
\newcommand{\rhi} {\ensuremath{\rm r_{\mathrm{H\textsc{i}}}}}        
\newcommand{\vhi} {\ensuremath{\rm V_{\mathrm{H\textsc{i}}}}}        
\newcommand{\mhi} {\ensuremath{\rm M_{\mathrm{H\textsc{i}}}}}        
\newcommand{\nhi}  {\ensuremath{\rm N_{\mathrm{H\textsc{i}}}}}       
\newcommand{\jykms}{\ensuremath{Jy\times \textrm{km~s}^{-1}}}        
\newcommand{\arcs}{$^{\prime\prime}$}                                

\title[\HI\ properties of BDDGs]{\HI\ Observations of Baryon-Dominated Dwarf Galaxy Candidates}

\author[Mirashi et al.]{
Atharva Mirashi$^{1}$\orcidlink{0009-0009-7106-2464}\thanks{E-mail: phd2301121001@iiti.ac.in},
Abhinav Narayan$^{1}$\orcidlink{0009-0006-4598-6522},
K. Keerthi$^{1}$\orcidlink{0009-0001-8441-3835},
Saurabh Kadawla$^{1,3}$,
Harshal Raut$^{1}$\orcidlink{0009-0009-2347-8772},
\newauthor
Narendra Nath Patra$^{1}$,
Nirupam Roy$^{2,4}$,
Prerana Biswas$^{3}$,
Mousumi Das$^{3}$,
Juliana Saponara$^{5}$
\\ \\
$^{1}$Indian Institute of Technology Indore, Indore 453552, India.\\
$^{2}$Indian Institute of Science, Bengaluru 560012, India.\\
$^{3}$Indian Institute of Astrophysics, Koramangala, Bengaluru 560034, India.\\
$^{4}$Department of Physics, New Mexico Institute of Mining and Technology, Socorro, NM 87801, USA.\\
$^{5}$Instituto Argentino de Radioastronom\'ia, CONIICET–CICPBA–UNLP, CC5 (1897) Villa Elisa, Prov. de Buenos Aires, Argentina.\\
}

\date{Accepted 2026 April 22. Received 2026 April 22; in original form 2026 January 29}

\pubyear{\the\year{}}

\begin{document}
\maketitle
\label{firstpage}
\pagerange{1--18}

\begin{abstract}
   We present resolved \HI\ observations of six dwarf galaxies drawn from a sample of baryon-dominated dwarf galaxy (BDDG) candidates previously identified using global \HI\ spectra from ALFALFA and optical inclinations from SDSS, both of which suffer from systematic uncertainties in irregular dwarf galaxies. Using uGMRT interferometric observations, we obtain high-resolution \HI\ cubes that enable more reliable determination of their geometry, circular velocity, and dynamical mass. We find that optical axial ratios systematically underestimate true disc thickness, inflating inclinations and underestimating rotation velocities in earlier work. Our \HI-derived axial ratios and kinematic position angles yield larger inclination corrections and hence larger dynamical masses. Four of these galaxies, UGC~6438, UGC~7983, AGC~191707, and AGC~733302, appear dark-matter deficient. The latter three of these four exhibit baryon enhancement efficiency factor (ratio of baryonic mass accumulated by a halo to the maximum expected value for its halo mass) exceeding 50\%, with AGC~191707 appearing formally super-efficient. Only UGC~9500 and AGC~220901 are consistent with being dark-matter dominated. Two of these high-efficiency galaxies lie in relatively isolated environments, showing no clear signatures of tidal disturbance or stripping, making their dark matter deficiency difficult to reconcile with standard $\Lambda$CDM expectations for low-mass halos. Our results underscore the importance of resolved \HI\ kinematics in confirming genuine BDDGs and suggest that more such systems may exist. Identifying a larger sample is essential for assessing their implications for baryon–halo coupling and structure formation within the $\Lambda$CDM paradigm.
\end{abstract}

\begin{keywords}
dark matter -- galaxies: dwarf -- radio lines: galaxies -- techniques: interferometric
\end{keywords}

\section{Introduction}
\label{sec:introduction}

Lambda-Cold Dark Matter model, ($\Lambda$CDM), the current standard model of cosmology, gives stringent estimates about the distribution of matter-energy density in the composition of the Universe. Our best measurements from Cosmic Microwave Background data put the matter density at $\approx$ 31\% of the total matter-energy density of the Universe. The visible baryonic matter makes up $\approx$ 5\% of the total matter-energy density, so the fraction of baryons in total matter, i.e. $f_{bar}\ =\ \Omega_{bar}/\Omega_{M}$ $\sim$ 0.17 \citep[][]{planck16}. Out of these, a small portion is trapped inside the potential wells of dark matter halos as galaxies \citep[see, e.g.][]{madau14}, which implies that these halos are inefficient at accumulating baryons over cosmic time. The ability of halos to acquire baryons from their surroundings may be thought of as baryon enhancement factor or baryon enhancement efficiency, and it directly connects the dark matter content of the Universe to the observed number density of galaxies \citep[see, e.g.][]{sawala15}. 

One can then understand baryon enhancement efficiency as the ratio of baryons accumulated by a halo, $M_{\rm bar}$, from its surroundings to the theoretically expected maximum for its given halo mass \citep[see, e.g.][]{white93}. This expected maximum baryonic mass within the halo's virial radius can be given by $f_{\rm bar}$\ $\times\ M_{200}$, where $M_{\rm 200}$ is the mass within the galaxy's virial radius (the radius within which the average density is 200 times the critical density of the Universe, $\rho_{\rm crit}$) \citep[see, e.g.][and references therein; although they use the term "galaxy formation efficiency" for this quantity]{oman16}. Thus, the baryon enhancement efficiency factor is theoretically defined as
\begin{equation}
    f_{\rm eff} = \frac{M_{\rm bar}}{ f_{\rm bar}\ \times\ M_{\rm 200} },
    \label{eqn:feff}
\end{equation}

\noindent Earlier observations have estimated that $f_{\rm eff}$ peaks to values $\sim\ 18\%$ in galaxies with stellar mass \mstar\ $\sim$ 3\ee{10} \msun\ \citep[see, e.g.][]{behroozi13a}. $f_{\rm eff}$ decreases on both sides of the mass scale and effectively reaches zero for a halo mass below $\sim$ \eten{9}\ \msun\ and above $\sim$ \eten{13}\ \msun\ \citep[see, e.g.][]{sawala16}.

This implies that there should be a lower limit on the dark matter halo mass for a galaxy with known baryonic mass, given a $f_{\rm eff}$. The lower the efficiency, the higher the dark matter halo mass.  Typically, at the lower end of the galaxy mass function (number density of galaxies as a function of their mass), i.e., for low-mass galaxies, the baryon enhancement efficiency is found to be poor. This efficiency can also be quantified by a mass-to-light ratio, i.e., the ratio of the total mass (DM+baryonic) of a galaxy to its luminosity (which comes only from baryons). The mass-to-light ratio in low-mass galaxies ($M_{\rm \ast}\ \sim\ \eten{6-9}\ \msun$) is found to be between 10--1000 \citep[][]{amorisco11, simon07}; indicating a significant dominance of dark matter over baryons. This is supported by numerical simulations as well \citep[see, e.g.][]{sawala15}, where they show only a small fraction of low-mass (< \eten{9.5}\msun) dark matter halos are able to efficiently accrete enough baryonic mass to host visible or luminous galaxies, leading to a high mass-to-light ratio overall at this mass range.  

Contrary to this theoretical and numerical understanding, \citet{guo20} have found several galaxies from the Arecibo Legacy Fast Arecibo L-band Feed Array (ALFALFA) $\alpha \rm.40$ data \citep{haynes18}, with considerably lower dark matter content than expected. From a sample of 324 galaxies in the ALFALFA survey, they identified 19 galaxies having baryonic mass comparable to or more than dark matter mass; such that, 
\begin{equation}
\begin{aligned}
     \frac{M_{\rm dynamic}(r\ \leq\ r_{\rm H \textsc i})}{M_{\rm bar}(r\ \leq\ r_{\rm H \textsc i})}\ =\ \frac{(M_{\rm bar}\ +\ M_{\rm DM})}{M_{\rm bar}}\ \leq\ 2; \\
     \therefore\ \frac{M_{\rm DM}}{M_{\rm bar}}\ \leq\ 1 \implies M_{\rm DM}\ \leq\ M_{\rm bar}
     \label{eqn:bddg_criterion}
\end{aligned}
\end{equation}

\noindent Where ${M_{\rm dynamic}}(r\ \leq\ r_{\rm H \textsc i})$ is the dynamic mass of the system determined using rotational width and assuming spherically symmetric rotation curve in the region up to $r_{\rm H \textsc i}$, which is the \HI-radius, defined by the distance from the centre where \HI-surface density ($\Sigma_{\rm H \textsc i}$) falls down to 1 \mspsqpc.

Here we strongly emphasize the fact that all the measurements of the dynamical mass are at $r_{\rm H \textsc i}$ and not up to virial radius $R_{\rm vir}$. To estimate the total dynamical mass at virial radius one would need to assume some dark matter radial density profile, \citep[as done by][for example, which suggested high baryon fractions up to $R_{\rm vir}$]{guo20} and carry out the analysis. However, the shape and structure of the dark matter halo itself may create large uncertainties especially for the galaxies in our sample. (see Section~\ref{sec:discussion} for relevant discussion). Since we cannot perform a complete mass modelling due to lack of enough independent beams across the major axes of these galaxies, we constrain ourselves to measuring the dynamical mass only up to $r_{\rm H \textsc i}$; as the criterion used by \citet{guo20} for quantifying baryon-dominance in these galaxies is also based on masses within $r_{\rm H \textsc i}$. In local Universe, the general trend seems to indicate dark matter dominance even within the $r \leq\ r_{\rm e}$ region in dwarf galaxies \citep[see, e.g.][]{strigari08, walker13inbook, scott21}. For all the galaxies in our sample, $r_{\rm H \textsc i}$ extends far beyond optical half light radius $r_{\rm e}$. However, there have been a few recent studies where dwarf galaxies with a much stronger baryon dominance was reported \citep[see, e.g.][]{vandokkum18, vandokkum19, mancera20, oh15} based on analysis of \HI\ gas disks. Therefore the 19 Baryon-Dominated Dwarf Galaxies (BDDGs) reported by \citet{guo20} seem worth investigating further to understand the wider picture.

Out of these 19 BDDGs, 14 galaxies are found to be isolated field galaxies, without any group or cluster association, as they are located at distances farther than 3 times the virial radii of any surrounding groups/clusters. Their results suggest that these galaxies either accreted more baryons than expected or have had their dark matter stripped away due to some interaction. However, given that these galaxies are located in isolated environments, both of these possibilities seem difficult, posing a significant challenge in understanding the galaxy formation process and baryon enhancement efficiency under $\Lambda$CDM.

\citet{guo20} used \HI~and optical observations to estimate the baryonic and dark matter mass in these galaxies. They used global \HI~spectra from Arecibo single-dish observation to calculate the spectral width, $w_{\rm 20}$. This spectral width is a measure of the total mass of the galaxy, after applying a correction for the inclination. They obtained the inclination from the optical observation of Sloan Digital Sky Survey \citep[SDSS;][]{abazajian09}. The use of single-dish spectra as a measure of the actual circular velocity of the galaxy and the adoption of optical inclination as the inclination of the \HI~disk pose significant limitations in the estimation of the dynamical mass. For example, single-dish spectra often get contaminated by stray radiation due to their large beam. Further, single-dish observation does not provide a resolved map of the galaxy. The optical inclination of a galaxy may not match that of the \HI~disk \citep[see, e.g.][]{yasin23, biswas23}. These shortcomings can significantly influence the estimation of the dynamical mass in these galaxies.

The limitations can affect the mass estimation primarily in two ways. Firstly, single-dish spectra can overestimate the \HI~flux coming from a galaxy, overestimating its baryonic component. Secondly, the spectral width can get altered due to the distortion of the spectra, leading to an error in estimating the dynamical mass. Additionally, one needs to assign an \HI~radius corresponding to the measured circular velocity to calculate the dynamical mass. As the single-dish observation does not provide a resolved \HI~map, \citet{guo20} used the mass-size relation from \citet{wang16} to estimate \HI~radius of each galaxy. An error in estimating this radius can lead to a wrong estimation of the dynamical mass. For example, the scatter in the \citet{wang16} (their figure 1), is about 0.5 dex. With this scatter, the estimated \HI~radius from this scaling relation can change by a factor of two. Moreover, the assumption that these BDDGs will follow the same relation may itself be false.

Another factor worth noting is that \citet{guo20} used optical inclination as measured using SDSS r-band data to estimate the circular velocities from the spectral widths. It should be noted here that the inclination ($i$) is one of the most sensitive parameters in the calculation of dynamical mass from the \HI\ spectral width ($M_{\rm dyn}\ \propto\ \frac{1}{sin^2\ (i)}$). This implies that a small variation in inclination can translate into a substantial change in the dynamical mass. Further, observationally, it is found that the kinematic inclination of the \HI~disk can deviate considerably from that of the optical disk, especially for dwarf galaxies \citep[see, e.g.][]{oh15}. In these cases, estimation of circular velocity from the global spectra is not straightforward and could be erroneous. At an inclination of 30\degree\ a $\pm$ 5\degree~difference in inclination, for example, can change the dynamic mass by up to 20\%.

Resolved interferometric observations of the \HI~disks in these galaxies can overcome these limitations. For example, total intensity \HI~images obtained through interferometric observation can provide the accurate gas mass of the galaxy. Further, the line-of-sight velocity of the disk can be estimated directly from the resolved velocity map (moment one map). This velocity can then be corrected for inclination by using the inclination of the \HI~disk, which can be directly determined using the total intensity map (moment zero map). Aiming for this, we observed six galaxies from the sample of 19 galaxies of \citet{guo20} using the upgraded Giant Metrewave Radio Telescope (uGMRT) and imaged their \HI~disks to investigate if the dark matter content in these galaxies is lower than what is expected in a $\Lambda$CDM Universe.
In following sections, we describe the chosen sample of galaxies and their properties, the observation strategies and the subsequent data reduction and imaging, our results and detailed analysis to determine the dynamic mass, and finally conclude with the relevant discussion about our results in the broader picture of galaxy structure formation.

\section{Sample}
\label{sec:sample}

\begin{table*}
    \centering
    \caption{Sample Galaxy Properties-- columns 1,2,3: name and position of the galaxy in equatorial co-ordinates, column 4: distance from observer in Mpc, column 5: receding velocity as measured in a barycentric frame, column 6: integrated \HI\ flux in \jykms\ \citep[position, distance, velocity and \HI\ flux are based on measurements by the ALFALFA group;][]{haynes18}, column 7: r-band magnitude from SDSS, taken directly from \citet{guo20}, column 8: optical $\frac{b}{a}$ obtained in r-band by \citet{guo20}, and column 9: the ratio of distance from the nearest group/cluster ($dist_{\rm gr}$) to the virial radius ($R_{\rm vir,gr}$) of that group/cluster.}
    \label{tab:sample_table}
    \begin{tabular}{|c|c|c|c|c|c|c|c|c|}
        \hline
         Galaxy Name & RA (J2000) & DEC (J2000) & Dist & $V_{\rm bary}$ & $S_{\rm int}$ & $M_{\rm r}$ & $\frac{b}{a}$ & $\frac{dist_{\rm gr}}{R_{\rm vir,gr}}$ \\
         & (hh:mm:ss) & (\degree\ \arcmin\ \arcsec) & Mpc & \kms\ & \jykms & (mag) &  & \\ \hline
         UGC 6438 & 11h25m53.5s & +09d59m14.9s & 20.4 & 1156 & 3.87 & -17.75 & 0.524 & 2.707 \\
         UGC 7983 & 12h49m47.0s & +03d50m32.3s & 16.6 & 694 & 6.71 & -15.21 & 0.600 & 1.885 \\
         UGC 9500 & 14h45m21.5s & +07d51m46.0s & 27.6 & 1690 & 9.02 & -17.56 & 0.444 & 4.761 \\
         AGC 191707 & 09h37m47.6s & +27d33m57.7s & 25.1 & 1595 & 2.94 & -16.31 & 0.570 & 5.408 \\
         AGC 220901 & 12h39m58.6s & +13d46m50.0s & 17.0 & 1005 & 4.23 & -14.07 & 0.486 & 0.38 \\
         AGC 733302 & 14h57m39.3s & +26d39m53.8s & 23.0 & 1280 & 3.75 & -16.47 & 0.515 & 5.913 \\ \hline
    \end{tabular}
\end{table*}

The primary sample of \cite{guo20} came from the galaxies detected in the 40\% ALFALFA survey, $\alpha.40$ \citep[][]{haynes18}. They identified field dwarf galaxies from this $\alpha.40$ catalogue, having r-band magnitude $M_{\rm r}$ \textgreater $-18$, and position matches with the Seventh Data Release of the Sloan Digital Sky Survey \citep[SDSS DR7;][]{abazajian09} PhotoPrimary optical catalogue. Further, for genuine detections, an SNR\textgreater $10$ criterion was used on the Arecibo \HI~spectra of the selected galaxies. These galaxies were further filtered by inspecting their \HI~spectra, and optical, UV images for any suspicious signature. They also imposed a condition on the r-band (SDSS-DR7) axial ratio, $0.3 \geq b/a \geq 0.6$ for minimizing the inclination discrepancies between the optical and the \HI~disks. These criteria resulted in a total of 324 galaxies in their parent sample, for which they investigated the dark matter to baryon ratio. Out of these, they found 19 galaxies, having much higher baryonic fraction than expected, as per the criterion given in equation~(\ref{eqn:bddg_criterion}), terming them as Baryon-Dominated Dwarf Galaxies (BDDGs). Our pilot sample of six galaxies was selected out of these 19 galaxies for uGMRT observation. We carefully selected these galaxies, so as to have a significant \HI~flux (upto 1 \mspsqpc\ surface density) detectable by uGMRT in a reasonable observing time. In Table~\ref{tab:sample_table} we list our pilot sample galaxies and their basic properties. Columns 1, 2, and 3 show the names of the galaxies along with their J2000 coordinates, respectively. Column 4 presents the distance to the galaxies, and column 5 presents the recession velocity in the barycentric frame of reference. Column 6 represents the integrated \HI~flux as obtained from Arecibo observations. Columns 7 and 8 present the r-band magnitude (from SDSS-DR7) and the optical (r-band) axial ratio ($b/a$) of the sample galaxies, respectively. The last column (column 9) presents the ratio of the distance of the galaxy from the centre of the nearest group or cluster to the virial radius of that group or cluster. If this ratio $\geq$ 3, then the galaxy is an isolated or field dwarf galaxy.

\section{Observations and Data Analysis}
\label{sec:obs_analysis}

Our six target galaxies were observed as part of the 39th cycle of uGMRT under the proposal code 39\_064. Total observing time was $\sim 14$ hours per target, resulting in $\sim 9-11$ hours of on-source time (excluding overheads and time on calibrators). We used both the GMRT Software Backend (GSB) and GMRT Wideband Backend (GWB) for the observations (simultaneous data acquisition was available during this cycle). The GSB was configured to have a 4.17 MHz bandwidth with 512 spectral channels. This results in a spectral resolution of $\sim 8.1$ KHz (velocity resolution of $\sim $ 1.7 \kms). We chose primarily GSB for detecting \HI~emission from our sample galaxies. A narrow bandwidth with high spectral resolution is suitable for obtaining the velocity map and estimating the dynamical mass. 

For each observing run, standard flux calibrators (3C48, 3C147, or 3C286) available on the uGMRT sky were observed for 10-15 mins at the beginning and at the end of the observation. The phase calibrators were observed every 45 minutes between the target scans. For each target galaxy, a phase calibrator was chosen carefully from the VLA calibrator list\footnote{\label{foot:vlacal}\url{https://www.vla.nrao.edu/astro/calib/manual/csource.html}} having an angular distance $\leq$ 10$\degree$.

All the data were reduced using Common Astronomy Software Applications (\texttt{CASA}) \citep[][]{casa22} following standard flagging and calibration steps. Dead antennas were flagged from the data at the beginning of the analysis. Flux calibration was done using a primary calibrator, whereas solutions for the phase were obtained using phase calibrators. The same primary calibrators were used to perform the bandpass calibration. The flagged and calibrated source data were split into a separate file that does not contain calibrator data. The total observing time for each target was divided into multiple observing sessions (days). The basic data analysis for each session was performed independently before combining (concatenating) the UV data together. This combined UV cube is then used for spectral imaging. We used the \texttt{tclean} task in CASA for spectral imaging. To recover the faint diffuse emission from our target galaxies, we imaged them with UV tapering of $\sim 5 - 10$ kilo lambda with natural weighting. This results in a spatial resolution of $\sim 30$\arcs$-50$\arcs. The Doppler correction for the rotation of the Earth was applied on-the-fly during imaging using barycentric frame correction. We achieve an RMS of $\sim 1-1.5$ mJy per channel in final spectral cubes.

We further use \texttt{SoFiA2} \citep[][]{sofia21} to produce moment maps from these spectral cubes. \texttt{SoFiA2} uses a source finding algorithm that can generate masks and moment maps from a data cube using a smoothing and clipping technique. We used an s+c threshold of 5.0 mad to generate masks in our spectral cubes \citep[see][for more details]{sofia21}. Using this mask, \texttt{SoFiA2} produces different moment maps, providing total intensity (Mom-0), velocity field (Mom-1), and velocity dispersion (Mom-2). These maps, along with the spectral cube, are used to extract dynamical information about our target galaxies. 

\section{Results}
\label{sec:results}
\subsection{Gas Mass}
\begin{figure*}
        \begin{tabular}{ccc}
        \subcaptionbox{UGC 6438\label{fig:spec1}}{\includegraphics[width=0.30\linewidth]{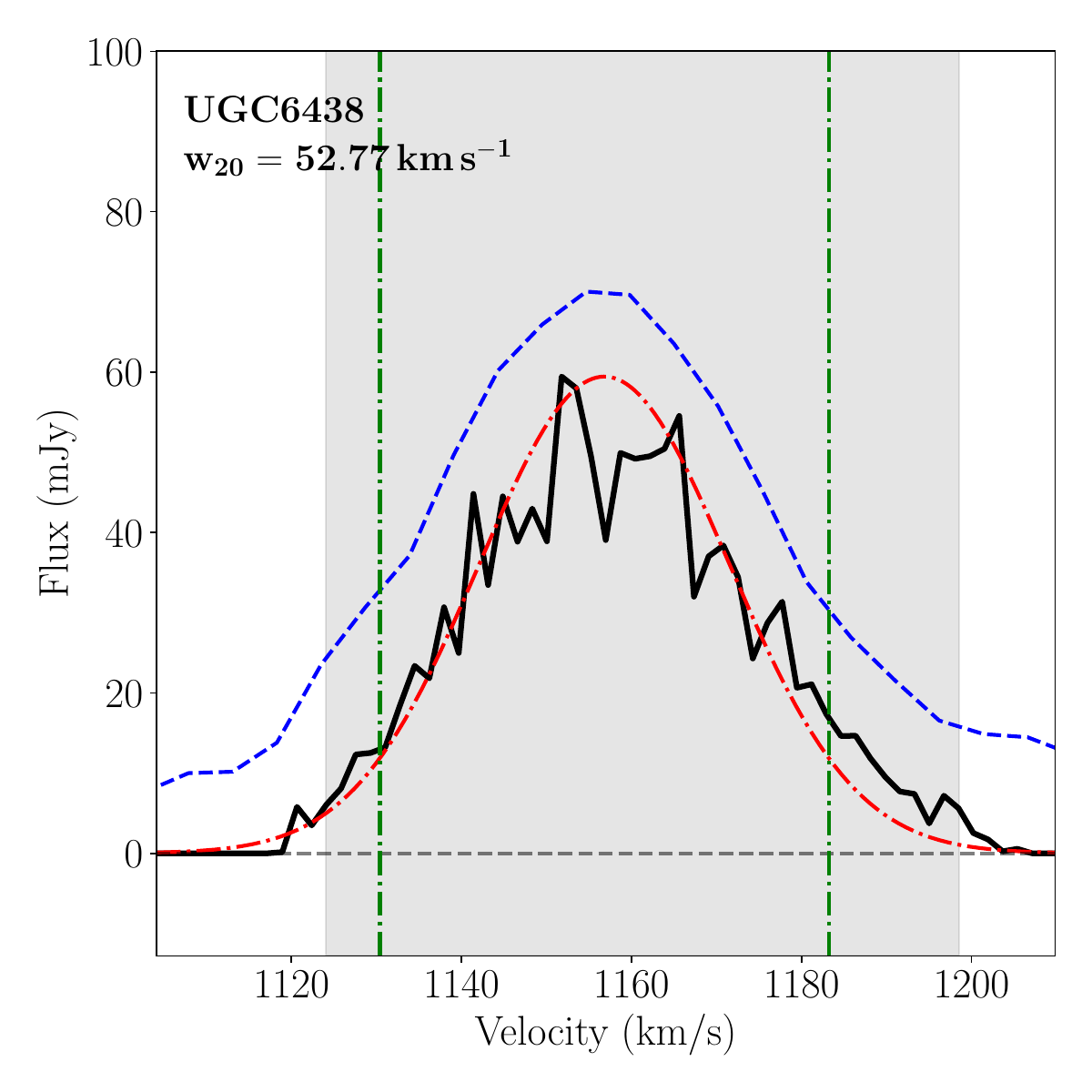}} &
        \subcaptionbox{UGC 7983\label{fig:spec2}}{\includegraphics[width=0.30\linewidth]{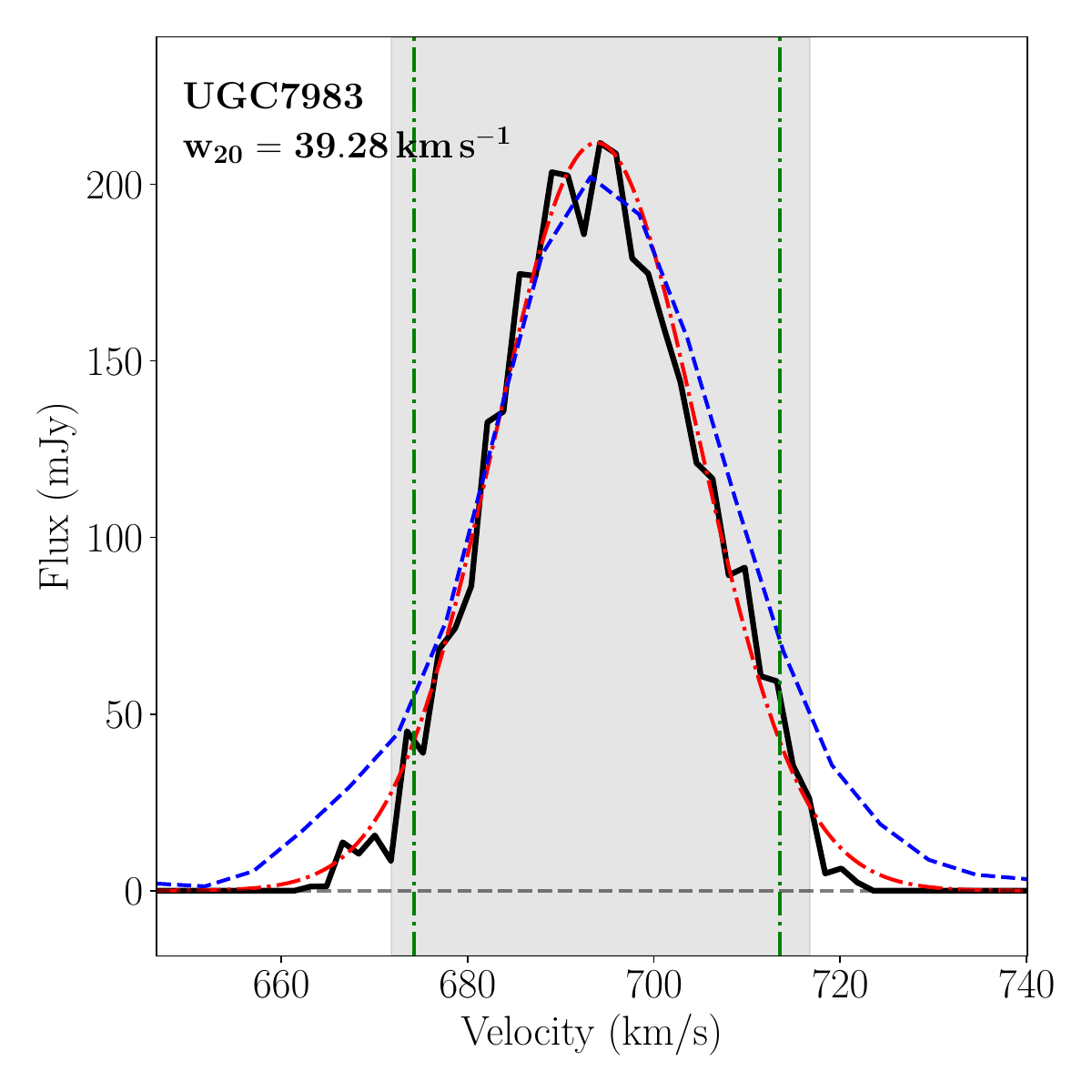}} &
        \subcaptionbox{UGC 9500\label{fig:spec3}}{\includegraphics[width=0.30\linewidth]{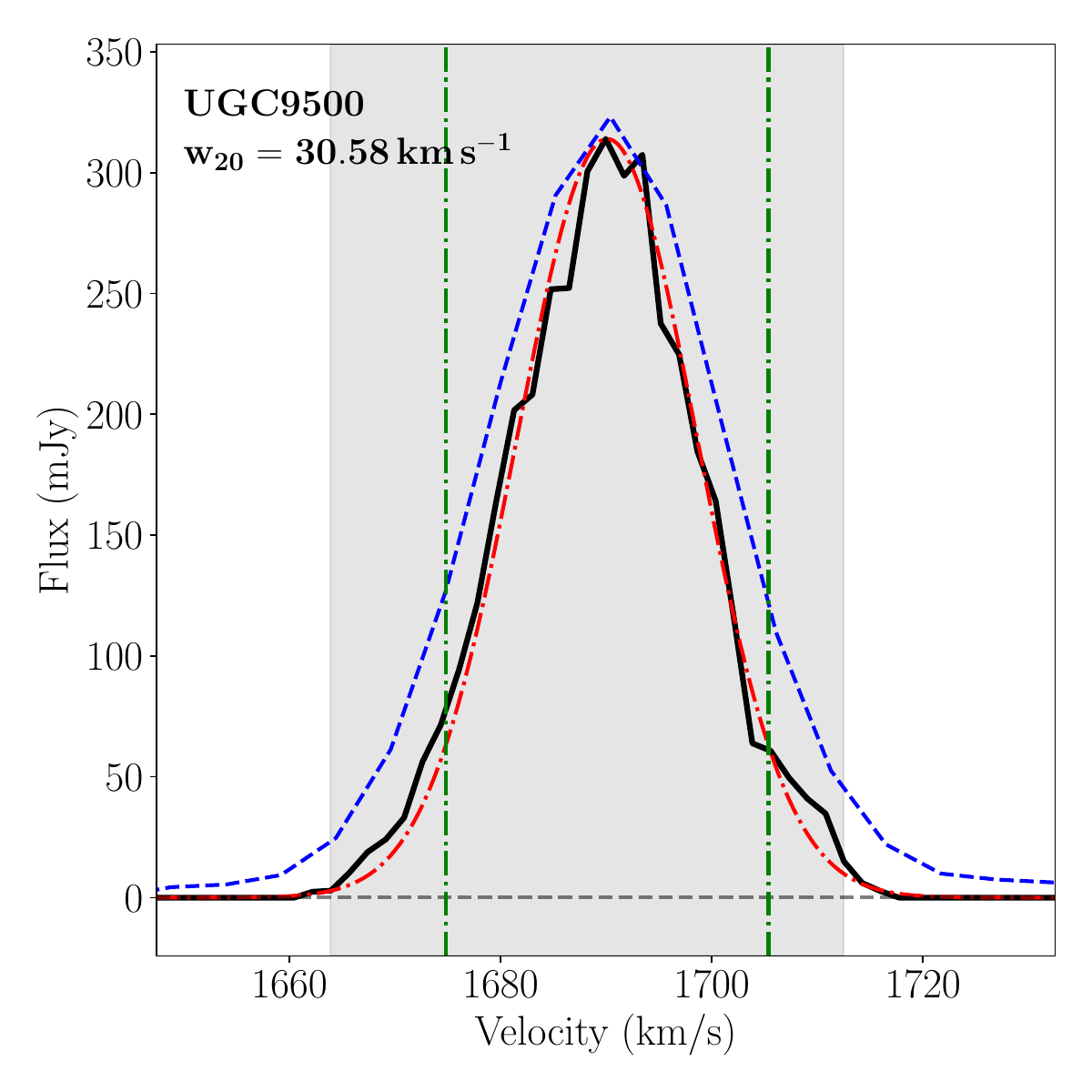}} \\
        
        \subcaptionbox{AGC 191707\label{fig:spec4}}{\includegraphics[width=0.30\linewidth]{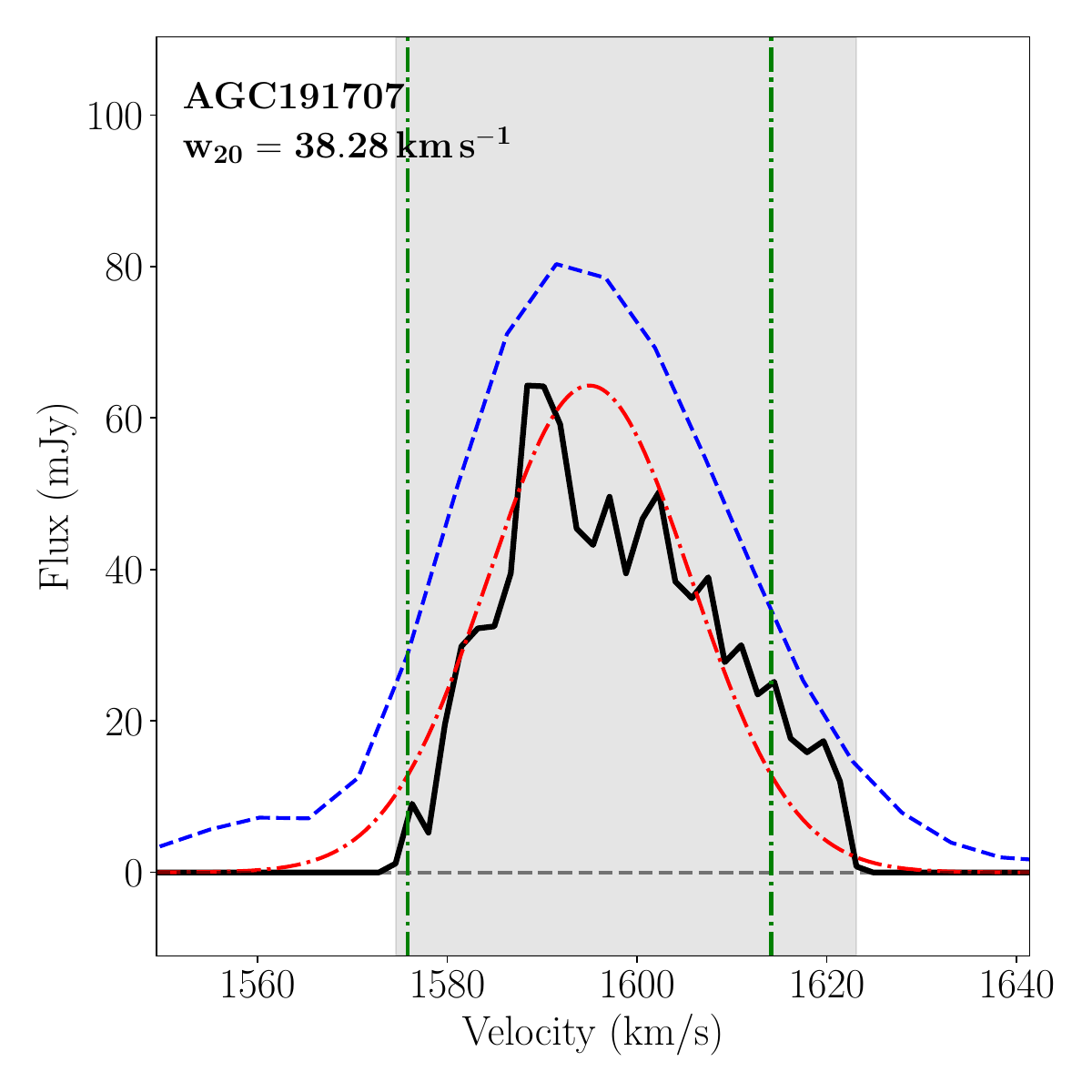}} &
        \subcaptionbox{AGC 220901\label{fig:spec5}}{\includegraphics[width=0.30\linewidth]{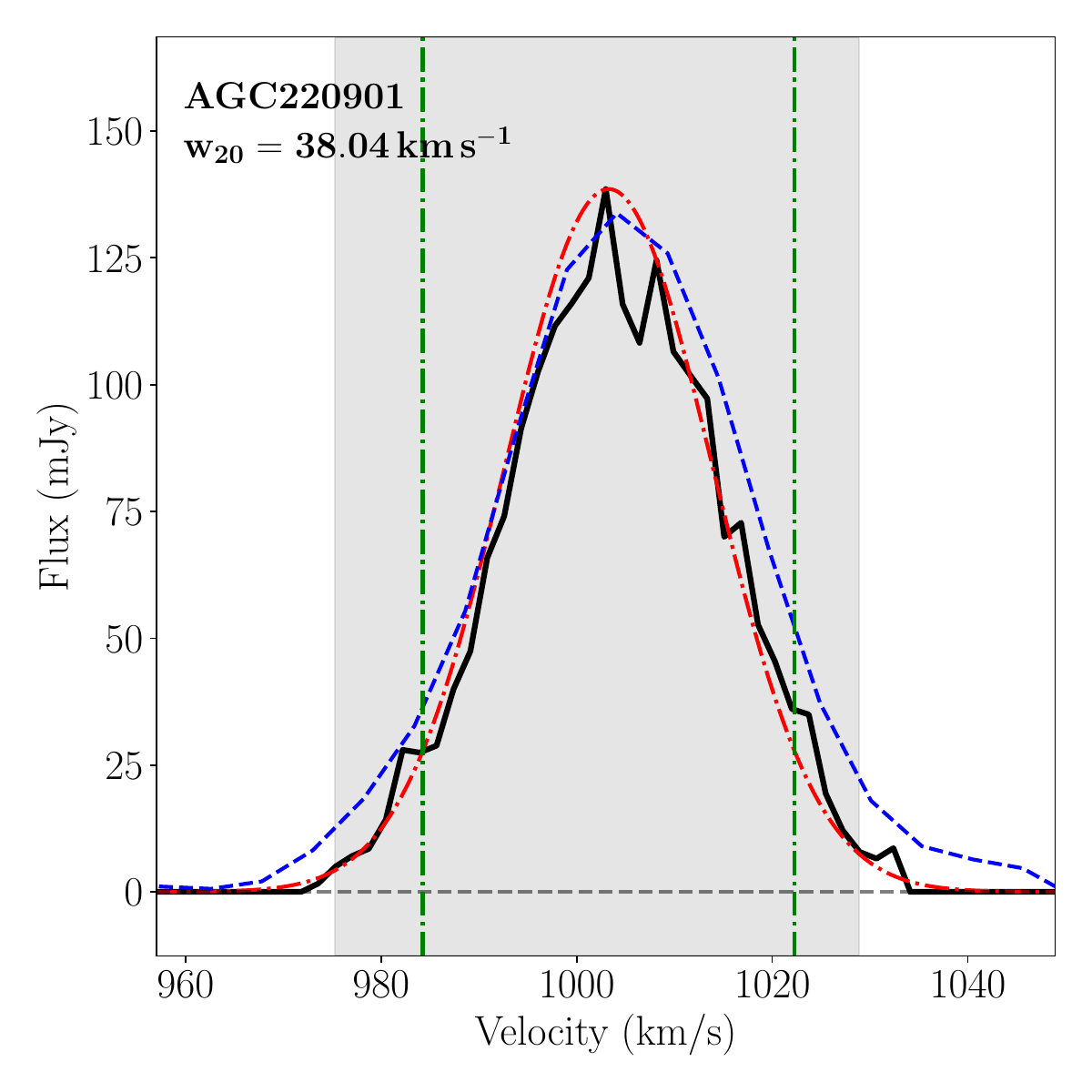}} &
        \subcaptionbox{AGC 733302\label{fig:spec6}}{\includegraphics[width=0.30\linewidth]{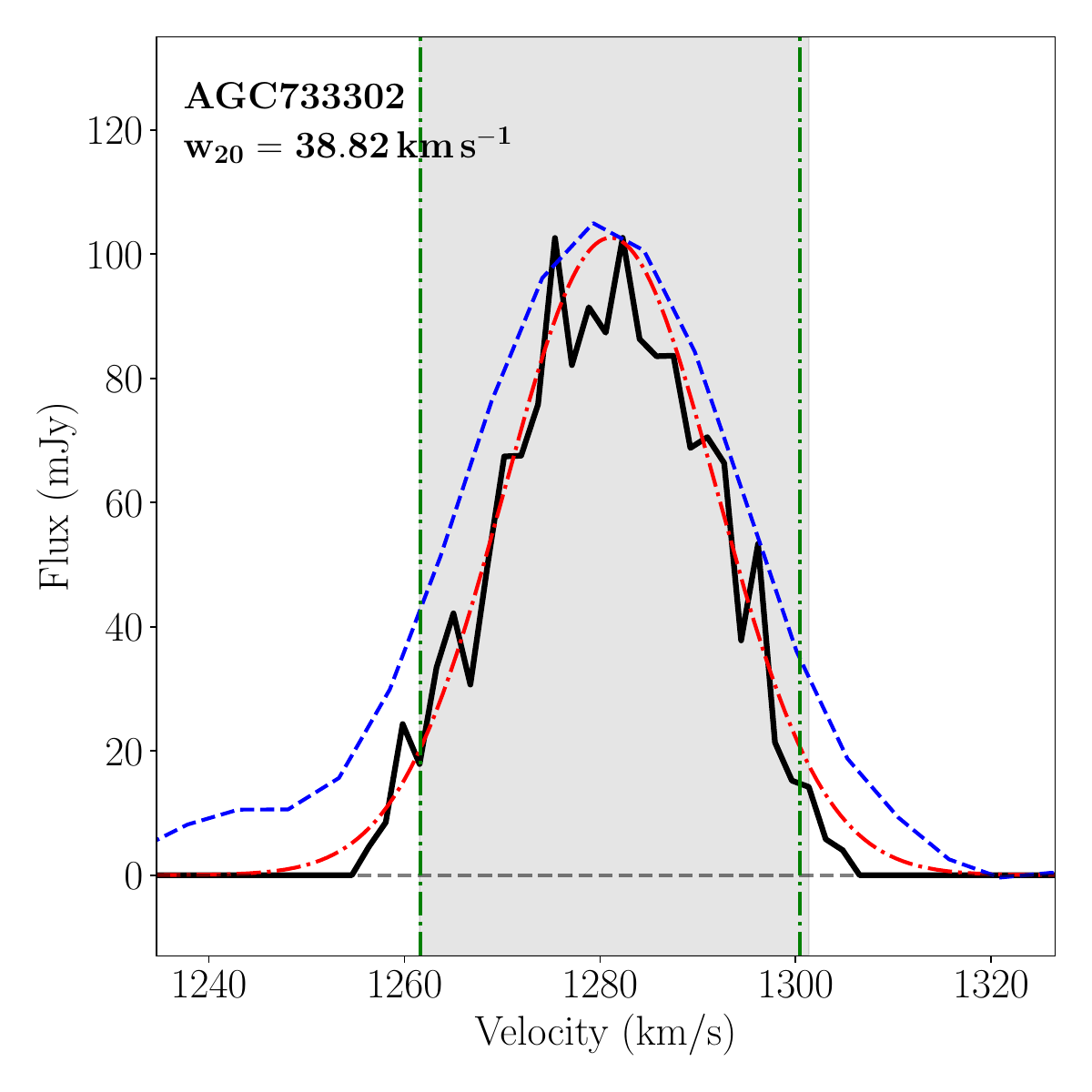}} \\
    \end{tabular}
    \centering\caption{Global \HI~spectra of our target galaxies. Black solid lines indicate the flux values obtained after summing over the masks in each channel. The red dashed lines in each panel indicate a Gaussian fit to the spectrum. The $w_{20}$ width for each spectrum is shown by the green vertical dashed lines, whereas the shaded region shows the full spectral width obtained from the channel maps. The blue dashed lines represent the spectra obtained from ALFALFA. See the text for more details.}
    \label{fig:combined_spectra}
\end{figure*}

We use the mask generated by \texttt{SoFiA2} to demarcate the regions of emission in the spectral cube. This mask identifies the emitting regions in each channel, as different amounts of gas appear at different line-of-sight velocities. We then sum the total emission channel by channel to obtain the global \HI~spectrum of the galaxy. In Fig.~\ref{fig:combined_spectra}, we plot the global spectra of our sample galaxies (solid black line).

For comparison, we also plot the single-dish \HI~spectra from the ALFALFA survey \citep[][]{haynes18} on top of our recovered spectra (dashed blue lines in each panel in Fig.~\ref{fig:combined_spectra}). As can be seen, our spectra are roughly consistent with those from single-dish observations; however, we do observe discrepancies in some cases. For example, in the spectrum of UGC~6438, the single-dish baseline lies significantly above zero, shifting the single-dish spectrum toward higher flux values. Similar behaviour is seen for AGC~191707 and AGC~733302. In all these systems, the baselines of the single-dish observations are elevated, leading to an apparently higher flux. In contrast, our interferometric observations do not show such baseline offsets.

We integrate the global \HI~spectra to obtain the total integrated flux for our target galaxies. In Table~\ref{tab:bary_masses_corrected} (second column), we list these integrated fluxes, and the corresponding \HI~masses are provided in the fourth column of the same table. To convert the integrated \HI~flux into \HI~mass, we use the standard relation \citep{roberts62}:

\begin{equation}
     \frac{\mhi}{\rm \msun} = 2.36\ee{5}\ \times\ \left( \frac{D}{\rm Mpc} \right)^2 \int_{\rm v} \frac{F_{\rm v}}{\rm Jy\ \kms} \,d \rm v\
     \label{eqn:flux_to_mhi}
 \end{equation}

\noindent where $D$ is the distance to the galaxy, and $F_{\rm v}$ is the detected flux density as a function of velocity (channel). The total gas mass was estimated by considering the contribution from Helium, i.e., $M_{\rm gas} = 1.33 \times M_{\rm H\textsc i}$.

\subsection{Stellar Mass}
A significant fraction of the baryonic mass in galaxies is contributed by their stellar component; therefore, estimating the stellar mass of our sample galaxies is essential for determining their total baryonic content. We obtain the stellar masses from \citet{guo20}, who derived them using SDSS DR7 optical images in the $g$ and $r$-bands. Following Bell’s prescription \citep{bell03}, the stellar mass was calculated using the relation
\begin{equation}
 \log_{10}(M_{\rm \ast}/L_{\rm r}) = a_{\rm \lambda} + b_{\rm \lambda} \times (g - r)
 \end{equation}
\noindent where $M_{\rm \ast}$ is the stellar mass, $L_{\rm r}$ is the r-band luminosity (derived from the r-band magnitude), and $(g-r)$ is the colour. A Kroupa initial mass function was assumed \citep[][]{starkenburg19}, for which $a_{\rm r} = -0.306 - 0.15 = - 0.456$ and $b_{\rm r} = 1.097$. The resulting stellar masses for our sample galaxies are listed in the first column of Table~\ref{tab:bary_masses_corrected}. We take into account the 0.1 dex method error as prescribed by \citep{bell03}.

It can be noted that these values for $a_{\rm \lambda}$ and $b_{\rm \lambda}$ in the $\rm colour-M_{\rm \ast}/L_{\rm r}$ relation from \citet{bell03} are not heavily affected by dust extinction \citep[see, e.g.][]{zhang17, jarrett23}. We also examined WISE data\footnote{\label{foot:wise}\url{https://irsa.ipac.caltech.edu/applications/wise}}; and especially the W1 - W2 colours, which suggest that there is negligible extinction due to intrinsic dust content \citep[based on analysis by][]{jarrett13}. This is expected as gas in dwarf galaxies is usually dust-deficient owing to low heavy-element abundances \citep[see, e.g.][and the references therein]{welty12, oh15}. There is also a possibility that due to their lower surface brightness the stellar mass may be underestimated due to detection limits. However this effect is more prominent only at lower mass scales ($\leq$ \eten{5} \msun) \citep[see, e.g.][and the references therein]{wheeler25}, leaving stellar mass estimates above this value, where all the galaxies in current sample lie, not affected by more tha 20\%. Additionally some of the galaxies from our sample have been studied in more recent works with different methods, \citep[][]{hoyer21, lin23, subramanian24, guo24} and their estimates of stellar mass do not differ from our estimates significantly. Given this, we assume a conservative error of 0.1 dex to account for the uncertainties in estimation of the stellar masses in our galaxies.

\subsection{Baryonic Mass}
The total baryonic mass is taken as the sum of the stellar and gas components (last column of Table~\ref{tab:bary_masses_corrected}). As our target systems are low-mass dwarf galaxies, they are not expected to host a significant reservoir of molecular gas due to their low metallicities, weak dust content, and inefficient H\textsubscript{2} formation \citep[e.g.][]{schruba12, cormier14, hunt20}. We therefore neglect the contribution from molecular gas when computing the total baryonic mass.
\begin{table*}
    \centering
    \caption{Stellar and \HI\ masses of the galaxies: Column 2,3,4,5,6 and 7 give the total integrated flux \SHI\ from the spectra, stellar mass taken from \citet{guo20}, \HI\ mass calculated from equation~(\ref{eqn:flux_to_mhi}) with values in brackets corresponding to those given by \citet{guo20}, the total baryonic mass with necessary corrections, and ratios of stellar and \HI\ masses with baryonic masses respectively.}
    \label{tab:bary_masses_corrected}
    \begin{tabular}{|c|c|c|c|c|c|c|} \hline 
         Galaxy Name & \SHI\ & \mstar\ & \mhi\ & $M_{\rm bar}$ & \mstar/$M_{\rm bar}$ & \mhi/$M_{\rm bar}$ \\ 
         & \jykms\ & \eten{8} \msun\ & \eten{8} \msun\ & \eten{8} \msun\ &  &  \\ 
         &  & \citep{guo20} & Ours \citep{guo20} & (1.33 $\times$ Our \mhi) + \mstar\ & & \\ \hline
         
UGC 6438 & 2.17 $\pm$ 0.02 & 8.93 $\pm$ 2.31 & 2.13 $\pm$ 0.02 (3.80) & 11.77 $\pm$ 2.31 & 0.76 $\pm$ 0.25 & 0.18 $\pm$ 0.04 \\
UGC 7983 & 5.63 $\pm$ 0.10 & 0.50 $\pm$ 0.13 & 3.66 $\pm$ 0.07 (4.37) & 5.37 $\pm$ 0.16 & 0.09 $\pm$ 0.02 & 0.68 $\pm$ 0.02 \\
UGC 9500 & 6.83 $\pm$ 0.22 & 4.65 $\pm$ 1.20 & 12.27 $\pm$ 0.40 (16.22) & 20.96 $\pm$ 1.31 & 0.22 $\pm$ 0.06 & 0.59 $\pm$ 0.04 \\
AGC 191707 & 1.59 $\pm$ 0.04 & 1.87 $\pm$ 0.49 & 2.36 $\pm$ 0.06 (4.37) & 5.01 $\pm$ 0.49 & 0.37 $\pm$ 0.10 & 0.47 $\pm$ 0.05 \\
AGC 220901 & 3.55 $\pm$ 0.11 & 0.08 $\pm$ 0.02 & 2.42 $\pm$ 0.08 (2.88) & 3.30 $\pm$ 0.10 & 0.02 $\pm$ 0.01 & 0.73 $\pm$ 0.03 \\
AGC 733302 & 2.60 $\pm$ 0.10 & 1.89 $\pm$ 0.49 & 3.24 $\pm$ 0.13 (4.68) & 6.20 $\pm$ 0.52 & 0.30 $\pm$ 0.08 & 0.52 $\pm$ 0.05 \\
         \hline
    \end{tabular}
\end{table*}

\subsection{$r_{\rm H \textsc i}$ and axial ratio}
The size of a galaxy is a crucial parameter when estimating its dynamical mass from the velocity width of the observed \HI~spectrum. In this analysis, we use the \HI~velocity width as a proxy for the dynamical mass. In such calculations, the radius at which the velocity width is assigned is critical, since the same rotational velocity at a larger radius implies a higher dynamical mass. Owing to the lack of resolved \HI~observations, \citet{guo20} employed the \HI~mass–size relation \citep[e.g.][]{wang16}:

\begin{equation}
 \log(r_{\rm H\textsc i}) = (0.506 \pm 0.003)\log(M_{\rm H\textsc i}) - (3.293 \pm 0.009) - \log(2).
 \label{eqn:size_mass_relation}
 \end{equation}

\noindent where, $r_{\rm H\textsc i}$ is in kpc and $M_{\rm H\textsc i}$ is in \msun; to infer the \HI~extent of their galaxies, and hence the radius associated with the observed \HI~velocity width. However, the \HI~mass–size relation exhibits substantial intrinsic scatter, and can easily introduce uncertainties that change the inferred radius by a factor of two, leading to dynamical mass estimates that are incorrect by up to a factor of two as well. Resolved \HI~maps are therefore essential for accurately determining the true \HI~extent of our sample galaxies and for deriving reliable dynamical masses.

\begin{figure*}
    \centering
    \caption{Combined images of Sample Galaxies: Left Column: DECam r-band images with radio contours (starting from $\Sigma_{\rm H\textsc i}$ = 1~\mspsqpc, with linear spacing upto the maximum emission) over-plotted. The red dashed ellipse shows the \HI\ disk predicted by \citet{guo20} using mass-size relation for radius and optical axial ratio from SDSS r-band, while the yellow dashed ellipse is our fit for to \nhi = 1 \mspsqpc\ surface density; Middle Column: Moment-0 and Right Column: Moment-1 map of the galaxy's \HI\ disk with the same ellipse fit along with major axis derived from moment-0 map shown in magenta over moment-1 map.}
    \label{tab:combined_images}
    \begin{tabular}{ccc}
        {\includegraphics[width=0.30\linewidth]{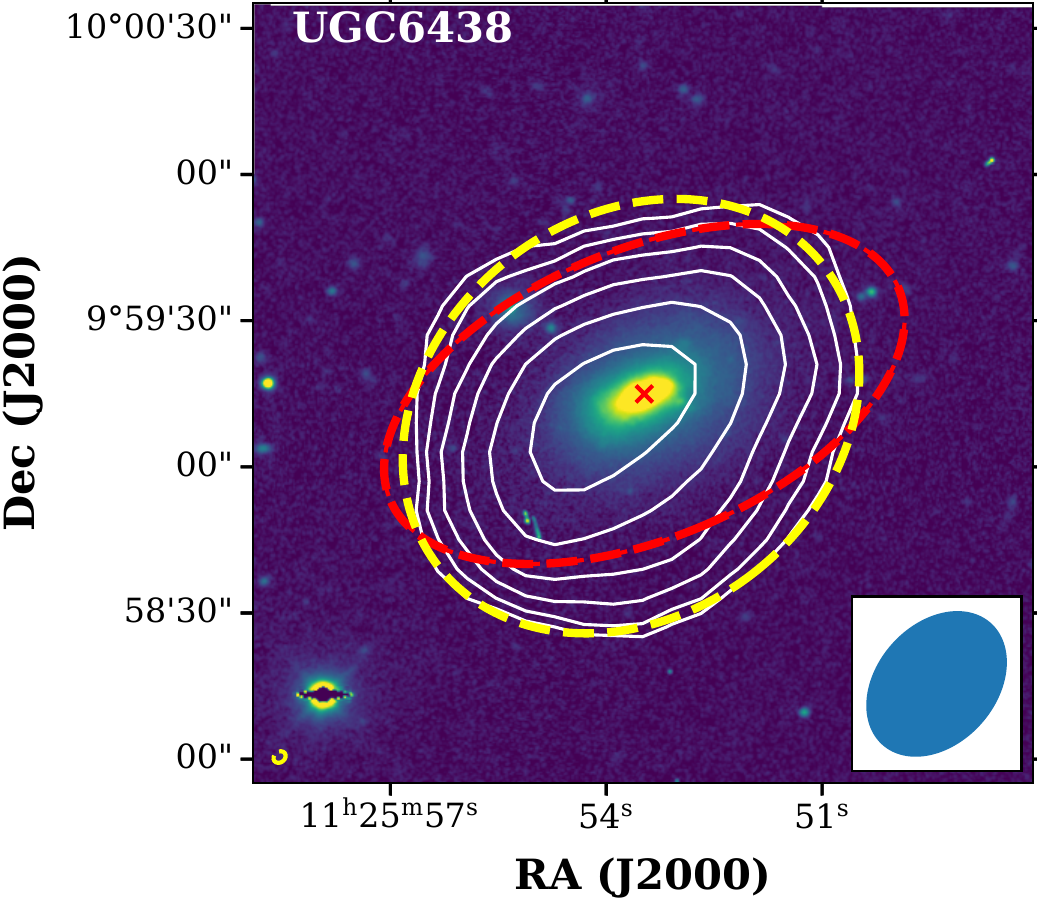}} &
        {\includegraphics[width=0.30\linewidth]{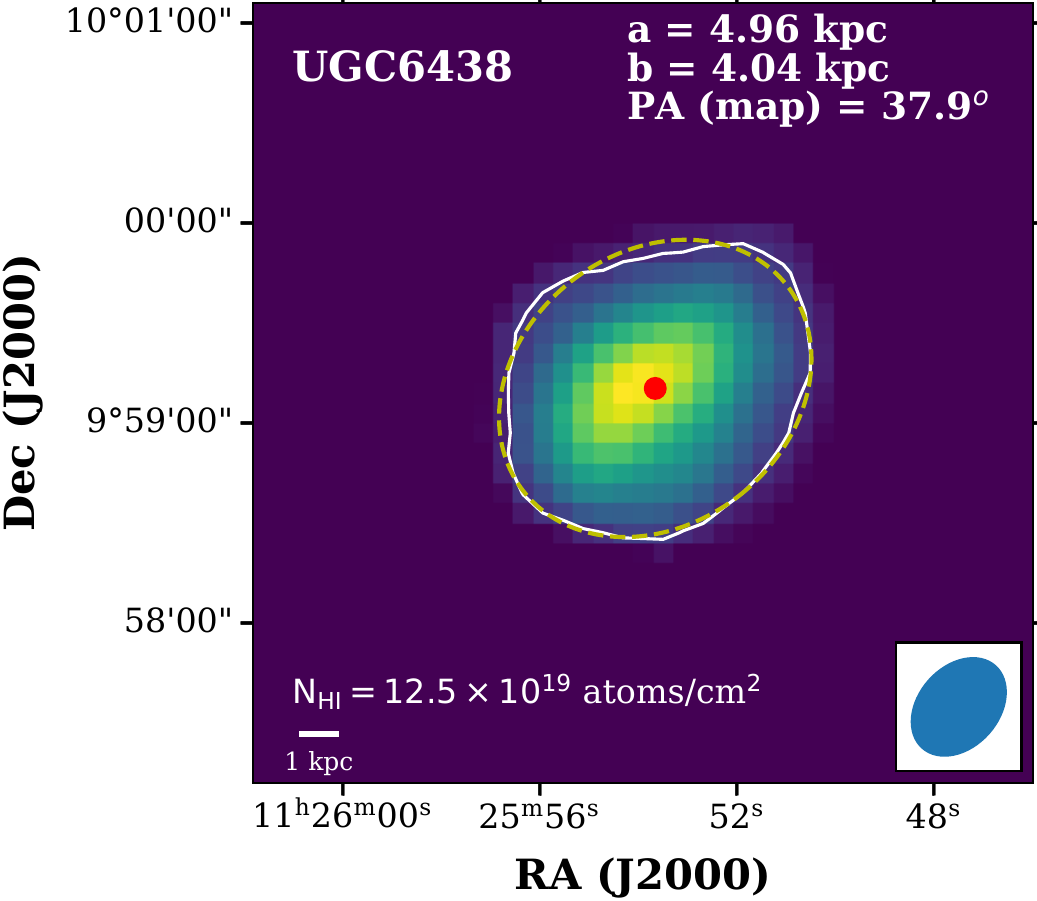}} &
        {\includegraphics[width=0.38\linewidth]{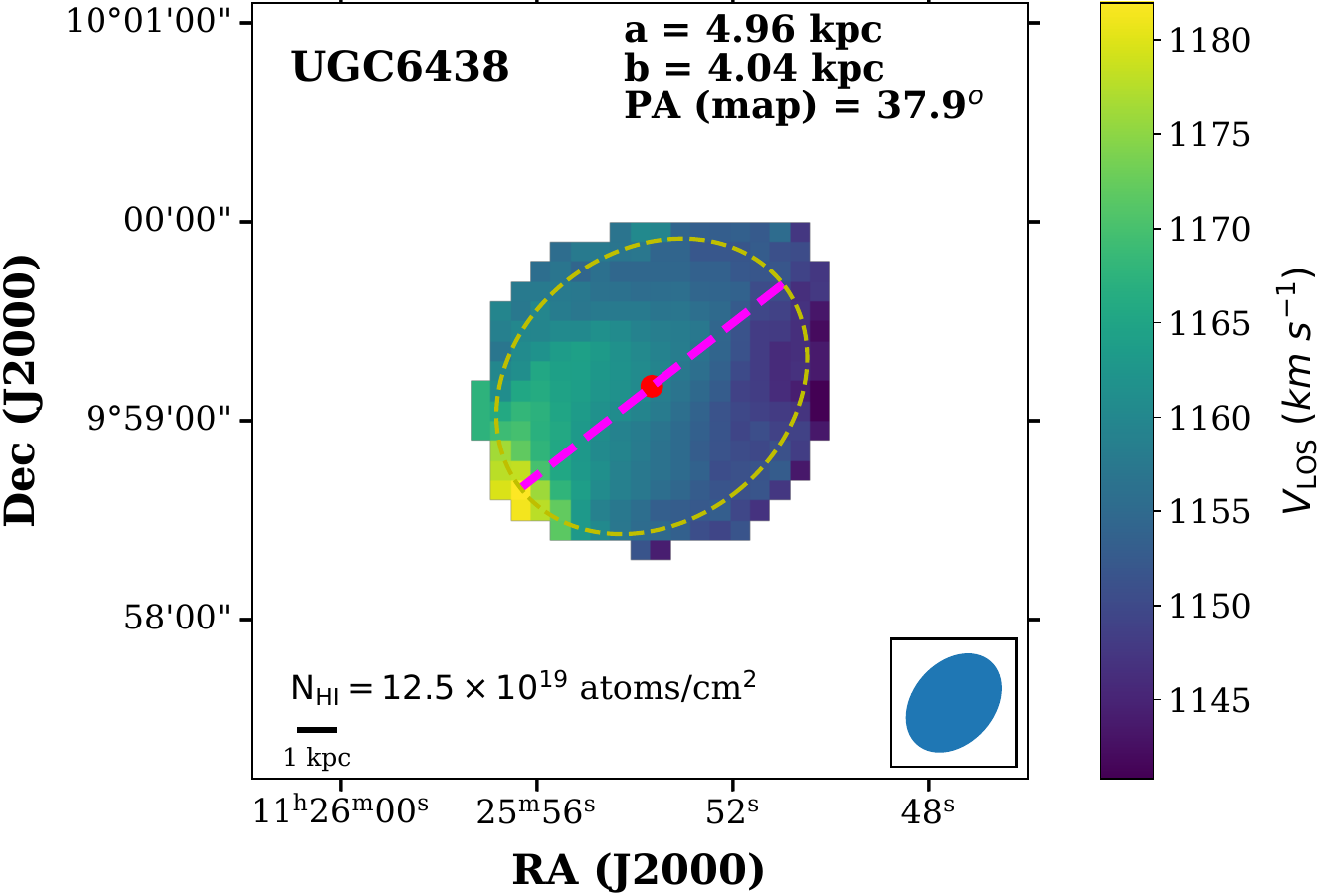}} \\

        {\includegraphics[width=0.30\linewidth]{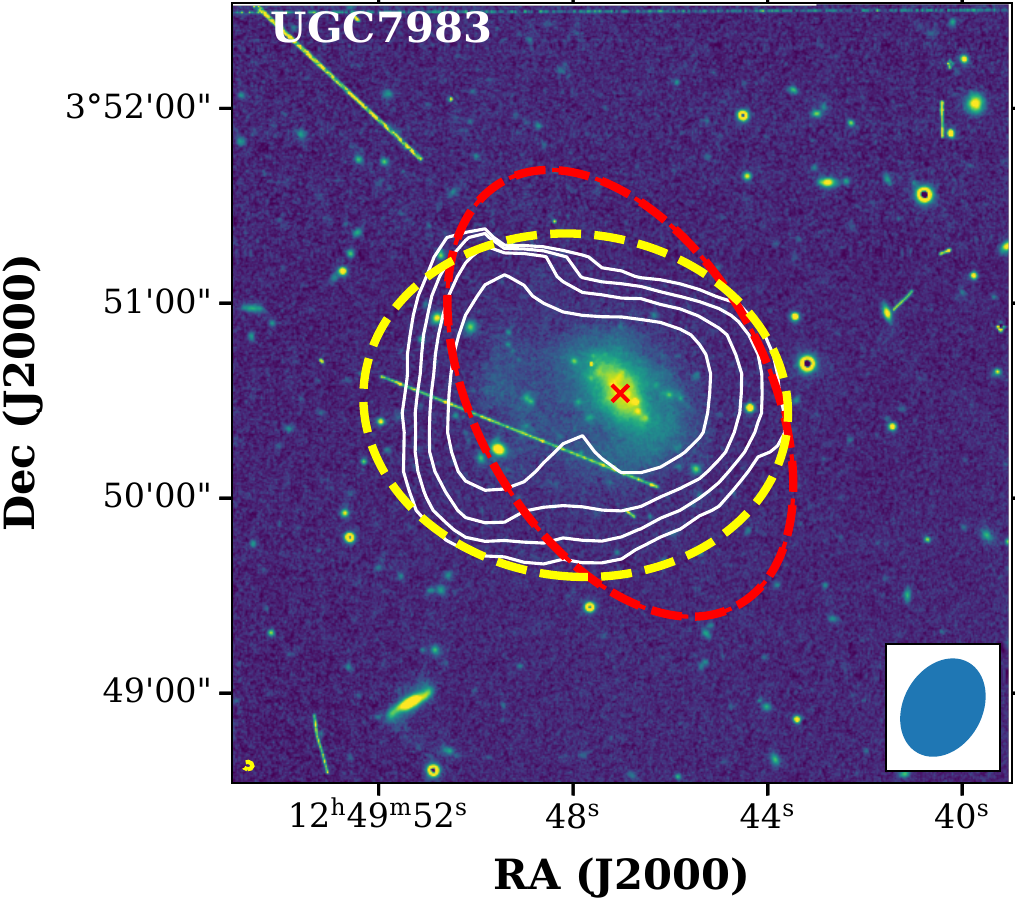}} &
        {\includegraphics[width=0.30\linewidth]{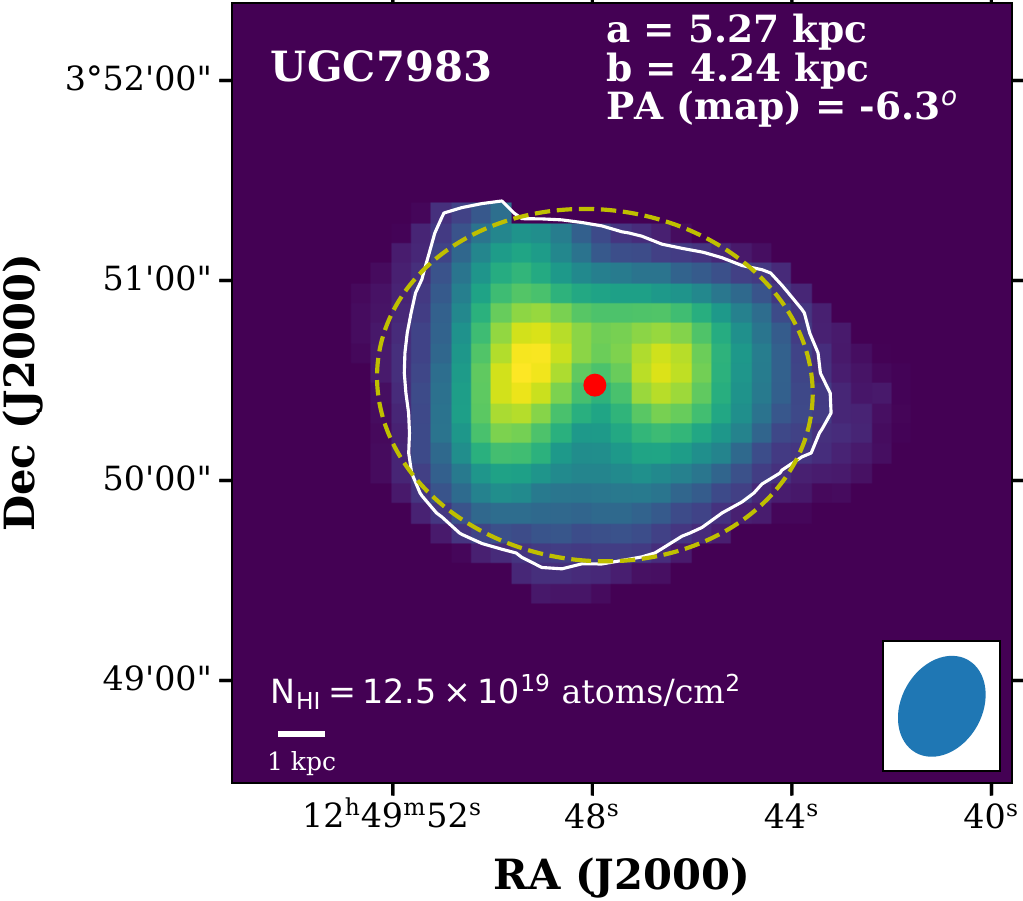}} &
        {\includegraphics[width=0.38\linewidth]{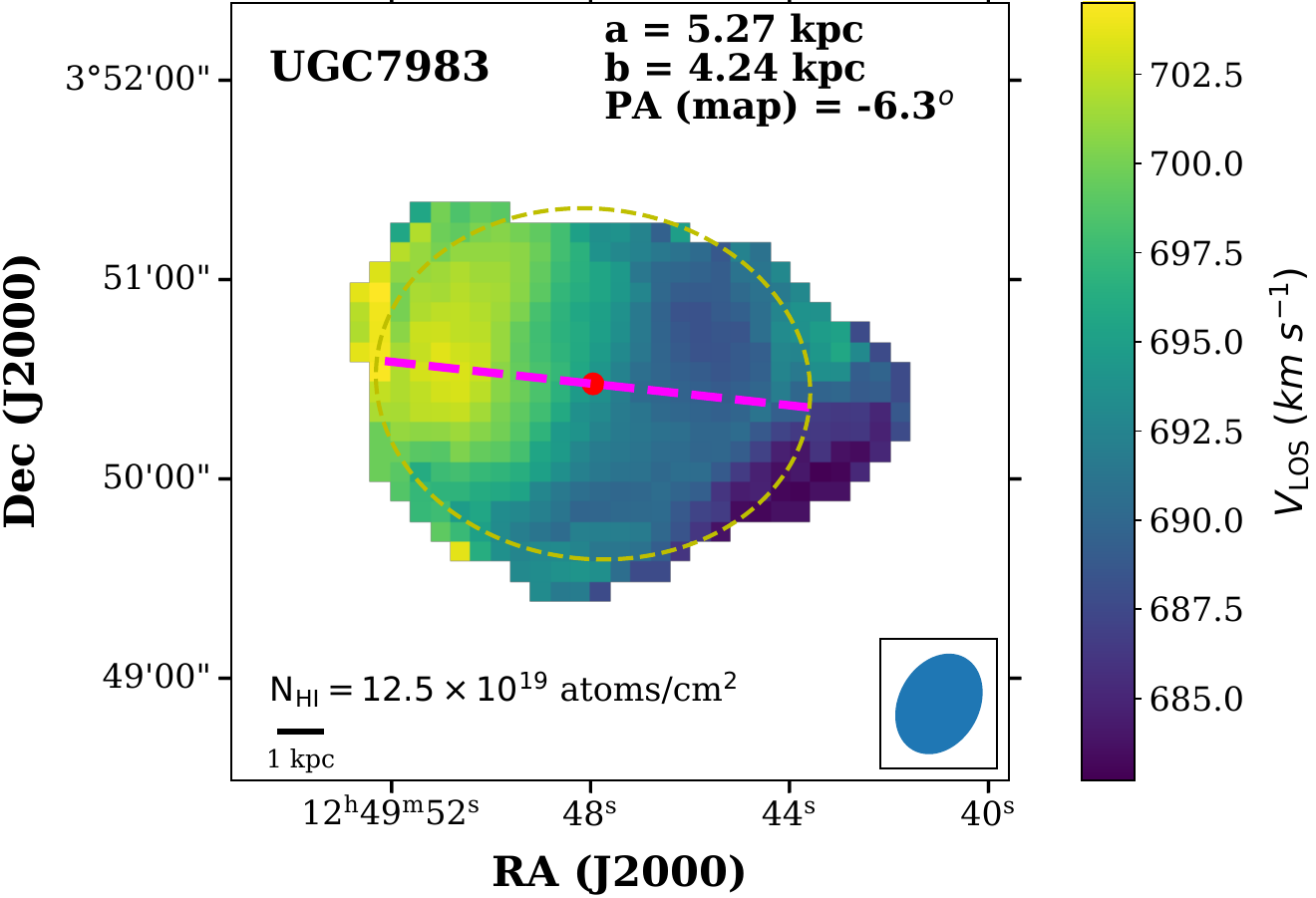}} \\

        {\includegraphics[width=0.30\linewidth]{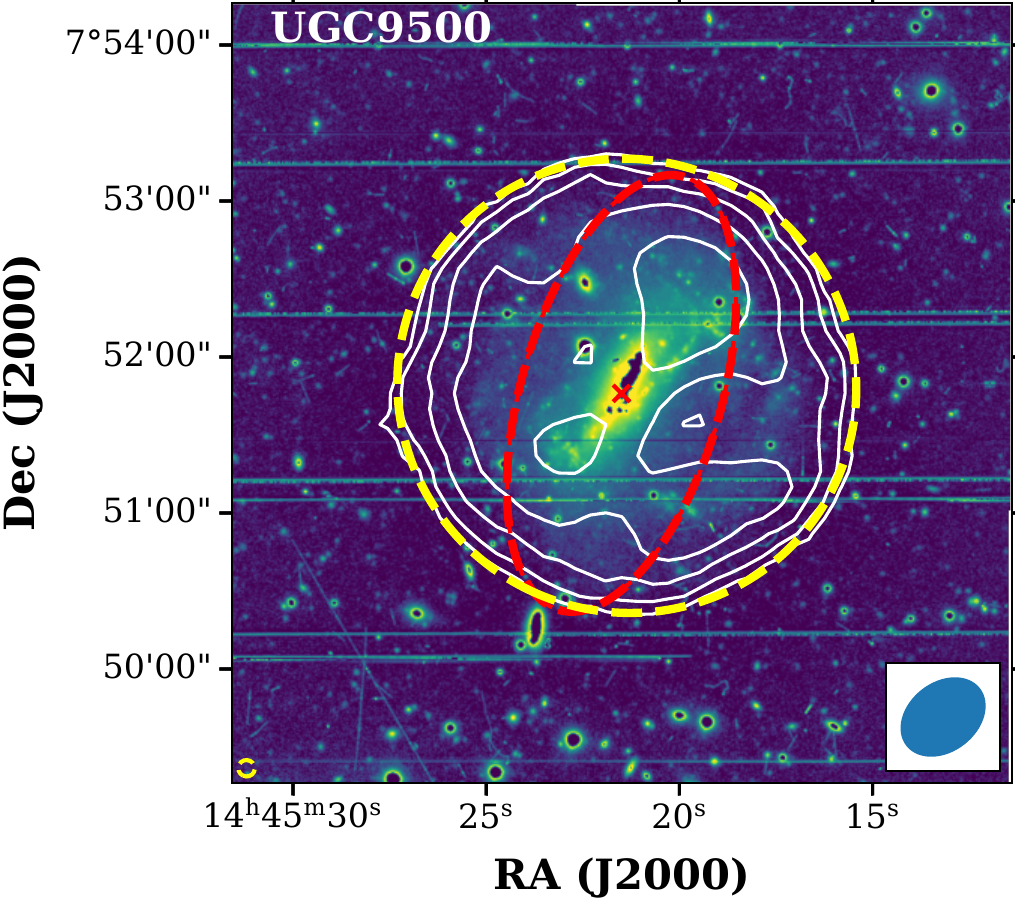}} &
        {\includegraphics[width=0.30\linewidth]{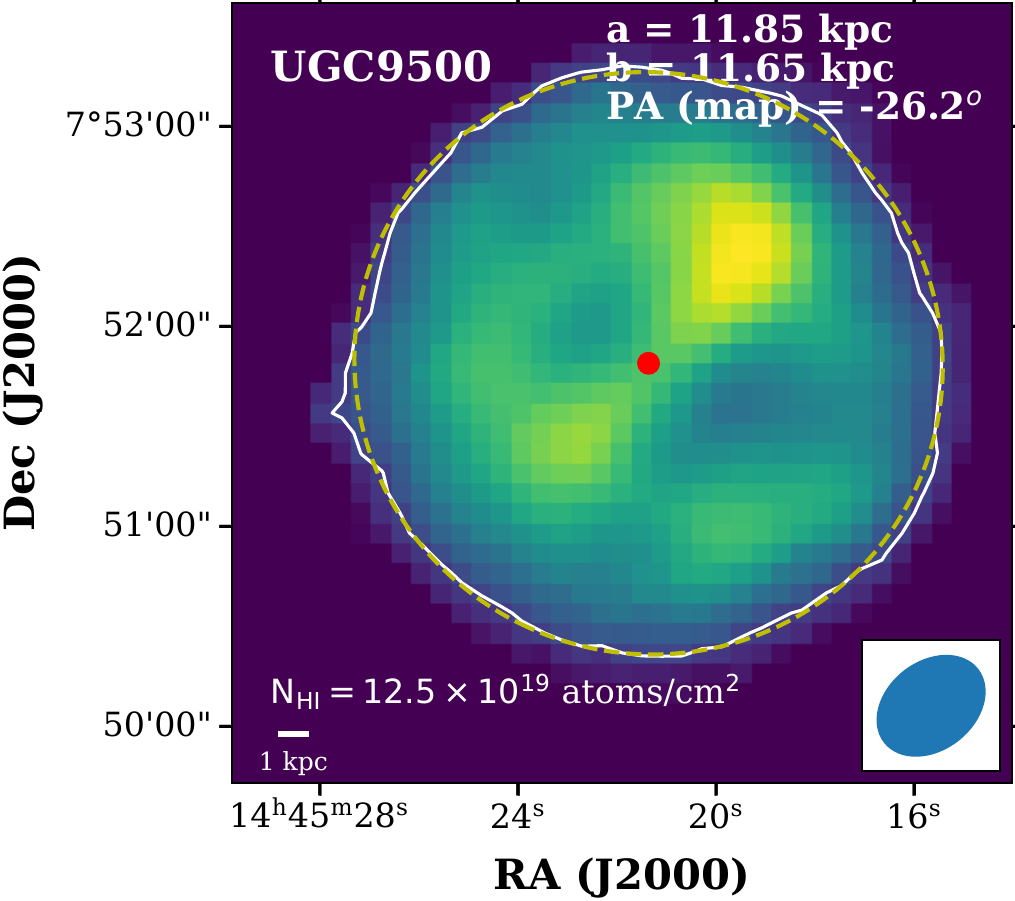}} &
        {\includegraphics[width=0.38\linewidth]{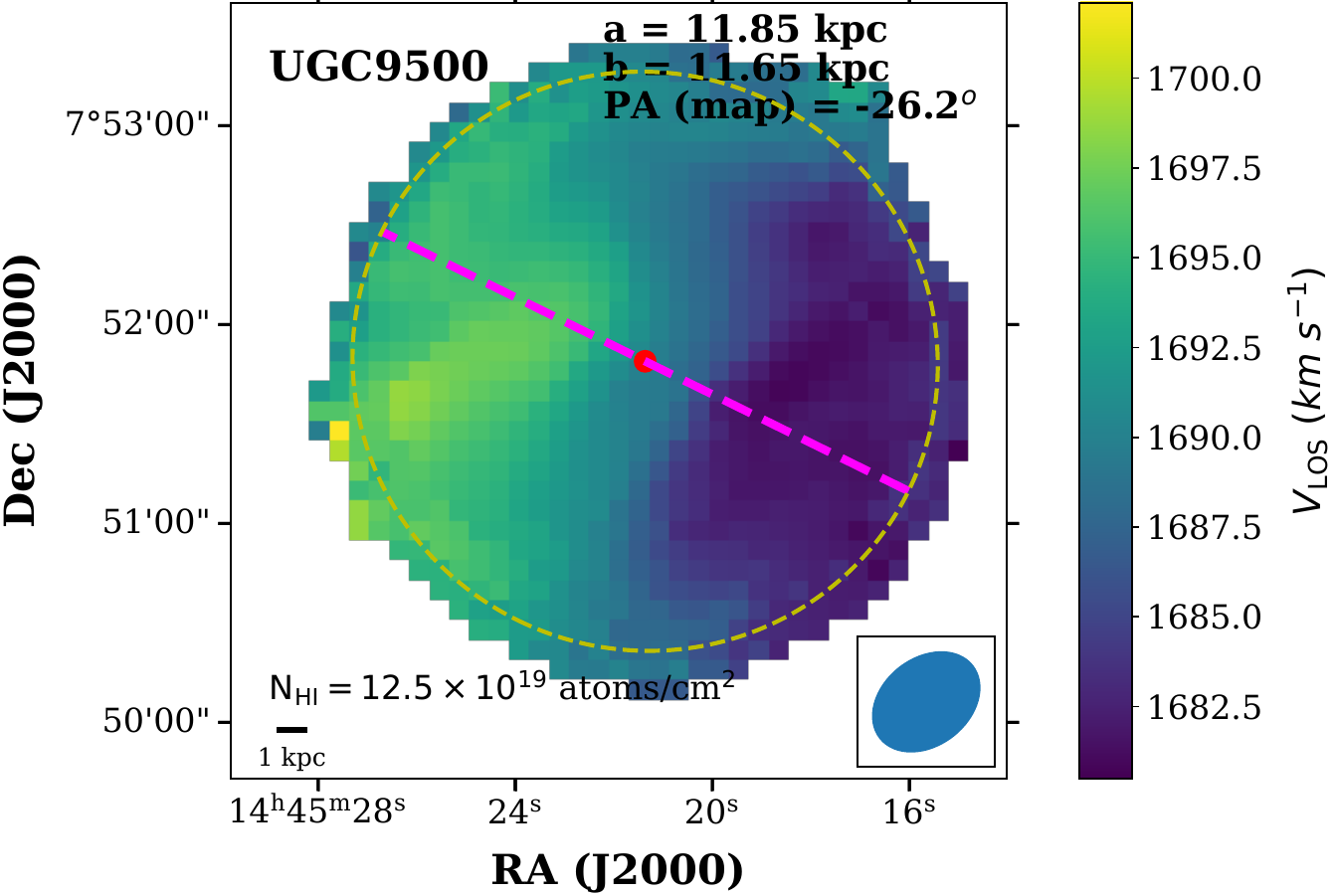}} \\

        {\includegraphics[width=0.30\linewidth]{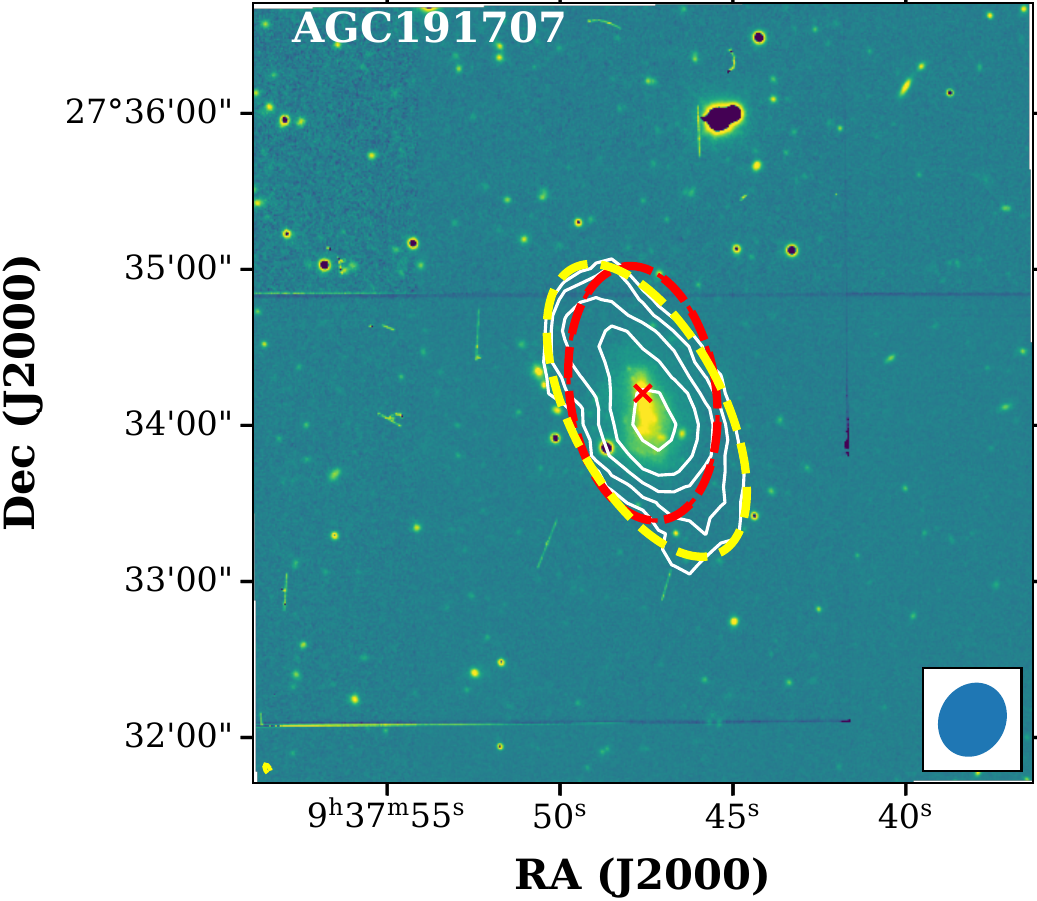}} &
        {\includegraphics[width=0.30\linewidth]{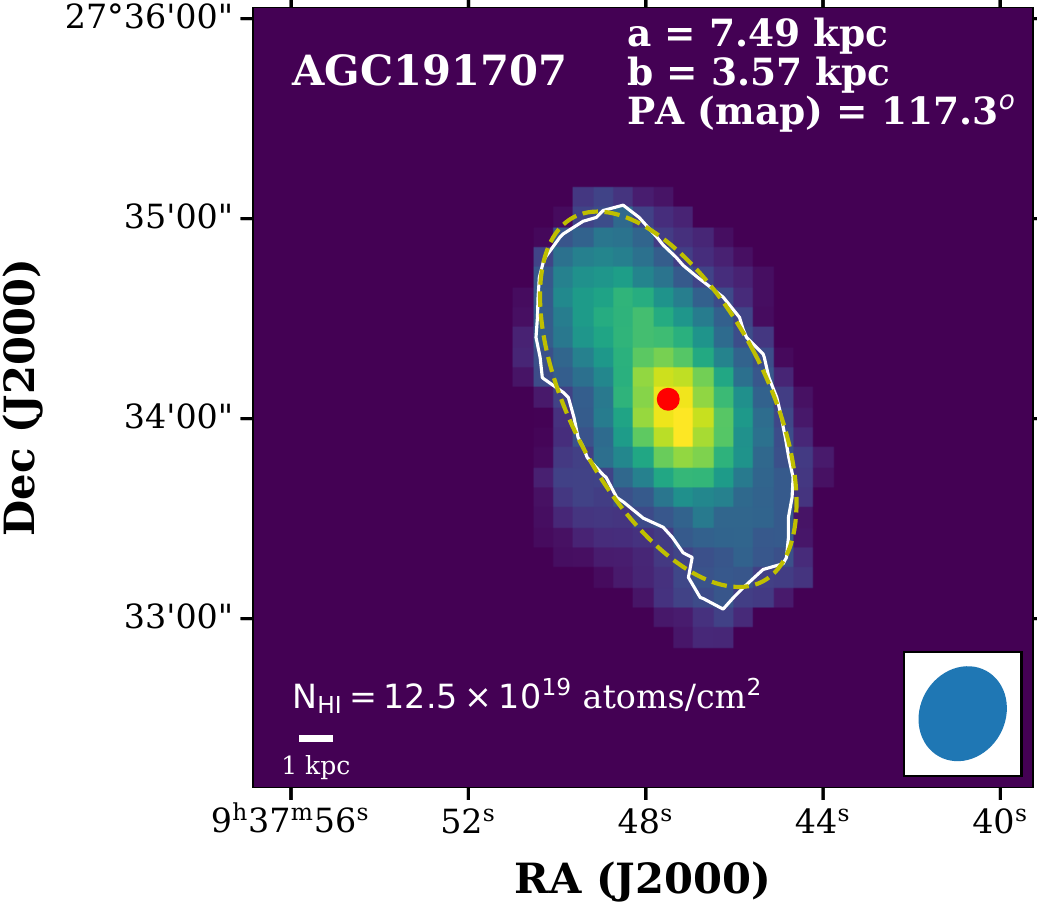}} &
        {\includegraphics[width=0.38\linewidth]{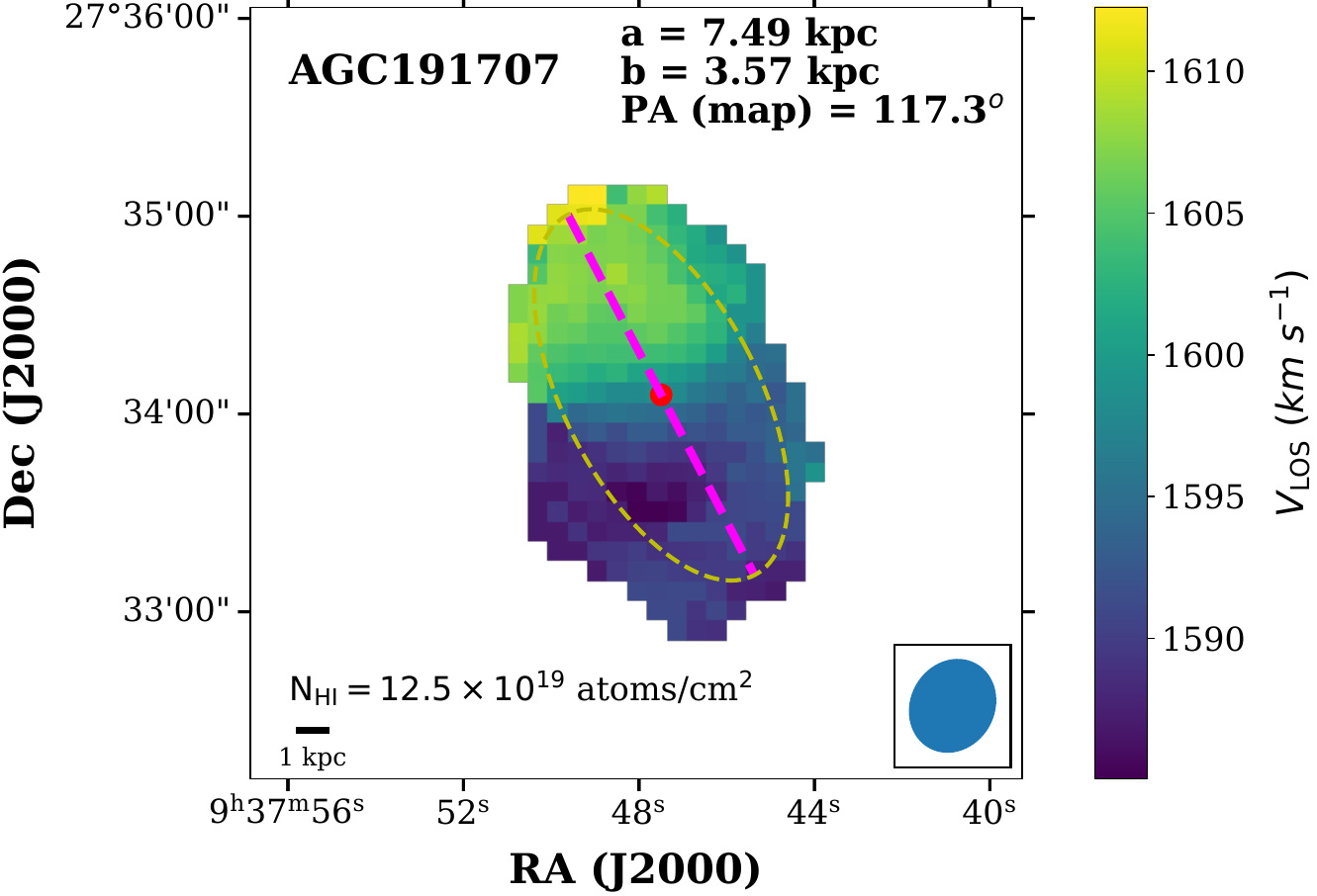}} \\
    \end{tabular}
\end{figure*}

\begin{figure*}
    \ContinuedFloat
    \centering
    \caption{Continued}
    \begin{tabular}{ccc}
        {\includegraphics[width=0.30\linewidth]{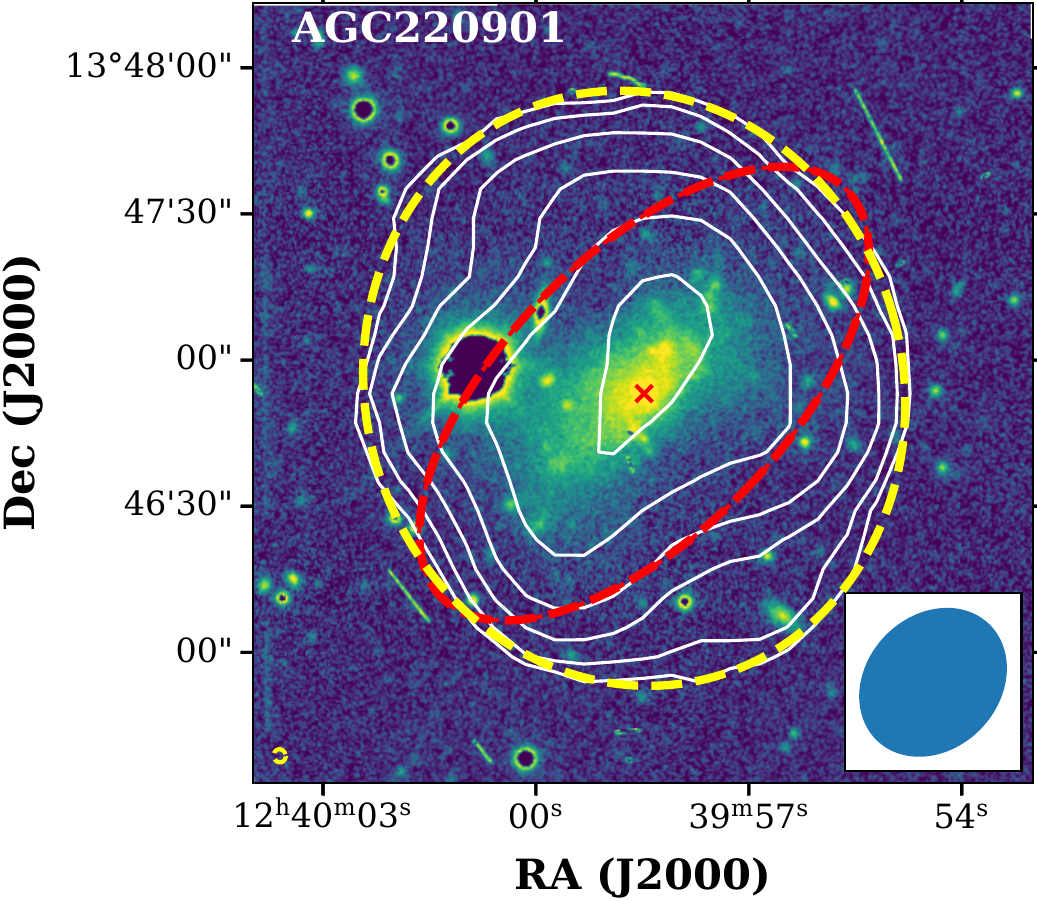}} &
        {\includegraphics[width=0.30\linewidth]{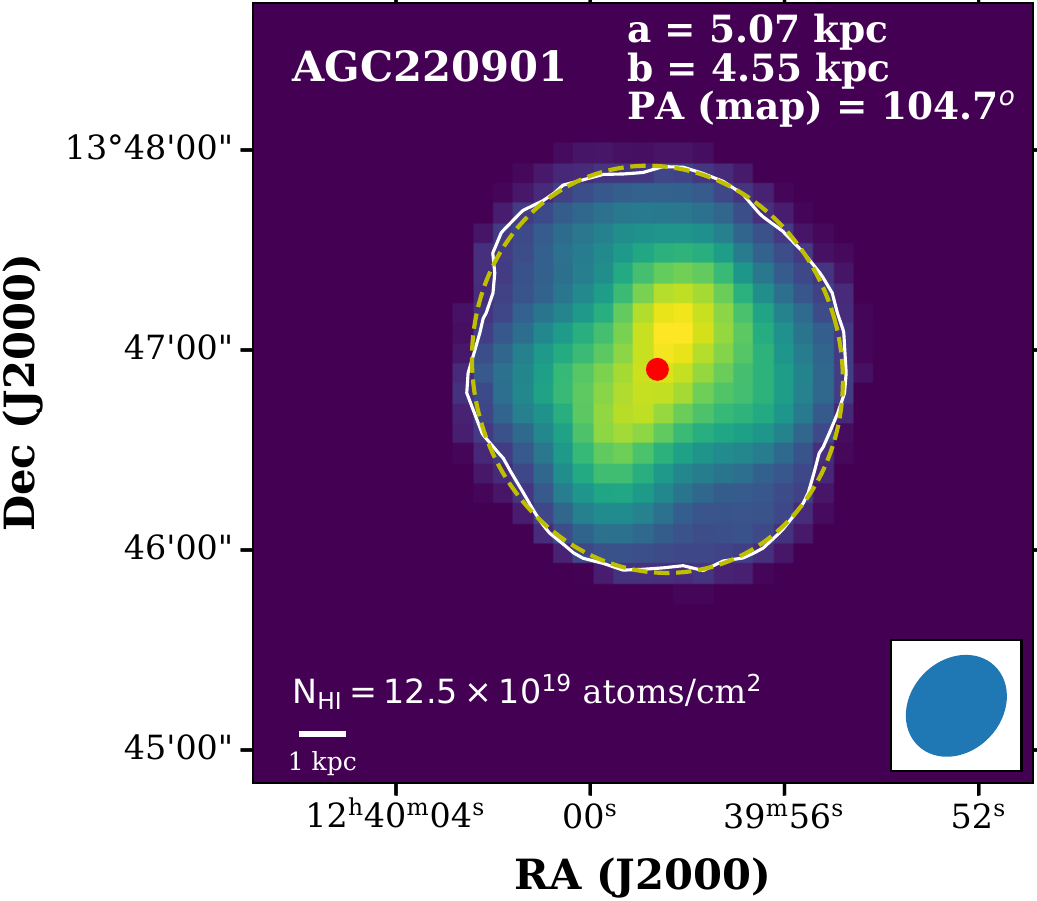}} &
        {\includegraphics[width=0.38\linewidth]{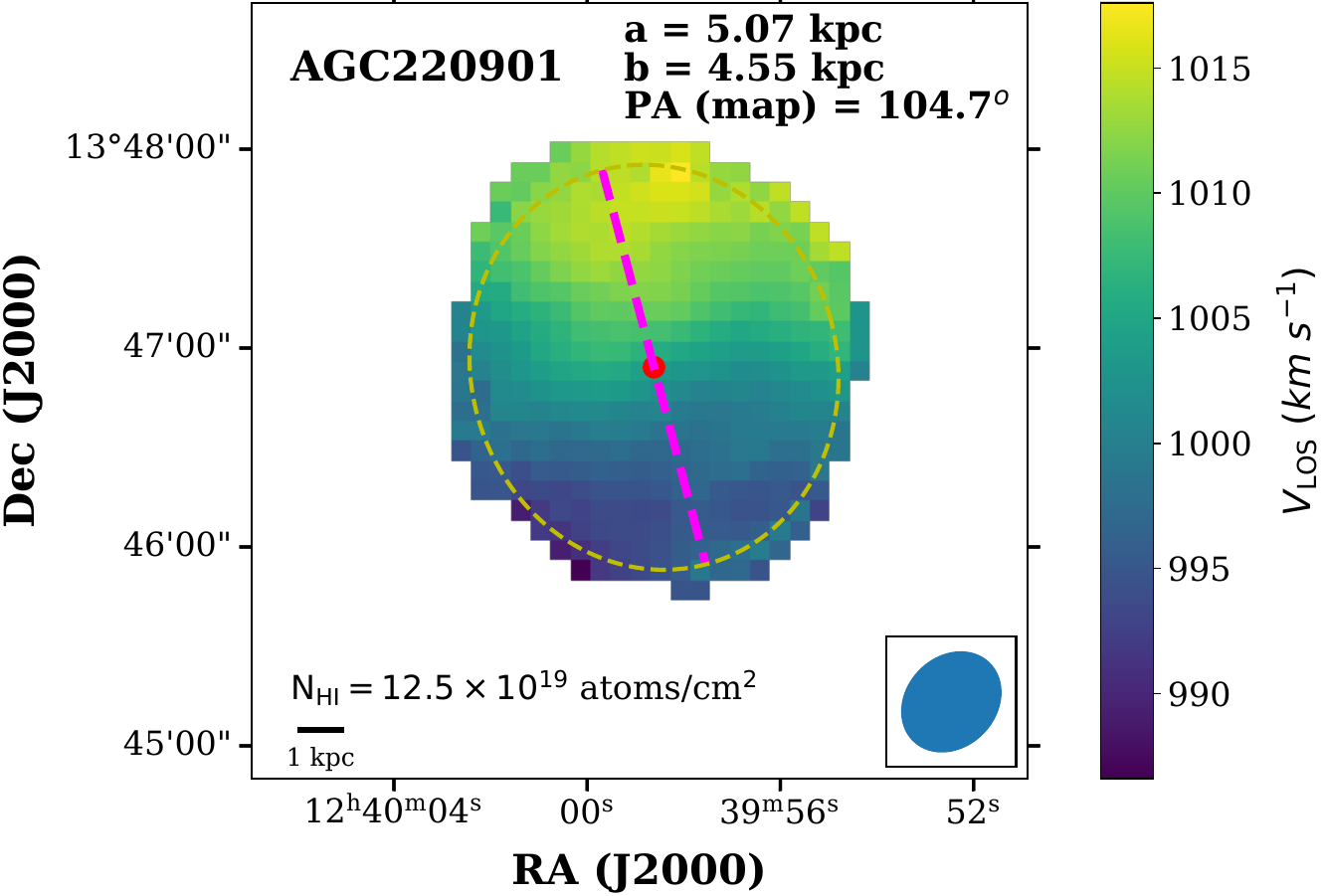}} \\

        {\includegraphics[width=0.30\linewidth]{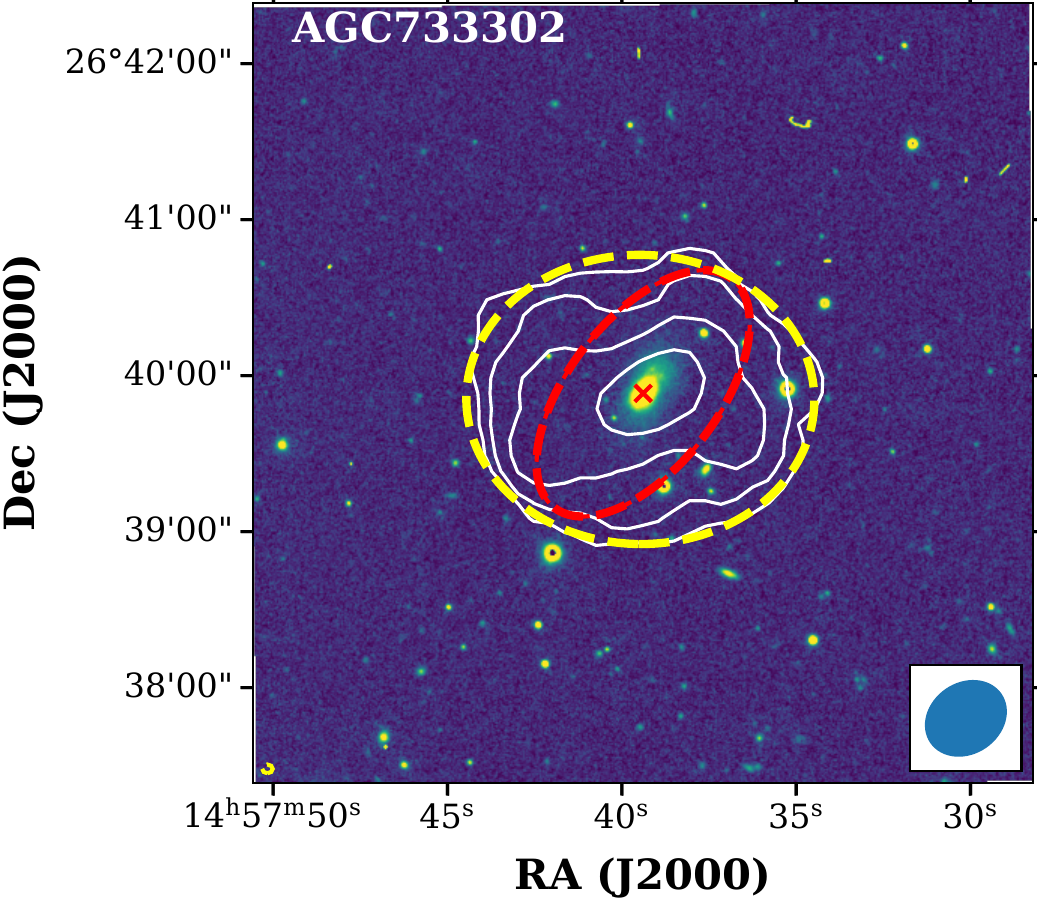}} &
        {\includegraphics[width=0.30\linewidth]{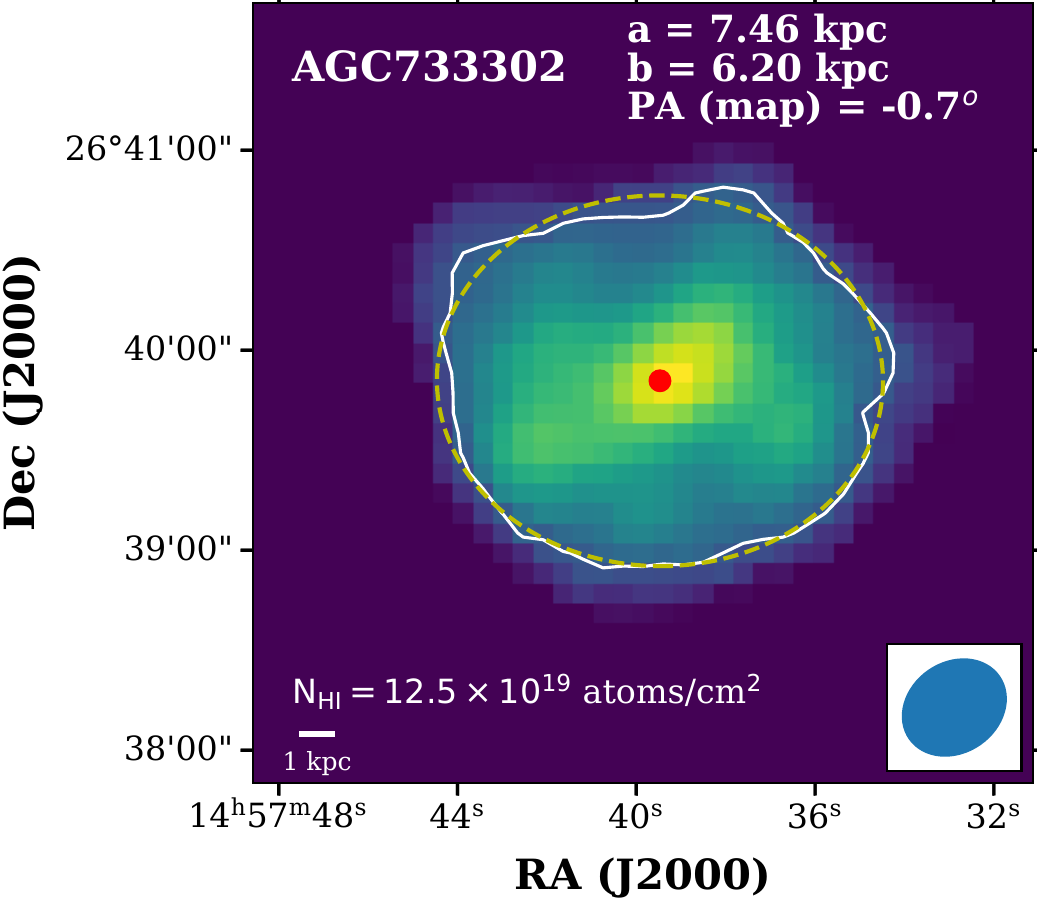}} &
        {\includegraphics[width=0.38\linewidth]{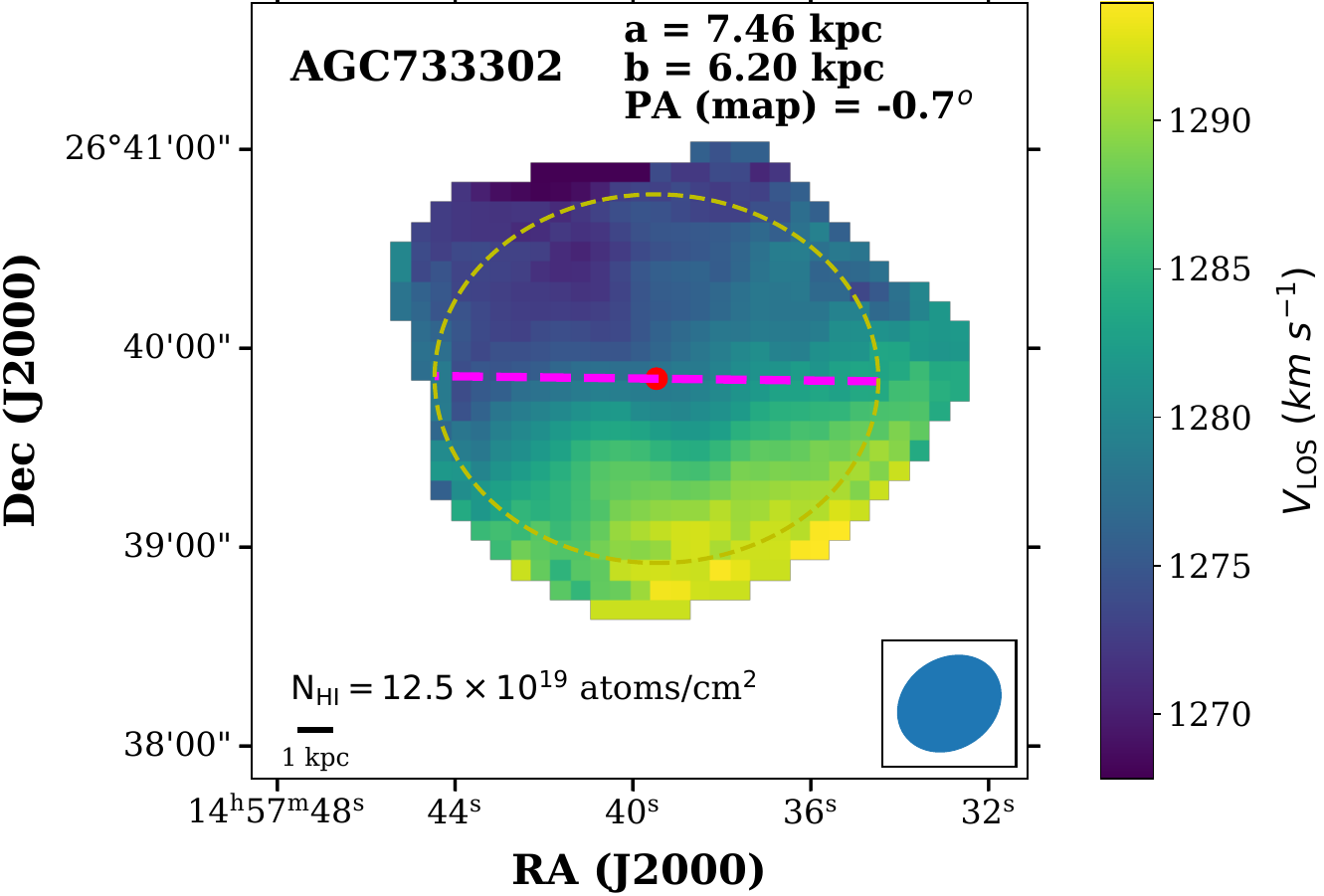}} \\
    \end{tabular}
\end{figure*}

To estimate the \HI~sizes of our sample galaxies, we use the \HI~total-intensity (moment-0) maps produced from the spectral cubes using the \texttt{SoFiA2} masks. Along with the total-intensity maps, we also generate the velocity field (moment-1) and velocity-dispersion (moment-2) maps. In Fig.~\ref{tab:combined_images}, we present the moment-0 maps for our sample galaxies (middle panels). We adopt an iso–surface-density level of 1~\mspsqpc\ (\nhi\ $= 1.25\times10^{20}$ atoms cm$^{-2}$) to define the \HI\ radius. This threshold is widely used in the literature as a standard measure of the \HI\ extent \citep[e.g.][]{broeils97, swaters02, wang13, wang16}.

To measure the \HI\ radius for each system, we fit the 1~\mspsqpc\ isophotal contour with an ellipse, assuming that a circular isophote in the galaxy plane appears as an ellipse when projected onto the sky. The major-axis length of the fitted ellipse is taken as the \HI\ diameter of the galaxy. The contours and corresponding fits are shown in Fig.~\ref{tab:combined_images} by the solid red and dashed yellow lines, respectively. As seen from the figure, the fitted ellipses capture the isophotal shapes well, indicating that the inferred \HI\ radii are reliable.

In Table~\ref{tab:corrected_radii}, we list the \HI\ radii derived from the resolved maps. For comparison, we also include (in parentheses) the radii estimated by \citet{guo20}.
As shown in the table, the \HI\ radii derived from the resolved maps differ slightly from those inferred via the mass–size relation.

\begin{table}
    \centering
    \caption{Galaxy geometry: Column 2 gives radii of the \HI\ disk upto 1 \mspsqpc\ boundary, taken from the major axes of ellipse fitted in Fig.~\ref{tab:combined_images}. Values in parenthesis denote the same as given by \citet{guo20} using the scaling relation. Column 3 gives the axial ratio $\frac{b}{a}$, the ratio of minor to major axis length as adapted from the fitting shown in Fig.~\ref{tab:combined_images}, with values in parenthesis giving the same ratio as used by \citet{guo20} obtained using optical images.}
    \label{tab:corrected_radii}
    \begin{tabular}{|c|c|c|} \hline 
         Galaxy Name & \rhi\ & $\frac{b}{a}$ \\ 
         & Ours \citep{guo20} & Ours \citep{guo20} \\ 
         & kpc & \\ \hline
         UGC 6438 & 4.96$\pm$1.2 (5.59) & 0.8145 (0.524) \\  
         UGC 7983 & 5.27$\pm$1.0 (6.0) & 0.8 (0.6) \\  
         UGC 9500 & 11.85$\pm$1.8 (11.6) & 0.98 (0.444) \\ 
         AGC 191707 & 7.49$\pm$1.6 (6.0) & 0.48 (0.57) \\ 
         AGC 220901 & 5.07$\pm$1.1 (4.86) & 0.817 (0.486) \\ 
         AGC 733302 & 7.46$\pm$1.5 (6.21) & 0.831 (0.515) \\ \hline
    \end{tabular}
\end{table}

\begin{figure}
    \centering
    \includegraphics[width=0.70\linewidth]{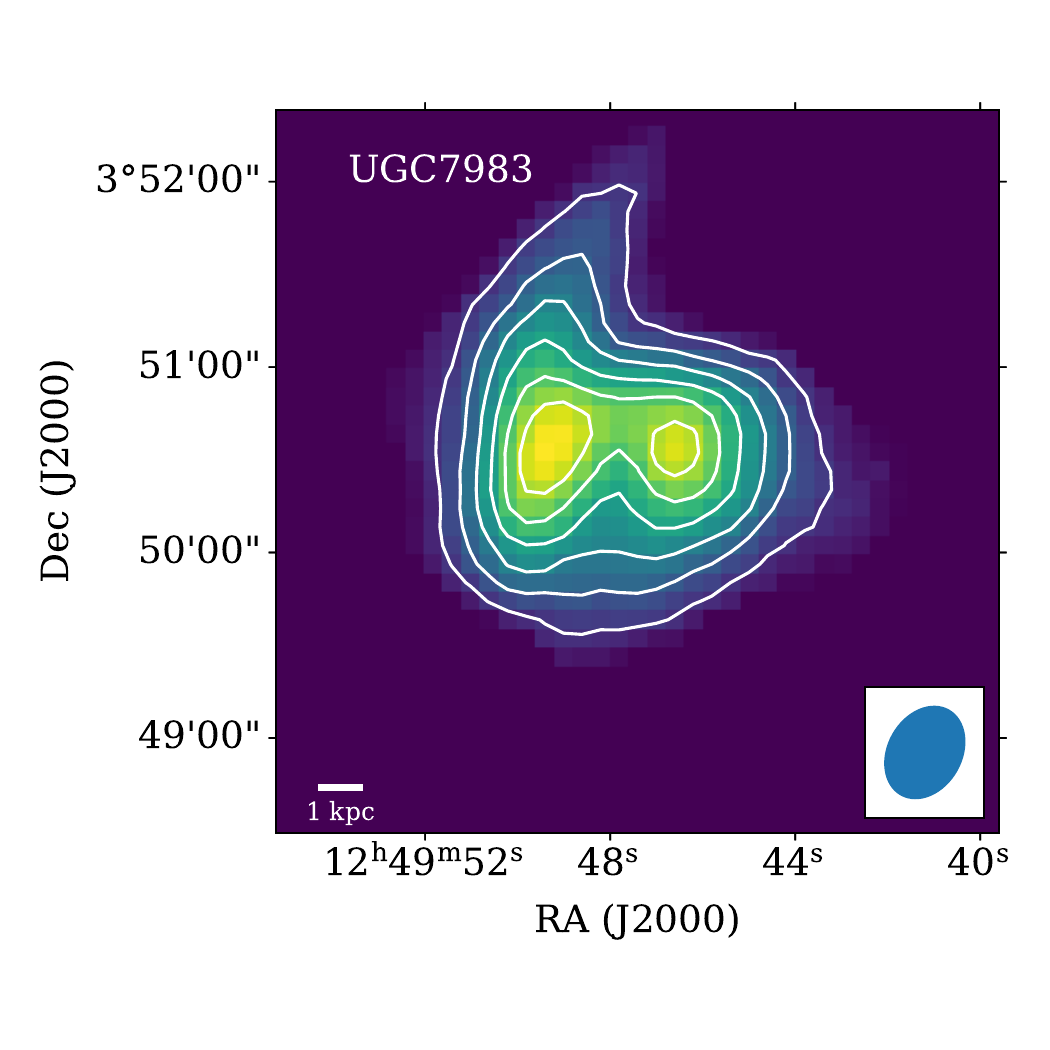}
    \caption{UGC 7983 Moment-0 map showing extended structure which was blanked out in Fig.~\ref{tab:combined_images} in order to determine ellipticity. }
    \label{fig:7983_full}
\end{figure}

It is worth noting here that UGC~7983 is a possible member of the Virgo Cluster \citep[][]{kim14a}, located near the NGC~4636 group, and shows clear signs of tidal disturbance, likely caused by interactions with NGC~4586 \citep[][]{karachentsev13, lin23}. To obtain a reliable estimate of its \HI\ size, we visually blanked the disturbed region of the moment-0 map and fitted ellipse to the isophotal contour of 1~\mspsqpc. The moment-0 map shown in Fig.~\ref{tab:combined_images} corresponds to the blanked version used for the size estimation, while Fig.~\ref{fig:7983_full} presents the full, unblanked moment-0 map for comparison where the contour levels plotted clearly indicate the shape more in line with that described by Fig.~\ref{tab:combined_images}.

In the left panels of Fig.~\ref{tab:combined_images}, we compare the \HI\ discs of our sample galaxies with their optical discs by overlaying the \HI\ contours (solid white lines; starting from $\Sigma_{\rm H\textsc i}$ = 1~\mspsqpc, with linear spacing upto the maximum emission) on the optical $r$-band DECam images. The 1~\mspsqpc\ (outermost) contours are fitted with ellipses shown as dashed yellow lines, while the dashed red lines represent the \HI\ disks estimated by \citet{guo20} with size determined by scaling relation, and inclination determined using axial ratios b/a given in parenthesis in the third column of Table~\ref{tab:corrected_radii}. It should be noted that \citet{guo20} assumed the geometric parameters of the \HI\ discs, such as the axial ratios and position angle, to be identical to those of the optical discs. However, numerous studies have shown that \HI\ discs can have geometric parameters that differ significantly from their optical counterparts \citep[e.g.][]{bosma81, begeman87phd, swaters02, wang13, lelli16a}. This behaviour is also evident in our sample: the yellow dashed and red dashed ellipses often show clear mismatches, indicating substantial differences between the inclinations and position angles derived from optical and \HI\ datasets.

Because these geometric parameters play a crucial role in determining the circular velocity and, consequently, the dynamical mass, such discrepancies must be treated carefully. To investigate this further, the rightmost column of Fig.~\ref{tab:combined_images} presents the \HI\ velocity fields (moment-1 maps) of our sample galaxies. The magenta dashed lines mark the major axes derived from our isophotal fits to the \HI\ total-intensity maps. As seen in the figure, the \HI-derived major axes (magenta) align well with the kinematic axes of the galaxies, while the optical major axes are considerably misaligned in several systems. However, for AGC~733302, the kinematic axis shows a notable offset from the major axis inferred from the \HI\ intensity contours, likely due to the patchy and irregular nature of the \HI\ emission, typical of dwarf irregular galaxies, which can distort isophotal shapes.

Nevertheless, we emphasize that our dynamical analysis is more sensitive to the ellipticity (and thus the inclination) of the \HI\ contours than to their position angles. Changes in ellipticity directly affect the inferred inclination, which in turn modifies the deprojected velocity width and hence the estimated dynamical mass.

\subsection{Rotational velocity}
To estimate the dynamical mass of a galaxy from \HI\ observations, one must dynamically model the \HI\ disc. A widely used approach is the tilted-ring method, in which the \HI\ disc is represented as a set of concentric rings, each characterised by its own geometric parameters (inclination, position angle, centre) and kinematic parameters (systemic velocity, rotational velocity, etc.). These parameters are then optimised to reproduce the observables. Traditionally, such modelling was performed using the 2D velocity field. However, more recently, full 3D modelling, where the entire spectral cube is fitted, is preferred, as it yields more robust kinematic parameters \citep[e.g.][]{diteodoro15,oh18,kamphuis15}. Ideally, a tilted-ring analysis using the full \HI\ cube provides the most reliable rotation curve and thus the dynamical mass at the outermost fitted radius.

In practice, however, such modelling requires high-quality data, i.e., good sensitivity and a sufficiently large number of independent beams across the major axis to constrain the ring parameters reliably. For example, \texttt{$^{3D}$Barolo} requires at least $\sim$3 beams across the semi-major axis to obtain stable fits, while \texttt{FAT} (Fully Automated TiRiFiC) requires $\geq$ 8 beams across major axis. Although our sample galaxies are expected to have large enough \HI\ discs when the size is calculated from the mass-size relation, in practice, we recovered only a few beams across due to their low surface brightness. Reducing the beam size results in resolving out the diffuse emission and a shrinking \HI\ disc. With the optimal resolution, all our attempts at tilted-ring modelling failed to converge satisfactorily.

We also attempted the recipe suggested by \citet{mancera20}, designed specifically for faint dwarf galaxies. Their method fixes several geometric parameters, such as the position angle, centre, and systemic velocity, using position–velocity slices extracted with \texttt{KPVSLICE} from the \texttt{KARMA} package \citep{gooch96ip}, followed by 3D fitting with \texttt{$^{3D}$Barolo}. This approach also failed to produce a meaningful kinematic model for our galaxies due to the patchy nature of diffuse gas.

We further attempted several 2D approaches with fewer free parameters, such as \texttt{DiskFit} \citep[][]{sellwood15}, which fits a global disc model, and \texttt{kinemetry} \citep[][]{krajnovic06}, but these methods also did not converge for our sample. One possible reason is the intrinsic thickness of dwarf-galaxy \HI\ discs, which are known to be substantially thicker than those of massive spirals (e.g. \citealt{roychowdhury10}). The thick disc produces strong line-of-sight projection effects that mix velocities from different heights, making it difficult to recover circular rotation curves using 2D projections.

It is important to note, however, that full kinematic modelling at all radii is only required to infer the radial distribution of the dynamical mass. To estimate the total dynamical mass, one needs only the circular velocity at the outer edge of the \HI\ disc, rather than a complete rotation curve. This can be derived directly from the global \HI\ spectrum or by identifying the velocity range over which \HI\ emission is detected in the channel maps.

To this end, we inspect the channel maps of each galaxy to determine the velocity interval over which \HI\ emission is present. This velocity range represents the projected rotational width of the disc. In Fig.~\ref{fig:combined_chanplots}, we show the channel maps of our sample galaxies. Contours begin at $3\sigma$ of the noise level. As can be seen, the \HI\ emission emerges in the lowest-velocity channels and fades away in the highest-velocity channels. The yellow ellipses mark the \HI\ disc as determined from the ellipse-fitting procedure. All detected emission lies within these ellipses, confirming that the channel-map emission originates from the \HI\ disc. We therefore define the projected velocity width of each galaxy as the full velocity range over which \HI\ emission is detected. This width depends upon the sensitivity of the image, particularly to diffuse and low surface-brightness emission; however, we find that the velocity width remains consistent for emission above $\Sigma_{\rm H\textsc i}\ = 1\ \mspsqpc$.

We mention these velocity ranges for each galaxy in the third column of Table~\ref{tab:vel_corrections} and show them as grey shaded region lines in Fig.~\ref{fig:combined_spectra}. As seen in the figure, the starting and ending velocity limits encompass nearly the entire global spectrum, indicating that the velocity width inferred from the channel maps is fully consistent with that obtained from the integrated \HI\ spectrum.

In several earlier studies, many authors have used the parameter $w_{\rm 20}$ to estimate the projected velocity width of galaxies. Here, $w_{\rm 20}$ is defined as the velocity width measured at the 20\% level of the peak flux of the global \HI\ spectrum. For instance, \citet{guo20} adopted $w_{\rm 20}$ as the velocity width for their sample, as resolved \HI\ spectral cubes were not available and the global spectrum provided the only measurable proxy.

For our galaxies, we also compute the $w_{\rm 20}$ values, which are shown as green vertical dashed lines in Fig.~\ref{fig:combined_spectra}. As is apparent, $w_{\rm 20}$ systematically traces a much smaller velocity range than the full width in the channel maps. This naturally leads to an underestimation of the rotational velocity and, therefore, an underestimation of the dynamical mass, potentially giving the false impression of baryon domination.

Furthermore, in several cases, the single-dish spectra exhibit noticeable baseline offsets (e.g., UGC~6438, AGC~191707, AGC~733302). For such systems, the measured $w_{\rm 20}$ can be artificially overestimated, which would inflate the inferred dynamical mass. Thus, both underestimation and overestimation are possible when relying solely on global spectra, highlighting the necessity of resolved \HI\ observations with well-calibrated flux scales.

Our velocity-width measurements, derived directly from the resolved channel maps, therefore provide a more reliable estimate of the projected rotational width for our sample galaxies. These measurements are listed in the fourth column of Table~\ref{tab:vel_corrections}. For comparison, we also list the $w_{\rm 20}$ values measured from our own global \HI\ spectra, which, as expected, are systematically lower than the channel-map–derived widths. Additionally, we show the velocity widths reported by \citet{guo20} from Arecibo spectra (values in parentheses). As can be seen, the velocity widths derived from the resolved channel maps differ considerably from those measured using global spectra, such as the $w_{\rm 20}$ width. 
The errors are given based on fainter emission seen in neighbouring channels which does not show geometrical continuity with the preceding channels, as this can originate from noise.

Next, to derive the dynamical mass of a galaxy, we must determine its circular velocity in the galaxy plane. This requires deprojecting the observed velocity width. The deprojected velocity width is obtained using

\begin{equation}
 \mathrm{Corrected}\ V_{\rm H\textsc i}
 = \frac{\mathrm{Uncorrected}\ V_{\rm H\textsc i}}{\sin(i)}
 = \frac{\mathrm{Uncorrected}\ V_{\rm H\textsc i}}
 {\sqrt{\frac{1 - (b/a)^2}{1 - {q_0}^2}}} ,
 \end{equation}

\noindent where the corrected $V_{\rm H\textsc i}$ is half of the full velocity width after deprojection, and the uncorrected value is half of the velocity width measured in the sky plane. Inclination is therefore a critical parameter in this calculation. The inclination can be estimated from the observed axial ratio of the galaxy using
\begin{equation}
 \sin (i) = \sqrt{ \frac{1 - (b/a)^2} {1 - q_0^2} } ,
 \end{equation}

\noindent where $b/a$ is the observed axial ratio, and $q_0$ is the intrinsic axial ratio of the disc. We estimate $b/a$ for our sample by fitting ellipses to the 1~\mspsqpc\ \HI\ iso-surface density contours (see, e.g., Fig.~\ref{tab:combined_images}). The resulting $b/a$ values are listed in the middle panels of the figure. These $b/a$ values are also listed in the third column of Table~\ref{tab:corrected_radii}. For comparison, we also list the $b/a$ values of \citet{guo20} in the parentheses. As can be seen, \citet{guo20} systematically underestimated the $b/a$. They used optical r-band image from SDSS DR7 to fit isophots and obtain $b/a$. 

It should be noted that \HI\ discs typically extend to much larger radii than optical discs, often by factors of 2–4 in dwarf galaxies \citep[e.g.][]{swaters02, hunter12}. This effect is especially pronounced in low-mass systems, where the optical emission is confined to the central regions. Moreover, because star formation in dwarf galaxies is highly patchy and stochastic, the central optical light often fails to trace the true shape of the underlying disc \citep[e.g.][]{roychowdhury10, hunter06ip}. A clear example from our sample is AGC~220901 (Fig.~\ref{tab:combined_images}, fifth row). For this galaxy, the $b/a$ values derived from \HI\ (0.817) and from optical imaging (0.486) differ substantially. As seen in the figure, the optical emission is limited to the central region and exhibits an elongated structure that resembles a bar. In contrast, the \HI\ distribution is considerably more extended and displays a roughly circular morphology.

\begin{figure}
    \centering
    \includegraphics[width=0.8\linewidth]{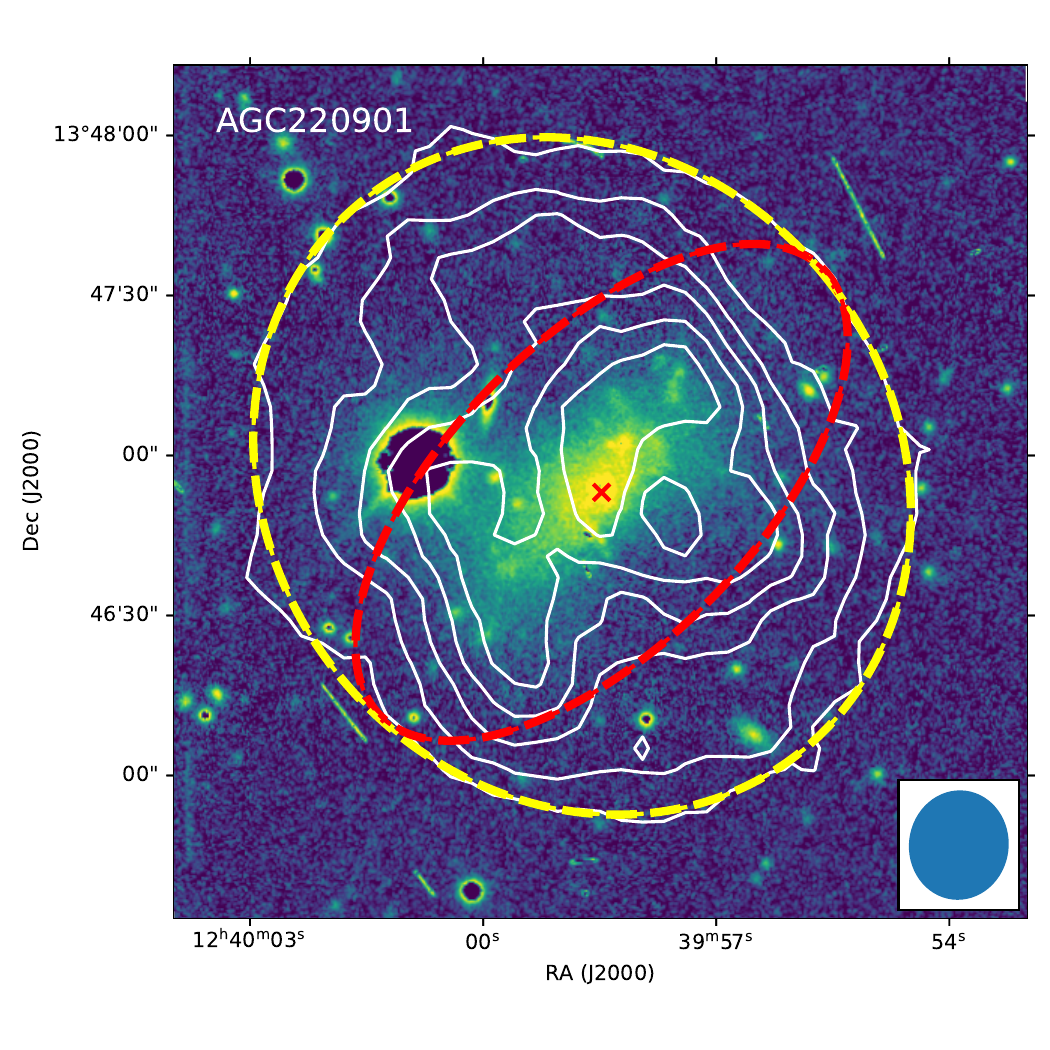}
    \caption{AGC220901: Contours from a high resolution radio image overlaid on top of DECAM r-band image with outermost contour fitted with an ellipse shown in dashed yellow line. The dashed red line indicates geometry suggested by \citet{guo20}.}
    \label{fig:AGC220901_highres}
\end{figure}

To investigate this further, we imaged AGC~220901 at a higher spatial resolution of $20\arcsec \times 19\arcsec$. The resulting image is shown in Fig.~\ref{fig:AGC220901_highres}. The high-resolution \HI\ contours in the inner region indeed trace the same elongated, star-forming structure seen in the optical. However, beyond this central zone, the contours recover the circular \HI\ disc visible in the lower-resolution map. This clearly demonstrates that optical imaging alone can bias the inferred axial ratio toward the geometry of the central, star-forming component. In contrast, the \HI\ disc traces the global disc structure more reliably and is therefore better suited for determining the inclination of the \HI\ disc \citep[see also][]{begum08b, ott12}. We note that \citet{guo20} systematically underestimated $b/a$ for all galaxies in our sample except AGC~191707 (a highly inclined system). Their underestimated axial ratios lead to overestimated inclinations, which in turn produce underestimated inclination-corrected velocity widths.

To compute the inclination from $b/a$, one must assume an intrinsic axial ratio $q_0$. \citet{guo20} used axial ratios from SDSS $r$-band images and adopted $q_0 = 0.2$, a typical value for massive spiral galaxies \citep[e.g.][]{tully98,giovanelli94}. They also assumed that the inclination of the \HI\ disc is identical to that of the optical disc, an assumption known to be problematic, as \HI\ discs often exhibit different geometric parameters than their stellar counterparts \citep[e.g.][]{begeman87phd, swaters02, wang13}.

Dwarf galaxies, in particular, are known to host significantly thicker \HI\ discs than spirals. Several works \citep[see, e.g.][]{roychowdhury10, johnson17, patra20} found intrinsic axial ratios as high as $q_0 \approx 0.6$ for dwarf \HI\ discs. To account for this uncertainty, we compute inclinations and, therefore, corrected velocity widths, using a range of intrinsic axial ratios: $q_0 = 0.2$, $0.4$, and $0.6$, spanning the range expected for these low mass dwarf galaxies.
We emphasize that inclination is a crucial quantity for estimating the true velocity width and hence the dynamical mass. Changing the assumed $q_0$ from 0.2 to 0.6 can alter the inferred dynamical mass by up to a factor of $\sim 1.5$. The inclination-corrected velocity widths for all three assumed intrinsic axial ratios are reported in the last three columns of Table~\ref{tab:vel_corrections}. 

For comparison, we also quote the inclination-corrected velocity widths obtained by \citet{guo20} (shown in parentheses). As can be seen, their values differ significantly from our estimates. This discrepancy arises from two effects. First, their velocity widths were measured from single-dish spectra, which differ from the widths we obtain using resolved channel maps. Second, the deprojection applied by \citet{guo20} is based on different axial ratios ($b/a$) derived from optical images and on an assumed intrinsic axial ratio ($q_0$) of 0.2. Both of these choices lead to inclination estimates that differ from those derived from our resolved \HI\ maps, and hence result in different corrected velocity widths.

\begin{table*}
    \centering
    \caption{Velocity Corrections: column 2 gives velocity width at 20\% the maximum flux in our analysis as shown in Fig.~\ref{fig:combined_spectra}; column 3 gives the end-to-end spectral extent in \kms\ used for total velocity width as observed from the channel maps in Fig.~\ref{fig:combined_chanplots}; column 4 gives total velocity width, column 5 is simply the half width of column 4, and the last three columns give corrections based on three different intrinsic axial ratios. Values in brackets; wherever they appear, correspond to the same quantity as measured by \citet{guo20}}.
    \label{tab:vel_corrections}
    \begin{tabular}{|c|c|c|c|c|c|c|c|} \hline
         Galaxy Name & $w_{\rm 20}$ & Spectral extent & Velocity width & $V_{\rm H\textsc i}$\ (Uncorrected) & \multicolumn{3}{c|}{$V_{\rm H\textsc i}$\ (Corrected) } \\ 
         & \kms\ &  \kms\ & \kms\ & \kms\ &  \multicolumn{3}{c|}{\kms\ } \\ \cline{6-8}
         & Ours \citep{guo20} &  &  &  & $q_0$ = 0.2 & $q_0$ = 0.4 & $q_0$ = 0.6 \\ \hline
         UGC 6438 & 52.77 (80.36) & 1124.1--1198.5 & 74.4$\pm$3.4 & 37.2$\pm$1.7 & 62.82$\pm$2.87 (46.22) & 58.77$\pm$2.69 & 51.3$\pm$2.34 \\
         UGC 7983 & 39.28 (46.12) & 671.8--716.7 & 44.9$\pm$3.5 & 22.45$\pm$1.75 & 36.66$\pm$2.86 (28.24) & 34.29$\pm$2.67 & 29.93$\pm$2.33 \\
         UGC 9500 & 30.58 (39.08) & 1663.9--1712.5 & 48.6$\pm$3.5 & 24.3$\pm$1.75 & 119.64$\pm$8.61 (21.37) & 111.92$\pm$8.06 & 97.69$\pm$7.04 \\
         AGC 191707 & 38.28 (49.27) & 1574.6--1623.1 & 48.5$\pm$1.8 & 24.25$\pm$0.9 & 27.08$\pm$1.0 (29.38) & 25.33$\pm$0.94 & - \\
         AGC 220901 & 38.04 (45.38) & 975.3--1028.9 & 53.6$\pm$3.5 & 26.8$\pm$1.75 & 45.54$\pm$2.97 (25.44) & 42.6$\pm$2.78 & 37.18$\pm$2.43 \\
         AGC 733302 & 38.82 (48.36) & 1261.5--1301.4 & 39.9$\pm$1.8 & 19.95$\pm$0.9 & 35.14$\pm$1.59 (27.64) & 32.87$\pm$1.48 & 28.69$\pm$1.29 \\ \hline
    \end{tabular}
\end{table*}

\subsection{Dynamical Mass}
Using the deprojected velocity widths ($V_{\rm H\textsc i}$) and the \HI\ radii ($r_{\rm H\textsc i}$) of our sample galaxies, we compute their dynamical masses enclosed within $r_{\rm H\textsc i}$ using

\begin{equation}
 \frac{M_{\rm dyn}}{M_\odot} =
 \left(\frac{V_{\rm H\textsc i}}{\rm km \thinspace s^{-1}}\right)^2
 \left(\frac{r_{\rm H\textsc i}}{\rm kpc}\right)
 \times 0.23\times10^{6}.
 \label{eqn:dyn_mass}
 \end{equation}

\noindent Equation~(\ref{eqn:dyn_mass}) comes from $M_{\rm dyn}(<r_{\rm H\textsc i})\ =\ \frac{{V_{\rm H\textsc i}}^2\ r_{\rm H\textsc i}}{G}$ with appropriate unit conversions.

The resulting dynamical masses are listed in Table~\ref{tab:dyn_masses_corrected}. As discussed earlier, we compute inclination–corrected velocity widths using a range of intrinsic axial ratios, $q_0 = 0.2,\ 0.4$, and $0.6$, in order to encompass the plausible geometric thickness of dwarf \HI\ discs. We therefore report the corresponding dynamical masses for all three assumed $q_0$ values. For our sample, the dynamical masses span the range $\sim 1.5\times10^{9}\ M_\odot$ to $\sim 3.2\times10^{10}\ M_\odot$. For comparison, we also list the dynamical masses reported by \citet{guo20} for the same galaxies. As can be seen, in all cases \citet{guo20} systematically underestimate the dynamical masses. This arises because their inclinations were overestimated, which in turn led to applying a smaller inclination correction to the observed velocity widths and therefore to lower inferred circular velocities and dynamical masses.

\begin{table*}
    \centering
    \caption{Dynamical Masses: column 2 gives baryonic masses as taken from Table ~\ref{tab:bary_masses_corrected}; the subsequent three columns give dynamic masses calculated using equation~(\ref{eqn:dyn_mass}) for the three different $V_{\rm H\textsc i}$ values given in Table ~\ref{tab:vel_corrections}. The values n brackets in column 3 are dynamic masses as given by \citet{guo20}}
    \label{tab:dyn_masses_corrected}
    \begin{tabular}{|c|c|c|c|c|} \hline
         Galaxy Name & $M_{\rm bary}$ & \multicolumn{3}{c|}{$M_{\rm dyn}$ } \\ 
         & \eten{9}\ \msun\ & \multicolumn{3}{c|}{\eten{9}\ \msun\ } \\ \cline{3-5}
         & & $q_0$ = 0.2 & $q_0$ = 0.4 & $q_0$ = 0.6 \\ 
         & & Ours \citep{guo20} & & \\ \hline
        UGC 6438 & 1.18$\pm$0.23 & 4.50$\pm$1.50 (2.78) & 3.94$\pm$1.31 & 3.00$\pm$1.00 \\
        UGC 7983 & 0.54$\pm$0.02 & 1.63$\pm$0.56 (1.11) & 1.43$\pm$0.49 & 1.09$\pm$0.38 \\
        UGC 9500 & 2.10$\pm$0.13 & 39.01$\pm$11.54 (1.24) & 34.14$\pm$10.10 & 26.01$\pm$7.70 \\
        AGC 191707 & 0.50$\pm$0.05 & 1.26$\pm$0.36 (1.20) & 1.11$\pm$0.32 & -- \\
        AGC 220901 & 0.33$\pm$0.01 & 2.42$\pm$0.84 (0.73) & 2.12$\pm$0.74 & 1.61$\pm$0.56 \\
        AGC 733302 & 0.62$\pm$0.05 & 2.12$\pm$0.62 (1.10) & 1.85$\pm$0.54 & 1.41$\pm$0.41 \\
        \hline
    \end{tabular}
\end{table*}

\subsection{$M_{\rm dyn}/M_{\rm bary}$ ratios}
\begin{figure}
    \centering
    \includegraphics[width=0.9999\linewidth]{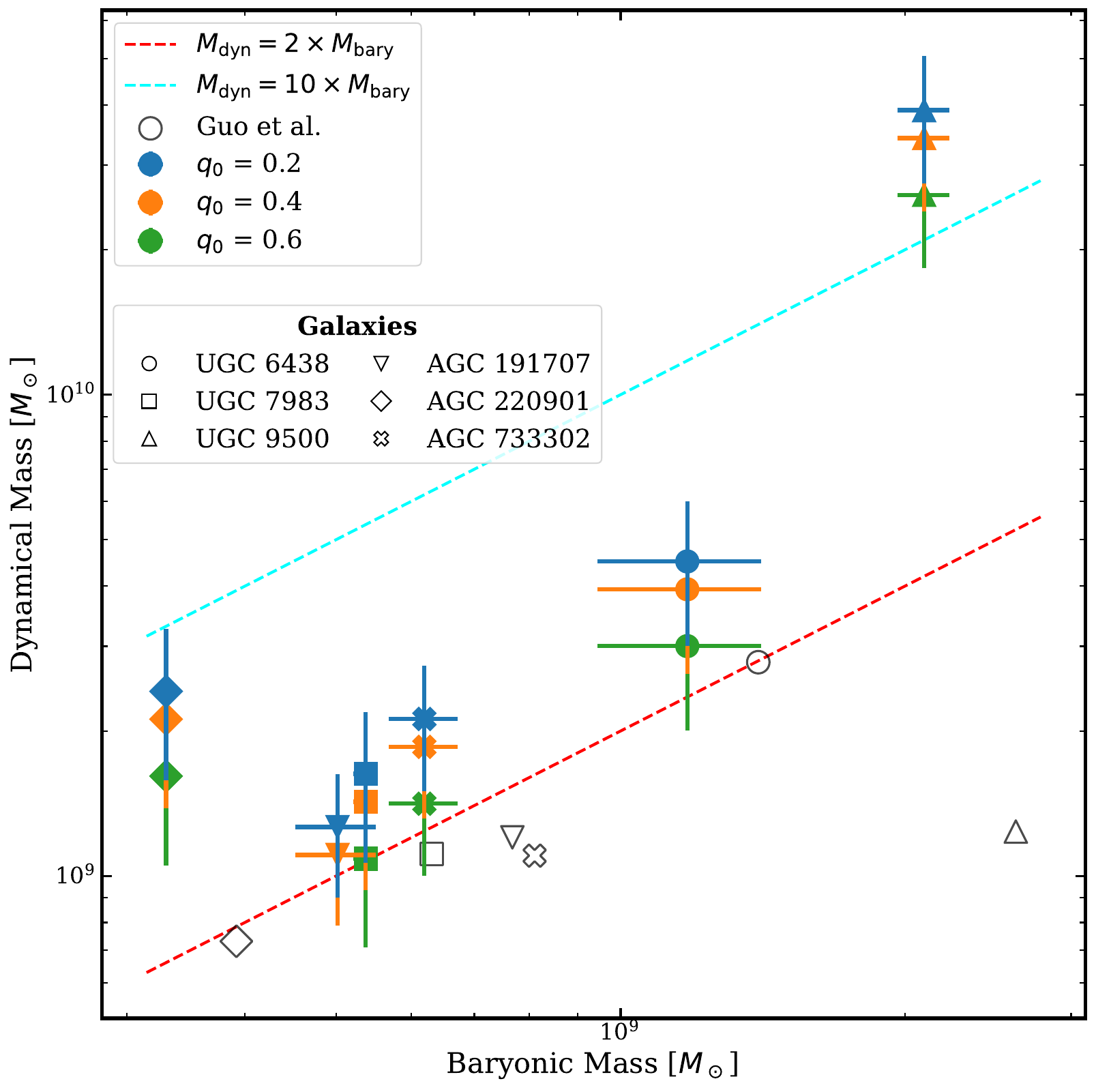}
    \caption{$M_{\rm dyn}$ as a function of $M_{\rm bary}$: The red dashed line shows the $M_{\rm dyn}/M_{\rm bary}\ =\ 2$ boundary, the open shapes correspond to the estimates of this ratio by \citet{guo20}; which all fall below the dashed line, putting the galaxies in BDDG regime. The filled shapes correspond to our measurements which are colour-coded according to the values adapted for $q_0$: blue for $q_0\ =\ 0.2$, orange for $q_0\ =\ 0.4$, and green for $q_0\ =\ 0.2$ respectively.}
    \label{fig:mdyn_by_mbary}
\end{figure}

Finally we compute the ratio of dynamical mass to baryonic mass to assess whether these galaxies are baryon dominated. These ratios, evaluated for each assumed value of $q_0$, are shown in fig.~\ref{fig:mdyn_by_mbary}. The figure shows the different estimates of the ratio $M_{\rm dyn}/M_{\rm bary}$ according to our analysis (filled shapes with error-bars in fig.~\ref{fig:mdyn_by_mbary}) as well as the same as given by \citet{guo20} in their analysis (empty shapes in fig.~\ref{fig:mdyn_by_mbary}). For AGC~191707, the observed axial ratio is 0.48, which is smaller than 0.6; therefore, an inclination cannot be computed for $q_0 = 0.6$. For this galaxy, we estimate the dynamical mass only for $q_0 = 0.2$ and $0.4$.

As can be seen from fig.~\ref{fig:mdyn_by_mbary}, for an assumed intrinsic axial ratio of $q_0 = 0.6$, four galaxies in our sample, i.e., UGC~6438, UGC~7983, AGC~191707, and AGC~733302, show significant baryonic contribution in total dynamical mass, with $M_{\rm dyn}/M_{\rm bary}$ values ranging from 2.02 to 2.55. We label the galaxies where the ratio is not strictly less than two (with error bars taken into account) as \textit{dark matter deficient galaxies} (since for BDDGs, this ratio is strictly $\leq$ 2; see equation~(\ref{eqn:bddg_criterion})). Such significant baryonic content is still exceptionally high when compared to other galaxies in similar mass scales. The dashed cyan line in fig.~\ref{fig:mdyn_by_mbary} represents $M_{\rm dyn}/M_{\rm bary}\ =\ 10$ limit which is the lower limit for this ratio in typical dwarf galaxies in the local universe (see, sec.~\ref{sec:introduction}). Five of total six galaxies lie below this limit. (See section~\ref{sec:discussion} for more discussion.)

For the remaining two galaxies, UGC~9500 and AGC~220901, the $M_{\rm dyn}/M_{\rm bary}$ ratios are 12.41$\pm$3.75 and 4.88$\pm$1.70, respectively (18.61 and 7.33 for $q_0 = 0.2$). These values clearly rule out the possibility that these systems are baryon dominated or dark matter deficient. For other galaxies, if we adopt a thinner \HI\ disk with $q_0 = 0.2$ (as used by \citet{guo20}), the $M_{\rm dyn}/M_{\rm bary}$ ratio increases, ranging from 2.52 to 3.83, putting them in the non-baryon-dominated category. In the following section we discuss the limitations and implications of our analysis and the surrounding context of galaxy formation and evolution.

\section{Discussion}
\label{sec:discussion}
It must be noted that the dark matter halos are assumed to have spherical symmetry in this analysis. Without a complete 3D modelling we cannot rule out the possibility of non-spherical halos. An oblate or prolate halo can impact the rotational velocity of the gas in galaxy plane significantly \citep[see, e.g.][]{das20, kumar22}. \citet{das23} showed that for most gas-rich  dwarf galaxies where baryonic mass is dominated by the gas component ($M_{\rm gas}/M_{\rm bary}\ \geq\ 0.5$), have oblate halos at $r_{\rm 25}$, the radius at which optical surface brightness of the galaxy drops to 25 ${mag}/{arcsec}^2$. From Table~\ref{tab:bary_masses_corrected}, we can see that only UGC~6438 is gas-poor galaxy while the remaining five dwarf galaxies are all gas-rich in our sample. For our galaxies we did not find $r_{\rm 25}$ in existing literature. It should be noted however, that an oblate halo with the same halo mass will result in slower rotation in the disk, thus lowering the total dynamical mass measured. In which case, our results of the $M_{\rm dyn}/M_{\rm bary}$ will be their lowest estimates and true values may be higher, making these galaxies no longer dark matter deficient.

The baryonic fractions we measure have important implications for understanding baryon enhancement efficiency in the broader cosmological context. In the $\Lambda$CDM framework, low-mass dark matter haloes are expected to host a large population of luminous dwarf galaxies \citep[e.g.][]{klypin99, moore99, bullock17, cristofari19}. Yet, observational surveys consistently report only a fraction of the predicted number \citep[e.g.][]{simon19, carlsten21b, drlicawagner20}. The widely discussed explanations suggest that many low-mass haloes exhibit extremely low baryon enhancement efficiencies and thus fail to accumulate sufficient baryons to be detected in optical surveys or other mechanisms such as feedback from supernovae drive out baryons, rendering such structures unable to produce enough stars and thus, dark \citep[see, e.g.][and the references therein]{mcconnachie12, smith19}. On the other hand, when baryonic processes are taken into account, they can reproduce the stellar mass function of dwarf galaxies on the relevant scales, eventually "solving" the missing-satellite problem and the too-big-to-fail problem \citep[see, e.g.][and the references therein]{sales22, garrisonkimmel19} but they do not predict any significant populations of dwarf galaxies where baryonic masses dominate. Several studies, in fact, shift the problem to the other extreme where they predict far less number of subhalos on dwarf galaxy mass scales to host luminous galaxies.  \citep[For a more thorough review we refer the reader to][and the references therein]{sales22}. Conversely, the dark matter deficient galaxies in our sample, particularly UGC~6438, UGC~7983, AGC~191707, and AGC~733302, represent the opposite extreme, where the efficiency of baryonic assembly appears unusually high relative to their dynamical mass. Notably, two these systems -- AGC~191707 and AGC~733302, reside in relatively normal field environments, suggesting that neither environmental quenching nor strong tidal interactions is responsible for their exceptionally high baryonic fractions. These findings highlight the diversity of baryon assembly histories in dwarf galaxies and underscore the need to understand the physical mechanisms that enable some haloes to accumulate and retain gas efficiently while others remain strongly baryon-poor.

\begin{figure}
    \centering
    \includegraphics[width=0.9999\linewidth]{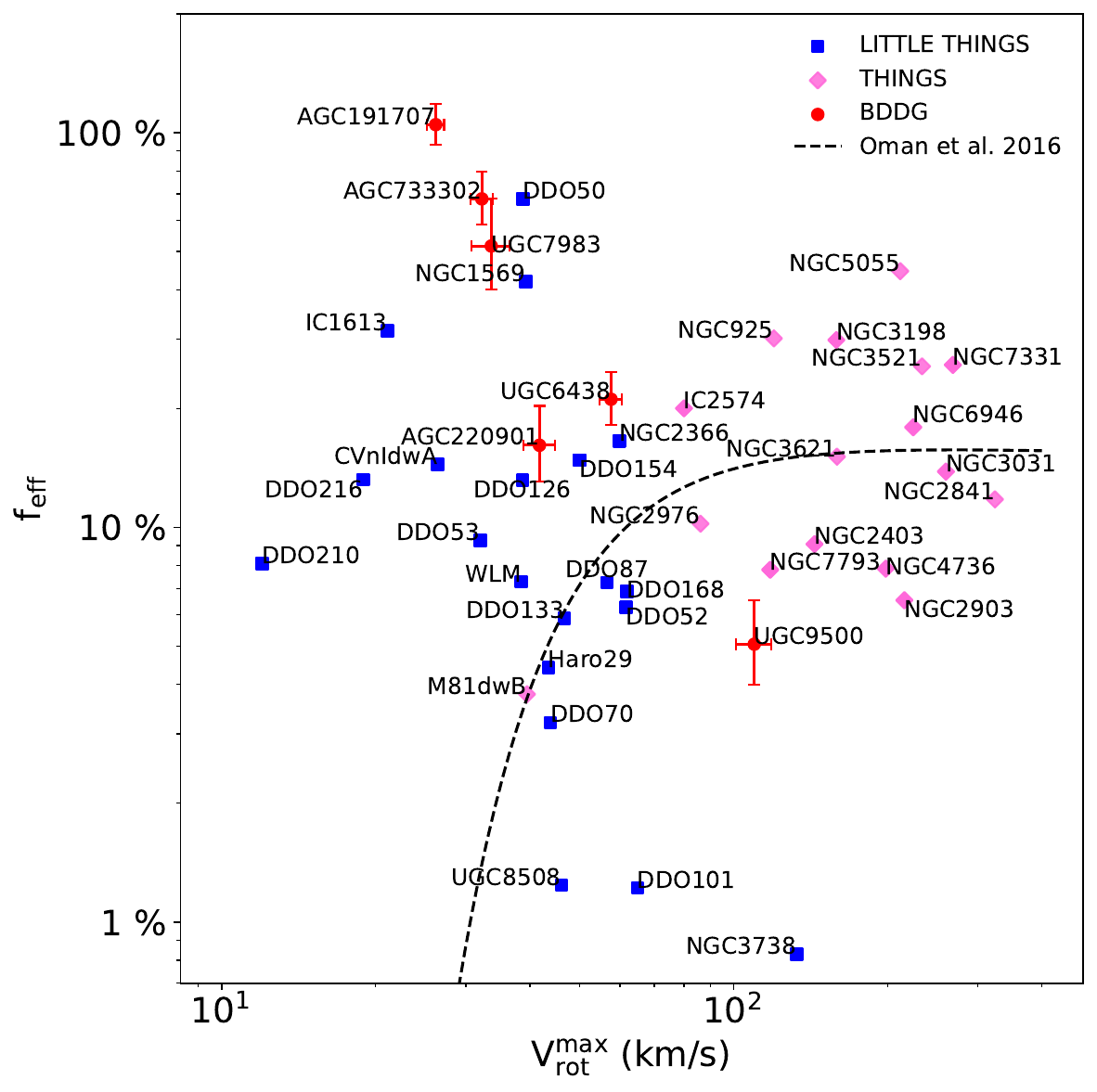}
    \caption{Baryon enhancement efficiency factor as a function of maximum rotational velocity. The red circles with error bars correspond to the sample of BDDG in current work, the blue squares to galaxies from LITTLE THINGS while pink squares correspond to THINGS galaxies. The dashed black line represents the best fit to the sample of APOSTLE simulations performed by \citet{oman16}}
    \label{fig:f_eff_vs_v_circ_max}
\end{figure}

To compare our galaxies with previously studied regular and dwarf systems, we also examine their baryon enhancement efficiencies as a function of dynamical mass as stated in section~\ref{sec:introduction}. In Fig.~\ref{fig:f_eff_vs_v_circ_max}, we compare the enhancement efficiencies of our sample with those of galaxies from other resolved \HI\ surveys, namely THINGS \citep[e.g.][]{walter08, deblok08} and LITTLE THINGS \citep[e.g.][]{hunter12, oh11a, oh15}. For equation~(\ref{eqn:feff}), a direct measurement of $M_{\rm 200}$ requires full dark matter halo modelling, which is not feasible for our galaxies due to the limited number of independent beams across their \HI\ discs. Instead, we estimate $M_{\rm 200}$ for our galaxies using the scaling relation between halo mass and maximum circular velocity from \citet{oman16}:

\begin{equation}
    \frac{M_{\rm 200}}{\mathrm{M_\odot}} 
    = 1.074 \times 10^{5} 
    \left( \frac{V_{\rm max}}{\mathrm{km\,s^{-1}}} \right)^{3.115}.
    \label{eqn:M200_vmax}
\end{equation}

We adopt the mean inclination-corrected \vhi\ values from Table~\ref{tab:vel_corrections} as proxies for $V_{\rm max}$. The dashed line in Fig.~\ref{fig:f_eff_vs_v_circ_max} shows the theoretically expected relation under $\Lambda$CDM as given by \cite{oman16} based on simulations. As seen in the figure, most galaxies from the THINGS and LITTLE THINGS samples lie close to the theoretical efficiency curve. However, a few LITTLE THINGS dwarf galaxies, such as DDO~50 and IC~1613, show unusually high efficiencies. Three of our galaxies, AGC~191707, AGC~733302, and UGC~7983, likewise exhibit exceptionally high baryon enhancement efficiencies ($\gtrsim 50\%$), while the typical observed peak efficiency is around $f_{\rm eff} \sim 18\%$ \citep{behroozi13a, oman16}. Notably, AGC~191707 exceeds even the theoretical maximum ($f_{\rm eff} > 1$), implying that it contains more baryonic mass than expected for a halo of its inferred dynamical mass.

For DDO~50 and IC~1613, the literature attributes their high efficiencies to weak stellar feedback and significant uncertainty in inclination estimates \citep[e.g.][]{iorio17, mcquinn22}. In contrast, for our sources, AGC~191707, AGC~733302, and UGC~7983, such factors are unlikely to account for the extreme values. These systems therefore pose a meaningful challenge to our understanding of baryon assembly in the $\Lambda$CDM framework, particularly at the low-mass end where baryon accumulation and galaxy formation is expected to be highly inefficient. For the remaining three galaxies in our sample, $f_{\rm eff}$ is consistent with values measured in other field dwarf galaxies.

The presence of these galaxies in our sample with unusually high baryon enhancement efficiencies has important implications for dwarf galaxy evolution and for the $\Lambda$CDM framework. Cosmological simulations predict that low-mass halos should be strongly inefficient at retaining baryons due to the combined effects of stellar feedback, photoionisation, and the shallow potential wells of dwarf systems \citep[see, e.g.][]{somerville15, hopkins14x, governato10, governato12, bullock17, okamoto08, fitts17}. In $\Lambda$CDM simulations, strong bursty outflows are expected to remove a substantial fraction of the gas from dwarf systems, driving their baryon fractions well below the cosmic mean. The unusually high baryon enhancement efficiencies inferred for AGC~191707, AGC~733302, and UGC~7983 therefore suggest that these galaxies may reside in atypical halo configurations, possibly halos with above-average concentrations, which are known to resist gas removal more effectively \citep[see, e.g.][]{dicintio17, jiang19}, or systems with unusually quiet assembly histories that promote long-term baryon retention \citep[see, e.g.][]{dutton16, fitts18, sawala16}. Regardless of the underlying mechanism, the existence of such high-efficiency dwarf galaxies highlights a growing tension at the low-mass end of the galaxy–halo connection, where the observed diversity of baryon fractions appears broader than predicted. Further resolved kinematic analysis and deeper multi-wavelength data will be essential to determine whether these systems represent statistical outliers or point toward a systematic shortcoming in current models of dwarf galaxy formation.

Previous attempts to understand the origin of such systems from a theoretical perspective have highlighted the role of environment and interactions, particularly the impact of tidal encounters and mergers that can efficiently strip away dark matter from low-mass galaxies. Several recent studies \citep[see, e.g.][]{moreno22, tau21, shin20} have investigated this issue using numerical and semi-analytic simulations within the \lcdm\ framework. These works demonstrate that even in standard \lcdm\ cosmology, dark-matter–deficient galaxies can arise naturally, especially as a consequence of high-velocity encounters with massive neighbours. However, such conditions are difficult to produce in isolated environments. Their simulations suggest that roughly one-third of massive central galaxies are expected to host at least one such dark-matter–deficient satellite. Furthermore, hydrodynamical simulations indicate that the fraction of such diffuse, dark-matter–deficient galaxies should be higher at earlier cosmic times, as early halo–halo encounters can remove a substantial fraction of the dark matter, followed by a more gradual stripping of the baryonic component. However, observationally confirming this prediction across a statistically significant sample remains challenging, given the low surface brightness of such systems and the sensitivity limits of current surveys.

\citet{ivleva24} further showed using high-resolution hydrodynamical simulations that mergers in cluster environments can produce dark-matter–deficient dwarf galaxies within tidal tails. Similar systems can also emerge via tidal stripping by massive neighbours or small groups \citep[see, e.g.][and references therein]{jackson21}, consistent with earlier analyses. In our sample, UGC~7983 shows signatures of possible ongoing interaction (Fig.~\ref{fig:7983_full} and channel maps in Fig.~\ref{fig:combined_chanplots}), whereas AGC~220901, despite its proximity to the Virgo cluster, appears to have retained sufficient dark matter to avoid falling into the baryon-dominated regime. This suggests that, at least for our sample, the environment plays a minimal role in determining whether low-mass dwarf galaxies become baryon-dominated.

The studies by \citet{mancera20, mancera19b}, which attempt to address this issue in the context of isolated dwarf galaxies, suggest that some small, ultra-diffuse systems may simply have experienced very low star-formation rates, resulting in weak stellar feedback. Such weak feedback would be insufficient to expel significant amounts of gas, allowing these galaxies to retain a larger fraction of their baryonic mass than their counterparts. In these cases, analysing the radio continuum emission associated with recent star formation may provide additional insights. It is therefore plausible that a complex interplay of feedback strength, star-formation history, and environmental interactions is at work, necessitating the comprehensive study of BDDGs. 

While a full dynamical reconstruction is challenging for individual systems, it appears that some of the candidates identified by \citet{guo20} may indeed be baryon-dominated dwarf galaxies. Importantly, some of these galaxies are ordinary field dwarf galaxies, not systems residing in extreme environments where strong tidal interactions or ram-pressure stripping are expected to dominate.

 Moreover, for some of them, including AGC~191707, and AGC~733302 there is no clear observational signature of the mechanisms typically invoked to explain dark-matter–deficient or baryon-dominated systems, such as strong tidal stripping, unusual merger histories, or exceptionally weak feedback. Their inferred baryon fractions, therefore, remain difficult to reconcile within the conventional $\Lambda$CDM framework.

These results thus pose a potentially significant tension: if such baryon-dominated dwarf galaxies genuinely exist in the field, they would require either atypical halo assembly histories or a revision of the standard assumptions regarding feedback and baryon retention in low-mass halos. Identifying additional BDDGs and conducting deeper, spatially resolved observations will therefore be essential for determining whether these systems represent rare outliers or a population that current galaxy-formation models have yet to fully account for.

\section{Summary}
\label{sec:summary}

We have presented resolved \HI~observations of six dwarf galaxies drawn from the sample of baryon-dominated dwarf galaxy candidates identified by \citet{guo20} in the ALFALFA survey where earlier results indicated $M_{\rm DM}(r\ \leq\ r_{\rm H \textsc i})\ \leq\ M_{\rm bar}(r\ \leq\ r_{\rm H \textsc i})$. Since estimating the total dynamical mass enclosed within $R_{\rm vir}$  of these systems is difficult and involves assumptions about the shape and dark matter density profile that may not be available due to lack of observational sensitivity, we determine whether they are baryon-dominated or not based on equation~\ref{eqn:bddg_criterion}. Their original claim of unusually high baryonic fractions was based entirely on single-dish \HI~spectra and optical inclinations. Both of these inputs suffer from systematic uncertainties, i.e., baseline offsets in Arecibo spectra, lack of spatial resolution, uncertain \HI~radii inferred from the mass–size relation, and potentially biased optical inclinations driven by central light concentration or irregular morphologies. Our uGMRT interferometric data overcome these limitations by providing (i) accurate integrated \HI~fluxes free of baseline offsets, (ii) resolved \HI~discs from which we directly measure the \HI~extent, (iii) kinematically motivated position angles, and crucially, (iv) \HI-based axial ratios to derive reliable inclinations.

Using our resolved \HI~spectral cubes and moment maps, we re-evaluated the dynamical masses of all six galaxies. We find that the \HI~radii derived from the moment-0 maps differ from the values inferred by \citet{guo20} by up to $\sim$20\%. More importantly, the optical axial ratios reported in the earlier work systematically underestimated the true thickness of the galaxies, leading to artificially low inclinations and consequently underestimated circular velocities. Our \HI-derived axial ratios are significantly rounder for most of the sample, resulting in larger inclination corrections and higher dynamical masses.

We use the resolved \HI~channel maps to extract the full velocity extent of our galaxies for calculating their dynamical mass. We estimated the total baryonic mass by combining the gas mass and the stellar mass. We find that four of the six systems (UGC~6438, UGC~7983, AGC~191707, AGC~733302) have baryon fractions close to or exceeding the $\Lambda$CDM-predicted peak efficiency of $\sim$18\%. In three of these galaxies, the inferred baryon enhancement efficiency exceeds 50\%, with AGC~191707 appearing formally super-efficient (>100\%), implying that it has accumulated more baryons than allowed by the cosmic baryon fraction when referenced to its inferred halo mass. Only two systems (UGC~9500 and AGC~220901) are consistent with being dark-matter dominated once our revised dynamical masses are adopted.

The existence of such high-efficiency dwarf galaxies in isolated environments remains difficult to reconcile with standard $\Lambda$CDM expectations, which predict inefficient baryon retention in low-mass halos. Unlike cluster or group satellites where tidal stripping, high-velocity encounters, or ram-pressure effects can lower dark-matter content, our sample galaxies reside in low-density environments and show no clear morphological or kinematic signatures of strong environmental effects. Weak-feedback scenarios may explain parts of the trend, but cannot naturally produce the highest efficiencies we find. The absence of coherent tidal features, the lack of strong disturbances in their velocity fields, and the normal \HI~morphology in most systems argue against recent violent interactions capable of producing artificially low dynamical masses.

Taken together, our results suggest that a non-negligible fraction of dwarf galaxies may deviate significantly from the canonical baryon-retention trends implied by $\Lambda$CDM. Either these galaxies occupy unusually concentrated or otherwise atypical dark-matter halos, or they follow evolutionary pathways that allow them to accumulate or retain baryons at efficiencies far above the median. Establishing which of these possibilities is correct requires a larger statistical sample. Our pilot uGMRT study demonstrates the importance of resolved \HI~kinematics in reliably identifying baryon-dominated dwarf galaxies. Expanding the sample of confirmed systems is therefore imperative for understanding whether such galaxies pose a genuine challenge to $\Lambda$CDM or reveal new aspects of baryon–halo coupling in the low-mass regime.

\section*{Acknowledgements}
We sincerely thank the anonymous referee for their helpful feedback and suggestions, which have greatly improved the quality of this work. We thank the staff of the Giant Metrewave Radio Telescope (GMRT) who have made these observations possible. GMRT is run by the National Centre for Radio Astrophysics of the Tata Institute of Fundamental Research. AM acknowledges the support of Council of Scientific \& Industrial Research and University Grants Commission (CSIR-UGC) for the UGC Junior Research Fellowship. AN and NNP acknowledge support from the Science and Engineering Research Board (SERB) of the Department of Science and Technology (DST), Government of India, through the Startup Research Grant (SRG) No. SRG/2022/000917. NNP also acknowledges support from the Science and Engineering Research Board (SERB) of the Department of Science and Technology (DST), Government of India, through the DST-FIST grant. NR acknowledges support from the United States-India Educational Foundation through the Fulbright Program. MD and PB acknowledge the financial support of the Science and Engineering Research Board (SERB) Core Research Grant CRG/2022/004531. MD also acknowledges the financial support of the Department of Science and Technology (DST) grant DST/WIDUSHI-A/PM/2023/25(G) for this research.

\section*{Data Availability}
All the radio data used in this study are available in the GMRT Online Archive\footnote{\label{foot:goa}\url{https://naps.ncra.tifr.res.in/goa/data/search}} with proposal code 39\_064.

\bibliographystyle{mnras}
\bibliography{citations} 

\begin{thebibliography}{}
\makeatletter
\relax
\def\mn@urlcharsother{\let\do\@makeother \do\$\do\&\do\#\do\^\do\_\do\%\do\~}
\def\mn@doi{\begingroup\mn@urlcharsother \@ifnextchar [ {\mn@doi@}
  {\mn@doi@[]}}
\def\mn@doi@[#1]#2{\def\@tempa{#1}\ifx\@tempa\@empty \href
  {http://dx.doi.org/#2} {doi:#2}\else \href {http://dx.doi.org/#2} {#1}\fi
  \endgroup}
\def\mn@eprint#1#2{\mn@eprint@#1:#2::\@nil}
\def\mn@eprint@arXiv#1{\href {http://arxiv.org/abs/#1} {{\tt arXiv:#1}}}
\def\mn@eprint@dblp#1{\href {http://dblp.uni-trier.de/rec/bibtex/#1.xml}
  {dblp:#1}}
\def\mn@eprint@#1:#2:#3:#4\@nil{\def\@tempa {#1}\def\@tempb {#2}\def\@tempc
  {#3}\ifx \@tempc \@empty \let \@tempc \@tempb \let \@tempb \@tempa \fi \ifx
  \@tempb \@empty \def\@tempb {arXiv}\fi \@ifundefined
  {mn@eprint@\@tempb}{\@tempb:\@tempc}{\expandafter \expandafter \csname
  mn@eprint@\@tempb\endcsname \expandafter{\@tempc}}}

\bibitem[\protect\citeauthoryear{{Abazajian} et~al.,}{{Abazajian}
  et~al.}{2009}]{abazajian09}
{Abazajian} K.~N.,  et~al., 2009, \mn@doi [\apjs]
  {10.1088/0067-0049/182/2/543}, \href
  {http://adsabs.harvard.edu/abs/2009ApJS..182..543A} {182, 543}

\bibitem[\protect\citeauthoryear{{Amorisco} \& {Evans}}{{Amorisco} \&
  {Evans}}{2011}]{amorisco11}
{Amorisco} N.~C.,  {Evans} N.~W.,  2011, \mn@doi [\mnras]
  {10.1111/j.1365-2966.2010.17715.x}, \href
  {https://ui.adsabs.harvard.edu/abs/2011MNRAS.411.2118A} {411, 2118}

\bibitem[\protect\citeauthoryear{{Begeman}}{{Begeman}}{1987}]{begeman87phd}
{Begeman} K.~G.,  1987, PhD thesis, , Kapteyn Institute, (1987)

\bibitem[\protect\citeauthoryear{{Begum}, {Chengalur}, {Karachentsev},
  {Sharina}  \& {Kaisin}}{{Begum} et~al.}{2008}]{begum08b}
{Begum} A.,  {Chengalur} J.~N.,  {Karachentsev} I.~D.,  {Sharina} M.~E.,
  {Kaisin} S.~S.,  2008, \mn@doi [\mnras] {10.1111/j.1365-2966.2008.13150.x},
  \href {http://adsabs.harvard.edu/abs/2008MNRAS.386.1667B} {386, 1667}

\bibitem[\protect\citeauthoryear{{Behroozi}, {Wechsler}  \&
  {Conroy}}{{Behroozi} et~al.}{2013}]{behroozi13a}
{Behroozi} P.~S.,  {Wechsler} R.~H.,   {Conroy} C.,  2013, \mn@doi [\apj]
  {10.1088/0004-637X/770/1/57}, \href
  {https://ui.adsabs.harvard.edu/abs/2013ApJ...770...57B} {770, 57}

\bibitem[\protect\citeauthoryear{{Bell}, {McIntosh}, {Katz}  \&
  {Weinberg}}{{Bell} et~al.}{2003}]{bell03}
{Bell} E.~F.,  {McIntosh} D.~H.,  {Katz} N.,   {Weinberg} M.~D.,  2003, \mn@doi
  [\apjs] {10.1086/378847}, \href
  {https://ui.adsabs.harvard.edu/abs/2003ApJS..149..289B} {149, 289}

\bibitem[\protect\citeauthoryear{{Biswas}, {Kalinova}, {Roy}, {Patra}  \&
  {Tyulneva}}{{Biswas} et~al.}{2023}]{biswas23}
{Biswas} P.,  {Kalinova} V.,  {Roy} N.,  {Patra} N.~N.,   {Tyulneva} N.,  2023,
  \mn@doi [\mnras] {10.1093/mnras/stad2285}, \href
  {https://ui.adsabs.harvard.edu/abs/2023MNRAS.524.6213B} {524, 6213}

\bibitem[\protect\citeauthoryear{{Bosma}}{{Bosma}}{1981}]{bosma81}
{Bosma} A.,  1981, \mn@doi [\aj] {10.1086/113063}, \href
  {https://ui.adsabs.harvard.edu/abs/1981AJ.....86.1825B} {86, 1825}

\bibitem[\protect\citeauthoryear{{Broeils} \& {Rhee}}{{Broeils} \&
  {Rhee}}{1997}]{broeils97}
{Broeils} A.~H.,  {Rhee} M.-H.,  1997, \aap, \href
  {http://adsabs.harvard.edu/abs/1997A%26A...324..877B} {324, 877}

\bibitem[\protect\citeauthoryear{{Bullock} \& {Boylan-Kolchin}}{{Bullock} \&
  {Boylan-Kolchin}}{2017}]{bullock17}
{Bullock} J.~S.,  {Boylan-Kolchin} M.,  2017, \mn@doi [\araa]
  {10.1146/annurev-astro-091916-055313}, \href
  {https://ui.adsabs.harvard.edu/abs/2017ARA&A..55..343B} {55, 343}

\bibitem[\protect\citeauthoryear{{CASA Team} et~al.,}{{CASA Team}
  et~al.}{2022}]{casa22}
{CASA Team} et~al., 2022, \mn@doi [\pasp] {10.1088/1538-3873/ac9642}, \href
  {https://ui.adsabs.harvard.edu/abs/2022PASP..134k4501C} {134, 114501}

\bibitem[\protect\citeauthoryear{{Carlsten}, {Greene}, {Greco}, {Beaton}  \&
  {Kado-Fong}}{{Carlsten} et~al.}{2021}]{carlsten21b}
{Carlsten} S.~G.,  {Greene} J.~E.,  {Greco} J.~P.,  {Beaton} R.~L.,
  {Kado-Fong} E.,  2021, \mn@doi [\apj] {10.3847/1538-4357/ac2581}, \href
  {https://ui.adsabs.harvard.edu/abs/2021ApJ...922..267C} {922, 267}

\bibitem[\protect\citeauthoryear{{Cormier} et~al.,}{{Cormier}
  et~al.}{2014}]{cormier14}
{Cormier} D.,  et~al., 2014, \mn@doi [\aap] {10.1051/0004-6361/201322096},
  \href {http://adsabs.harvard.edu/abs/2014A%26A...564A.121C} {564, A121}

\bibitem[\protect\citeauthoryear{{Cristofari} \& {Ostriker}}{{Cristofari} \&
  {Ostriker}}{2019}]{cristofari19}
{Cristofari} P.,  {Ostriker} J.~P.,  2019, \mn@doi [\mnras]
  {10.1093/mnras/sty2966}, \href
  {https://ui.adsabs.harvard.edu/abs/2019MNRAS.482.4364C} {482, 4364}

\bibitem[\protect\citeauthoryear{{Das}, {McGaugh}, {Ianjamasimanana},
  {Schombert}  \& {Dwarakanath}}{{Das} et~al.}{2020}]{das20}
{Das} M.,  {McGaugh} S.~S.,  {Ianjamasimanana} R.,  {Schombert} J.,
  {Dwarakanath} K.~S.,  2020, \mn@doi [\apj] {10.3847/1538-4357/ab5fcd}, \href
  {https://ui.adsabs.harvard.edu/abs/2020ApJ...889...10D} {889, 10}

\bibitem[\protect\citeauthoryear{{Das}, {Ianjamasimanana}, {McGaugh},
  {Schombert}  \& {Dwarakanath}}{{Das} et~al.}{2023}]{das23}
{Das} M.,  {Ianjamasimanana} R.,  {McGaugh} S.~S.,  {Schombert} J.,
  {Dwarakanath} K.~S.,  2023, \mn@doi [\apjl] {10.3847/2041-8213/acc10e}, \href
  {https://ui.adsabs.harvard.edu/abs/2023ApJ...946L...8D} {946, L8}

\bibitem[\protect\citeauthoryear{{Di Cintio}, {Brook}, {Dutton}, {Macci{\`o}},
  {Obreja}  \& {Dekel}}{{Di Cintio} et~al.}{2017}]{dicintio17}
{Di Cintio} A.,  {Brook} C.~B.,  {Dutton} A.~A.,  {Macci{\`o}} A.~V.,  {Obreja}
  A.,   {Dekel} A.,  2017, \mn@doi [\mnras] {10.1093/mnrasl/slw210}, \href
  {https://ui.adsabs.harvard.edu/abs/2017MNRAS.466L...1D} {466, L1}

\bibitem[\protect\citeauthoryear{{Di Teodoro} \& {Fraternali}}{{Di Teodoro} \&
  {Fraternali}}{2015}]{diteodoro15}
{Di Teodoro} E.~M.,  {Fraternali} F.,  2015, \mn@doi [\mnras]
  {10.1093/mnras/stv1213}, \href
  {https://ui.adsabs.harvard.edu/abs/2015MNRAS.451.3021D} {451, 3021}

\bibitem[\protect\citeauthoryear{{Drlica-Wagner} et~al.,}{{Drlica-Wagner}
  et~al.}{2020}]{drlicawagner20}
{Drlica-Wagner} A.,  et~al., 2020, \mn@doi [\apj] {10.3847/1538-4357/ab7eb9},
  \href {https://ui.adsabs.harvard.edu/abs/2020ApJ...893...47D} {893, 47}

\bibitem[\protect\citeauthoryear{{Dutton} et~al.,}{{Dutton}
  et~al.}{2016}]{dutton16}
{Dutton} A.~A.,  et~al., 2016, \mn@doi [\mnras] {10.1093/mnras/stw1537}, \href
  {https://ui.adsabs.harvard.edu/abs/2016MNRAS.461.2658D} {461, 2658}

\bibitem[\protect\citeauthoryear{{Fitts} et~al.,}{{Fitts}
  et~al.}{2017}]{fitts17}
{Fitts} A.,  et~al., 2017, \mn@doi [\mnras] {10.1093/mnras/stx1757}, \href
  {https://ui.adsabs.harvard.edu/abs/2017MNRAS.471.3547F} {471, 3547}

\bibitem[\protect\citeauthoryear{{Fitts} et~al.,}{{Fitts}
  et~al.}{2018}]{fitts18}
{Fitts} A.,  et~al., 2018, \mn@doi [\mnras] {10.1093/mnras/sty1488}, \href
  {https://ui.adsabs.harvard.edu/abs/2018MNRAS.479..319F} {479, 319}

\bibitem[\protect\citeauthoryear{{Garrison-Kimmel} et~al.,}{{Garrison-Kimmel}
  et~al.}{2019}]{garrisonkimmel19}
{Garrison-Kimmel} S.,  et~al., 2019, \mn@doi [\mnras] {10.1093/mnras/stz1317},
  \href {https://ui.adsabs.harvard.edu/abs/2019MNRAS.487.1380G} {487, 1380}

\bibitem[\protect\citeauthoryear{{Giovanelli}, {Haynes}, {Salzer}, {Wegner},
  {da Costa}  \& {Freudling}}{{Giovanelli} et~al.}{1994}]{giovanelli94}
{Giovanelli} R.,  {Haynes} M.~P.,  {Salzer} J.~J.,  {Wegner} G.,  {da Costa}
  L.~N.,   {Freudling} W.,  1994, \mn@doi [\aj] {10.1086/117014}, \href
  {https://ui.adsabs.harvard.edu/abs/1994AJ....107.2036G} {107, 2036}

\bibitem[\protect\citeauthoryear{{Gooch}}{{Gooch}}{1996}]{gooch96ip}
{Gooch} R.,  1996, in {Jacoby} G.~H.,  {Barnes} J.,  eds,  Astronomical Society
  of the Pacific Conference Series Vol. 101, Astronomical Data Analysis
  Software and Systems V. p.~80

\bibitem[\protect\citeauthoryear{{Governato} et~al.,}{{Governato}
  et~al.}{2010}]{governato10}
{Governato} F.,  et~al., 2010, \mn@doi [\nat] {10.1038/nature08640}, \href
  {https://ui.adsabs.harvard.edu/abs/2010Natur.463..203G} {463, 203}

\bibitem[\protect\citeauthoryear{{Governato} et~al.,}{{Governato}
  et~al.}{2012}]{governato12}
{Governato} F.,  et~al., 2012, \mn@doi [\mnras]
  {10.1111/j.1365-2966.2012.20696.x}, \href
  {http://adsabs.harvard.edu/abs/2012MNRAS.422.1231G} {422, 1231}

\bibitem[\protect\citeauthoryear{{Guo} et~al.,}{{Guo} et~al.}{2020}]{guo20}
{Guo} Q.,  et~al., 2020, \mn@doi [Nature Astronomy]
  {10.1038/s41550-019-0930-9}, \href
  {https://ui.adsabs.harvard.edu/abs/2020NatAs...4..246G} {4, 246}

\bibitem[\protect\citeauthoryear{{Guo}, {Sengupta}, {Scott}, {Lagos}  \&
  {Luo}}{{Guo} et~al.}{2024}]{guo24}
{Guo} Y.,  {Sengupta} C.,  {Scott} T.~C.,  {Lagos} P.,   {Luo} Y.,  2024,
  \mn@doi [\mnras] {10.1093/mnras/stae390}, \href
  {https://ui.adsabs.harvard.edu/abs/2024MNRAS.528.6593G} {528, 6593}

\bibitem[\protect\citeauthoryear{{Haynes} et~al.,}{{Haynes}
  et~al.}{2018}]{haynes18}
{Haynes} M.~P.,  et~al., 2018, \mn@doi [\apj] {10.3847/1538-4357/aac956}, \href
  {https://ui.adsabs.harvard.edu/abs/2018ApJ...861...49H} {861, 49}

\bibitem[\protect\citeauthoryear{{Hopkins}}{{Hopkins}}{2014}]{hopkins14x}
{Hopkins} P.~F.,  2014, \mn@doi [\nat] {10.1038/516044a}, \href
  {https://ui.adsabs.harvard.edu/abs/2014Natur.516...44H} {516, 44}

\bibitem[\protect\citeauthoryear{{Hoyer}, {Neumayer}, {Georgiev}, {Seth}  \&
  {Greene}}{{Hoyer} et~al.}{2021}]{hoyer21}
{Hoyer} N.,  {Neumayer} N.,  {Georgiev} I.~Y.,  {Seth} A.~C.,   {Greene} J.~E.,
   2021, \mn@doi [\mnras] {10.1093/mnras/stab2277}, \href
  {https://ui.adsabs.harvard.edu/abs/2021MNRAS.507.3246H} {507, 3246}

\bibitem[\protect\citeauthoryear{{Hunt}, {Tortora}, {Ginolfi}  \&
  {Schneider}}{{Hunt} et~al.}{2020}]{hunt20}
{Hunt} L.~K.,  {Tortora} C.,  {Ginolfi} M.,   {Schneider} R.,  2020, \mn@doi
  [\aap] {10.1051/0004-6361/202039021}, \href
  {https://ui.adsabs.harvard.edu/abs/2020A&A...643A.180H} {643, A180}

\bibitem[\protect\citeauthoryear{{Hunter}, {Elmegreen}  \& {Anderson}}{{Hunter}
  et~al.}{2006}]{hunter06ip}
{Hunter} D.~A.,  {Elmegreen} B.~G.,   {Anderson} E.,  2006, in American
  Astronomical Society Meeting Abstracts. p. 167.05

\bibitem[\protect\citeauthoryear{{Hunter} et~al.,}{{Hunter}
  et~al.}{2012}]{hunter12}
{Hunter} D.~A.,  et~al., 2012, \mn@doi [\aj] {10.1088/0004-6256/144/5/134},
  \href {http://adsabs.harvard.edu/abs/2012AJ....144..134H} {144, 134}

\bibitem[\protect\citeauthoryear{{Iorio}, {Fraternali}, {Nipoti}, {Di Teodoro},
  {Read}  \& {Battaglia}}{{Iorio} et~al.}{2017}]{iorio17}
{Iorio} G.,  {Fraternali} F.,  {Nipoti} C.,  {Di Teodoro} E.,  {Read} J.~I.,
  {Battaglia} G.,  2017, \mn@doi [\mnras] {10.1093/mnras/stw3285}, \href
  {https://ui.adsabs.harvard.edu/abs/2017MNRAS.466.4159I} {466, 4159}

\bibitem[\protect\citeauthoryear{{Ivleva}, {Remus}, {Valenzuela}  \&
  {Dolag}}{{Ivleva} et~al.}{2024}]{ivleva24}
{Ivleva} A.,  {Remus} R.-S.,  {Valenzuela} L.~M.,   {Dolag} K.,  2024, \mn@doi
  [\aap] {10.1051/0004-6361/202449605}, \href
  {https://ui.adsabs.harvard.edu/abs/2024A&A...687A.105I} {687, A105}

\bibitem[\protect\citeauthoryear{{Jackson} et~al.,}{{Jackson}
  et~al.}{2021}]{jackson21}
{Jackson} R.~A.,  et~al., 2021, \mn@doi [\mnras] {10.1093/mnras/stab093}, \href
  {https://ui.adsabs.harvard.edu/abs/2021MNRAS.502.1785J} {502, 1785}

\bibitem[\protect\citeauthoryear{{Jarrett} et~al.,}{{Jarrett}
  et~al.}{2013}]{jarrett13}
{Jarrett} T.~H.,  et~al., 2013, \mn@doi [\aj] {10.1088/0004-6256/145/1/6},
  \href {https://ui.adsabs.harvard.edu/abs/2013AJ....145....6J} {145, 6}

\bibitem[\protect\citeauthoryear{{Jarrett}, {Cluver}, {Taylor}, {Bellstedt},
  {Robotham}  \& {Yao}}{{Jarrett} et~al.}{2023}]{jarrett23}
{Jarrett} T.~H.,  {Cluver} M.~E.,  {Taylor} E.~N.,  {Bellstedt} S.,  {Robotham}
  A.~S.~G.,   {Yao} H.~F.~M.,  2023, \mn@doi [\apj] {10.3847/1538-4357/acb68f},
  \href {https://ui.adsabs.harvard.edu/abs/2023ApJ...946...95J} {946, 95}

\bibitem[\protect\citeauthoryear{{Jiang}, {Dekel}, {Freundlich}, {Romanowsky},
  {Dutton}, {Macci{\`o}}  \& {Di Cintio}}{{Jiang} et~al.}{2019}]{jiang19}
{Jiang} F.,  {Dekel} A.,  {Freundlich} J.,  {Romanowsky} A.~J.,  {Dutton}
  A.~A.,  {Macci{\`o}} A.~V.,   {Di Cintio} A.,  2019, \mn@doi [\mnras]
  {10.1093/mnras/stz1499}, \href
  {https://ui.adsabs.harvard.edu/abs/2019MNRAS.487.5272J} {487, 5272}

\bibitem[\protect\citeauthoryear{{Johnson}, {Hunter}, {Kamphuis}  \&
  {Wang}}{{Johnson} et~al.}{2017}]{johnson17}
{Johnson} M.~C.,  {Hunter} D.~A.,  {Kamphuis} P.,   {Wang} J.,  2017, \mn@doi
  [\mnras] {10.1093/mnrasl/slw203}, \href
  {https://ui.adsabs.harvard.edu/abs/2017MNRAS.465L..49J} {465, L49}

\bibitem[\protect\citeauthoryear{{Kamphuis}, {J{\'o}zsa}, {Oh}, {Spekkens},
  {Urbancic}, {Serra}, {Koribalski}  \& {Dettmar}}{{Kamphuis}
  et~al.}{2015}]{kamphuis15}
{Kamphuis} P.,  {J{\'o}zsa} G.~I.~G.,  {Oh} S. .~H.,  {Spekkens} K.,
  {Urbancic} N.,  {Serra} P.,  {Koribalski} B.~S.,   {Dettmar} R.~J.,  2015,
  \mn@doi [\mnras] {10.1093/mnras/stv1480}, \href
  {https://ui.adsabs.harvard.edu/abs/2015MNRAS.452.3139K} {452, 3139}

\bibitem[\protect\citeauthoryear{{Karachentsev}, {Makarov}  \&
  {Kaisina}}{{Karachentsev} et~al.}{2013}]{karachentsev13}
{Karachentsev} I.~D.,  {Makarov} D.~I.,   {Kaisina} E.~I.,  2013, \mn@doi [\aj]
  {10.1088/0004-6256/145/4/101}, \href
  {https://ui.adsabs.harvard.edu/abs/2013AJ....145..101K} {145, 101}

\bibitem[\protect\citeauthoryear{{Kim} et~al.,}{{Kim} et~al.}{2014}]{kim14a}
{Kim} S.,  et~al., 2014, \mn@doi [\apjs] {10.1088/0067-0049/215/2/22}, \href
  {https://ui.adsabs.harvard.edu/abs/2014ApJS..215...22K} {215, 22}

\bibitem[\protect\citeauthoryear{{Klypin}, {Kravtsov}, {Valenzuela}  \&
  {Prada}}{{Klypin} et~al.}{1999}]{klypin99}
{Klypin} A.,  {Kravtsov} A.~V.,  {Valenzuela} O.,   {Prada} F.,  1999, \mn@doi
  [\apj] {10.1086/307643}, \href
  {http://adsabs.harvard.edu/abs/1999ApJ...522...82K} {522, 82}

\bibitem[\protect\citeauthoryear{{Krajnovi{\'c}}, {Cappellari}, {de Zeeuw}  \&
  {Copin}}{{Krajnovi{\'c}} et~al.}{2006}]{krajnovic06}
{Krajnovi{\'c}} D.,  {Cappellari} M.,  {de Zeeuw} P.~T.,   {Copin} Y.,  2006,
  \mn@doi [\mnras] {10.1111/j.1365-2966.2005.09902.x}, \href
  {https://ui.adsabs.harvard.edu/abs/2006MNRAS.366..787K} {366, 787}

\bibitem[\protect\citeauthoryear{{Kumar}, {Das}  \& {Kataria}}{{Kumar}
  et~al.}{2022}]{kumar22}
{Kumar} A.,  {Das} M.,   {Kataria} S.~K.,  2022, \mn@doi [\mnras]
  {10.1093/mnras/stab3019}, \href
  {https://ui.adsabs.harvard.edu/abs/2022MNRAS.509.1262K} {509, 1262}

\bibitem[\protect\citeauthoryear{{Lelli}, {McGaugh}  \& {Schombert}}{{Lelli}
  et~al.}{2016}]{lelli16a}
{Lelli} F.,  {McGaugh} S.~S.,   {Schombert} J.~M.,  2016, \mn@doi [\apjl]
  {10.3847/2041-8205/816/1/L14}, \href
  {https://ui.adsabs.harvard.edu/abs/2016ApJ...816L..14L} {816, L14}

\bibitem[\protect\citeauthoryear{{Lin} et~al.,}{{Lin} et~al.}{2023}]{lin23}
{Lin} X.,  et~al., 2023, \mn@doi [\apj] {10.3847/1538-4357/accea2}, \href
  {https://ui.adsabs.harvard.edu/abs/2023ApJ...956..148L} {956, 148}

\bibitem[\protect\citeauthoryear{{Madau} \& {Dickinson}}{{Madau} \&
  {Dickinson}}{2014}]{madau14}
{Madau} P.,  {Dickinson} M.,  2014, \mn@doi [\araa]
  {10.1146/annurev-astro-081811-125615}, \href
  {https://ui.adsabs.harvard.edu/abs/2014ARA&A..52..415M} {52, 415}

\bibitem[\protect\citeauthoryear{{Mancera Pi{\~n}a} et~al.,}{{Mancera Pi{\~n}a}
  et~al.}{2019}]{mancera19b}
{Mancera Pi{\~n}a} P.~E.,  et~al., 2019, \mn@doi [\apjl]
  {10.3847/2041-8213/ab40c7}, \href
  {https://ui.adsabs.harvard.edu/abs/2019ApJ...883L..33M} {883, L33}

\bibitem[\protect\citeauthoryear{{Mancera Pi{\~n}a} et~al.,}{{Mancera Pi{\~n}a}
  et~al.}{2020}]{mancera20}
{Mancera Pi{\~n}a} P.~E.,  et~al., 2020, \mn@doi [\mnras]
  {10.1093/mnras/staa1256}, \href
  {https://ui.adsabs.harvard.edu/abs/2020MNRAS.495.3636M} {495, 3636}

\bibitem[\protect\citeauthoryear{{McConnachie}}{{McConnachie}}{2012}]{mcconnachie12}
{McConnachie} A.~W.,  2012, \mn@doi [\aj] {10.1088/0004-6256/144/1/4}, \href
  {https://ui.adsabs.harvard.edu/abs/2012AJ....144....4M} {144, 4}

\bibitem[\protect\citeauthoryear{{McQuinn} et~al.,}{{McQuinn}
  et~al.}{2022}]{mcquinn22}
{McQuinn} K. B.~W.,  et~al., 2022, \mn@doi [\apj] {10.3847/1538-4357/ac9285},
  \href {https://ui.adsabs.harvard.edu/abs/2022ApJ...940....8M} {940, 8}

\bibitem[\protect\citeauthoryear{{Moore}, {Ghigna}, {Governato}, {Lake},
  {Quinn}, {Stadel}  \& {Tozzi}}{{Moore} et~al.}{1999}]{moore99}
{Moore} B.,  {Ghigna} S.,  {Governato} F.,  {Lake} G.,  {Quinn} T.,  {Stadel}
  J.,   {Tozzi} P.,  1999, \mn@doi [\apjl] {10.1086/312287}, \href
  {http://adsabs.harvard.edu/abs/1999ApJ...524L..19M} {524, L19}

\bibitem[\protect\citeauthoryear{{Moreno} et~al.,}{{Moreno}
  et~al.}{2022}]{moreno22}
{Moreno} J.,  et~al., 2022, \mn@doi [Nature Astronomy]
  {10.1038/s41550-021-01598-4}, \href
  {https://ui.adsabs.harvard.edu/abs/2022NatAs...6..496M} {6, 496}

\bibitem[\protect\citeauthoryear{{Oh}, {de Blok}, {Brinks}, {Walter}  \&
  {Kennicutt}}{{Oh} et~al.}{2011}]{oh11a}
{Oh} S.-H.,  {de Blok} W.~J.~G.,  {Brinks} E.,  {Walter} F.,   {Kennicutt} Jr.
  R.~C.,  2011, \mn@doi [\aj] {10.1088/0004-6256/141/6/193}, \href
  {http://adsabs.harvard.edu/abs/2011AJ....141..193O} {141, 193}

\bibitem[\protect\citeauthoryear{{Oh} et~al.,}{{Oh} et~al.}{2015}]{oh15}
{Oh} S.-H.,  et~al., 2015, \mn@doi [\aj] {10.1088/0004-6256/149/6/180}, \href
  {https://ui.adsabs.harvard.edu/abs/2015AJ....149..180O} {149, 180}

\bibitem[\protect\citeauthoryear{{Oh}, {Staveley-Smith}, {Spekkens}, {Kamphuis}
   \& {Koribalski}}{{Oh} et~al.}{2018}]{oh18}
{Oh} S.-H.,  {Staveley-Smith} L.,  {Spekkens} K.,  {Kamphuis} P.,
  {Koribalski} B.~S.,  2018, \mn@doi [\mnras] {10.1093/mnras/stx2304}, \href
  {https://ui.adsabs.harvard.edu/abs/2018MNRAS.473.3256O} {473, 3256}

\bibitem[\protect\citeauthoryear{{Okamoto}}{{Okamoto}}{2008}]{okamoto08}
{Okamoto} T.,  2008, in {Frebel} A.,  {Maund} J.~R.,  {Shen} J.,   {Siegel}
  M.~H.,  eds,  Astronomical Society of the Pacific Conference Series Vol. 393,
  New Horizons in Astronomy. p.~111 (\mn@eprint {arXiv} {0712.0086}),
  \mn@doi{10.48550/arXiv.0712.0086}

\bibitem[\protect\citeauthoryear{{Oman}, {Navarro}, {Sales}, {Fattahi},
  {Frenk}, {Sawala}, {Schaller}  \& {White}}{{Oman} et~al.}{2016}]{oman16}
{Oman} K.~A.,  {Navarro} J.~F.,  {Sales} L.~V.,  {Fattahi} A.,  {Frenk} C.~S.,
  {Sawala} T.,  {Schaller} M.,   {White} S. D.~M.,  2016, \mn@doi [\mnras]
  {10.1093/mnras/stw1251}, \href
  {https://ui.adsabs.harvard.edu/abs/2016MNRAS.460.3610O} {460, 3610}

\bibitem[\protect\citeauthoryear{{Ott} et~al.,}{{Ott} et~al.}{2012}]{ott12}
{Ott} J.,  et~al., 2012, \mn@doi [\aj] {10.1088/0004-6256/144/4/123}, \href
  {http://adsabs.harvard.edu/abs/2012AJ....144..123O} {144, 123}

\bibitem[\protect\citeauthoryear{{Patra}}{{Patra}}{2020}]{patra20}
{Patra} N.~N.,  2020, \mn@doi [\mnras] {10.1093/mnras/staa1353}, \href
  {https://ui.adsabs.harvard.edu/abs/2020MNRAS.495.2867P} {495, 2867}

\bibitem[\protect\citeauthoryear{{Planck Collaboration} et~al.,}{{Planck
  Collaboration} et~al.}{2016}]{planck16}
{Planck Collaboration} et~al., 2016, \mn@doi [\aap]
  {10.1051/0004-6361/201525830}, \href
  {https://ui.adsabs.harvard.edu/abs/2016A&A...594A..13P} {594, A13}

\bibitem[\protect\citeauthoryear{{Roberts}}{{Roberts}}{1962}]{roberts62}
{Roberts} M.~S.,  1962, \mn@doi [\aj] {10.1086/108752}, \href
  {https://ui.adsabs.harvard.edu/abs/1962AJ.....67..437R} {67, 437}

\bibitem[\protect\citeauthoryear{{Roychowdhury}, {Chengalur}, {Begum}  \&
  {Karachentsev}}{{Roychowdhury} et~al.}{2010}]{roychowdhury10}
{Roychowdhury} S.,  {Chengalur} J.~N.,  {Begum} A.,   {Karachentsev} I.~D.,
  2010, \mn@doi [\mnras] {10.1111/j.1745-3933.2010.00835.x}, \href
  {https://ui.adsabs.harvard.edu/abs/2010MNRAS.404L..60R} {404, L60}

\bibitem[\protect\citeauthoryear{{Sales}, {Wetzel}  \& {Fattahi}}{{Sales}
  et~al.}{2022}]{sales22}
{Sales} L.~V.,  {Wetzel} A.,   {Fattahi} A.,  2022, \mn@doi [Nature Astronomy]
  {10.1038/s41550-022-01689-w}, \href
  {https://ui.adsabs.harvard.edu/abs/2022NatAs...6..897S} {6, 897}

\bibitem[\protect\citeauthoryear{{Sawala} et~al.,}{{Sawala}
  et~al.}{2015}]{sawala15}
{Sawala} T.,  et~al., 2015, \mn@doi [\mnras] {10.1093/mnras/stu2753}, \href
  {https://ui.adsabs.harvard.edu/abs/2015MNRAS.448.2941S} {448, 2941}

\bibitem[\protect\citeauthoryear{{Sawala} et~al.,}{{Sawala}
  et~al.}{2016}]{sawala16}
{Sawala} T.,  et~al., 2016, \mn@doi [\mnras] {10.1093/mnras/stv2597}, \href
  {https://ui.adsabs.harvard.edu/abs/2016MNRAS.456...85S} {456, 85}

\bibitem[\protect\citeauthoryear{{Schruba} et~al.,}{{Schruba}
  et~al.}{2012}]{schruba12}
{Schruba} A.,  et~al., 2012, \mn@doi [\aj] {10.1088/0004-6256/143/6/138}, \href
  {http://adsabs.harvard.edu/abs/2012AJ....143..138S} {143, 138}

\bibitem[\protect\citeauthoryear{{Scott}, {Sengupta}, {Lagos}, {Chung}  \&
  {Wong}}{{Scott} et~al.}{2021}]{scott21}
{Scott} T.~C.,  {Sengupta} C.,  {Lagos} P.,  {Chung} A.,   {Wong} O.~I.,  2021,
  \mn@doi [\mnras] {10.1093/mnras/stab390}, \href
  {https://ui.adsabs.harvard.edu/abs/2021MNRAS.503.3953S} {503, 3953}

\bibitem[\protect\citeauthoryear{{Sellwood} \& {Spekkens}}{{Sellwood} \&
  {Spekkens}}{2015}]{sellwood15}
{Sellwood} J.~A.,  {Spekkens} K.,  2015, \mn@doi [arXiv e-prints]
  {10.48550/arXiv.1509.07120}, \href
  {https://ui.adsabs.harvard.edu/abs/2015arXiv150907120S} {p. arXiv:1509.07120}

\bibitem[\protect\citeauthoryear{{Shin}, {Jung}, {Kwon}, {Kim}, {Lee}, {Jo}  \&
  {Oh}}{{Shin} et~al.}{2020}]{shin20}
{Shin} E.-j.,  {Jung} M.,  {Kwon} G.,  {Kim} J.-h.,  {Lee} J.,  {Jo} Y.,   {Oh}
  B.~K.,  2020, \mn@doi [\apj] {10.3847/1538-4357/aba434}, \href
  {https://ui.adsabs.harvard.edu/abs/2020ApJ...899...25S} {899, 25}

\bibitem[\protect\citeauthoryear{{Simon}}{{Simon}}{2019}]{simon19}
{Simon} J.~D.,  2019, \mn@doi [\araa] {10.1146/annurev-astro-091918-104453},
  \href {https://ui.adsabs.harvard.edu/abs/2019ARA&A..57..375S} {57, 375}

\bibitem[\protect\citeauthoryear{{Simon} \& {Geha}}{{Simon} \&
  {Geha}}{2007}]{simon07}
{Simon} J.~D.,  {Geha} M.,  2007, \mn@doi [\apj] {10.1086/521816}, \href
  {http://adsabs.harvard.edu/abs/2007ApJ...670..313S} {670, 313}

\bibitem[\protect\citeauthoryear{{Smith}, {Sijacki}  \& {Shen}}{{Smith}
  et~al.}{2019}]{smith19}
{Smith} M.~C.,  {Sijacki} D.,   {Shen} S.,  2019, \mn@doi [\mnras]
  {10.1093/mnras/stz599}, \href
  {https://ui.adsabs.harvard.edu/abs/2019MNRAS.485.3317S} {485, 3317}

\bibitem[\protect\citeauthoryear{{Somerville} \& {Dav{\'e}}}{{Somerville} \&
  {Dav{\'e}}}{2015}]{somerville15}
{Somerville} R.~S.,  {Dav{\'e}} R.,  2015, \mn@doi [\araa]
  {10.1146/annurev-astro-082812-140951}, \href
  {https://ui.adsabs.harvard.edu/abs/2015ARA&A..53...51S} {53, 51}

\bibitem[\protect\citeauthoryear{{Starkenburg}, {Sales}, {Genel},
  {Manzano-King}, {Canalizo}  \& {Hernquist}}{{Starkenburg}
  et~al.}{2019}]{starkenburg19}
{Starkenburg} T.~K.,  {Sales} L.~V.,  {Genel} S.,  {Manzano-King} C.,
  {Canalizo} G.,   {Hernquist} L.,  2019, \mn@doi [\apj]
  {10.3847/1538-4357/ab2128}, \href
  {https://ui.adsabs.harvard.edu/abs/2019ApJ...878..143S} {878, 143}

\bibitem[\protect\citeauthoryear{{Strigari}, {Bullock}, {Kaplinghat}, {Simon},
  {Geha}, {Willman}  \& {Walker}}{{Strigari} et~al.}{2008}]{strigari08}
{Strigari} L.~E.,  {Bullock} J.~S.,  {Kaplinghat} M.,  {Simon} J.~D.,  {Geha}
  M.,  {Willman} B.,   {Walker} M.~G.,  2008, \mn@doi [\nat]
  {10.1038/nature07222}, \href
  {http://adsabs.harvard.edu/abs/2008Natur.454.1096S} {454, 1096}

\bibitem[\protect\citeauthoryear{{Subramanian}, {Mondal}  \&
  {Kalari}}{{Subramanian} et~al.}{2024}]{subramanian24}
{Subramanian} S.,  {Mondal} C.,   {Kalari} V.,  2024, \mn@doi [\aap]
  {10.1051/0004-6361/202346536}, \href
  {https://ui.adsabs.harvard.edu/abs/2024A&A...681A...8S} {681, A8}

\bibitem[\protect\citeauthoryear{{Swaters}, {van Albada}, {van der Hulst}  \&
  {Sancisi}}{{Swaters} et~al.}{2002}]{swaters02}
{Swaters} R.~A.,  {van Albada} T.~S.,  {van der Hulst} J.~M.,   {Sancisi} R.,
  2002, \mn@doi [\aap] {10.1051/0004-6361:20011755}, \href
  {https://ui.adsabs.harvard.edu/abs/2002A&A...390..829S} {390, 829}

\bibitem[\protect\citeauthoryear{{Tau} \& {Sc{\'o}ccola}}{{Tau} \&
  {Sc{\'o}ccola}}{2021}]{tau21}
{Tau} E.~A.,  {Sc{\'o}ccola} C.~G.,  2021, Boletin de la Asociacion Argentina
  de Astronomia La Plata Argentina, \href
  {https://ui.adsabs.harvard.edu/abs/2021BAAA...62..237T} {62, 237}

\bibitem[\protect\citeauthoryear{{Tully}, {Pierce}, {Huang}, {Saunders},
  {Verheijen}  \& {Witchalls}}{{Tully} et~al.}{1998}]{tully98}
{Tully} R.~B.,  {Pierce} M.~J.,  {Huang} J.-S.,  {Saunders} W.,  {Verheijen} M.
  A.~W.,   {Witchalls} P.~L.,  1998, \mn@doi [\aj] {10.1086/300379}, \href
  {https://ui.adsabs.harvard.edu/abs/1998AJ....115.2264T} {115, 2264}

\bibitem[\protect\citeauthoryear{Walker}{Walker}{2013}]{walker13inbook}
Walker M.,  2013, Dark Matter in the Galactic Dwarf Spheroidal Satellites.
Springer Netherlands, Dordrecht, pp 1039--1089,
  \mn@doi{10.1007/978-94-007-5612-0_20}, \url
  {https://doi.org/10.1007/978-94-007-5612-0_20}

\bibitem[\protect\citeauthoryear{{Walter}, {Brinks}, {de Blok}, {Bigiel},
  {Kennicutt}, {Thornley}  \& {Leroy}}{{Walter} et~al.}{2008}]{walter08}
{Walter} F.,  {Brinks} E.,  {de Blok} W.~J.~G.,  {Bigiel} F.,  {Kennicutt} Jr.
  R.~C.,  {Thornley} M.~D.,   {Leroy} A.,  2008, \mn@doi [\aj]
  {10.1088/0004-6256/136/6/2563}, \href
  {http://adsabs.harvard.edu/abs/2008AJ....136.2563W} {136, 2563}

\bibitem[\protect\citeauthoryear{{Wang} et~al.,}{{Wang} et~al.}{2013}]{wang13}
{Wang} J.,  et~al., 2013, \mn@doi [\mnras] {10.1093/mnras/stt722}, \href
  {https://ui.adsabs.harvard.edu/abs/2013MNRAS.433..270W} {433, 270}

\bibitem[\protect\citeauthoryear{{Wang}, {Koribalski}, {Serra}, {van der
  Hulst}, {Roychowdhury}, {Kamphuis}  \& {Chengalur}}{{Wang}
  et~al.}{2016}]{wang16}
{Wang} J.,  {Koribalski} B.~S.,  {Serra} P.,  {van der Hulst} T.,
  {Roychowdhury} S.,  {Kamphuis} P.,   {Chengalur} J.~N.,  2016, \mn@doi
  [\mnras] {10.1093/mnras/stw1099}, \href
  {https://ui.adsabs.harvard.edu/abs/2016MNRAS.460.2143W} {460, 2143}

\bibitem[\protect\citeauthoryear{{Welty}, {Xue}  \& {Wong}}{{Welty}
  et~al.}{2012}]{welty12}
{Welty} D.~E.,  {Xue} R.,   {Wong} T.,  2012, \mn@doi [\apj]
  {10.1088/0004-637X/745/2/173}, \href
  {http://adsabs.harvard.edu/abs/2012ApJ...745..173W} {745, 173}

\bibitem[\protect\citeauthoryear{{Westmeier} et~al.,}{{Westmeier}
  et~al.}{2021}]{sofia21}
{Westmeier} T.,  et~al., 2021, \mn@doi [\mnras] {10.1093/mnras/stab1881}, \href
  {https://ui.adsabs.harvard.edu/abs/2021MNRAS.506.3962W} {506, 3962}

\bibitem[\protect\citeauthoryear{{Wheeler} et~al.,}{{Wheeler}
  et~al.}{2025}]{wheeler25}
{Wheeler} C.,  et~al., 2025, \mn@doi [\apj] {10.3847/1538-4357/ae1698}, \href
  {https://ui.adsabs.harvard.edu/abs/2025ApJ...995..162W} {995, 162}

\bibitem[\protect\citeauthoryear{{White}, {Navarro}, {Evrard}  \&
  {Frenk}}{{White} et~al.}{1993}]{white93}
{White} S. D.~M.,  {Navarro} J.~F.,  {Evrard} A.~E.,   {Frenk} C.~S.,  1993,
  \mn@doi [\nat] {10.1038/366429a0}, \href
  {https://ui.adsabs.harvard.edu/abs/1993Natur.366..429W} {366, 429}

\bibitem[\protect\citeauthoryear{{Yasin}, {Desmond}, {Devriendt}  \&
  {Slyz}}{{Yasin} et~al.}{2023}]{yasin23}
{Yasin} T.,  {Desmond} H.,  {Devriendt} J.,   {Slyz} A.,  2023, \mn@doi
  [\mnras] {10.1093/mnras/stad1183}, \href
  {https://ui.adsabs.harvard.edu/abs/2023MNRAS.526.5861Y} {526, 5861}

\bibitem[\protect\citeauthoryear{{Zhang}, {Puzia}  \& {Weisz}}{{Zhang}
  et~al.}{2017}]{zhang17}
{Zhang} H.-X.,  {Puzia} T.~H.,   {Weisz} D.~R.,  2017, \mn@doi [\apjs]
  {10.3847/1538-4365/aa937b}, \href
  {https://ui.adsabs.harvard.edu/abs/2017ApJS..233...13Z} {233, 13}

\bibitem[\protect\citeauthoryear{{de Blok}, {Walter}, {Brinks}, {Trachternach},
  {Oh}  \& {Kennicutt}}{{de Blok} et~al.}{2008}]{deblok08}
{de Blok} W.~J.~G.,  {Walter} F.,  {Brinks} E.,  {Trachternach} C.,  {Oh}
  S.-H.,   {Kennicutt} Jr. R.~C.,  2008, \mn@doi [\aj]
  {10.1088/0004-6256/136/6/2648}, \href
  {http://adsabs.harvard.edu/abs/2008AJ....136.2648D} {136, 2648}

\bibitem[\protect\citeauthoryear{{van Dokkum} et~al.,}{{van Dokkum}
  et~al.}{2018}]{vandokkum18}
{van Dokkum} P.,  et~al., 2018, \mn@doi [\nat] {10.1038/nature25767}, \href
  {https://ui.adsabs.harvard.edu/abs/2018Natur.555..629V} {555, 629}

\bibitem[\protect\citeauthoryear{{van Dokkum}, {Danieli}, {Abraham}, {Conroy}
  \& {Romanowsky}}{{van Dokkum} et~al.}{2019}]{vandokkum19}
{van Dokkum} P.,  {Danieli} S.,  {Abraham} R.,  {Conroy} C.,   {Romanowsky}
  A.~J.,  2019, \mn@doi [\apjl] {10.3847/2041-8213/ab0d92}, \href
  {https://ui.adsabs.harvard.edu/abs/2019ApJ...874L...5V} {874, L5}

\makeatother
\end{thebibliography}

\appendix

\section{Channel Maps}
Here we present the channel maps for the six galaxies analysed in this work. The velocity widths in column 4 of table~\ref{tab:vel_corrections} are obtained by inspecting the emission signal from the \HI\ line at least 3$\sigma$ or stronger within the galaxy's \HI\ radius ($r_{\rm H \textsc i}$) marked by the yellow dashed ellipses.
\begin{figure*}
    \caption{Channel plots of the six galaxies with contour levels starting at 3$\sigma$ with increments in multiples of 1.414. The yellow dashed ellipse represents the galaxy marked upto 1 \mspsqpc\ as described in Fig. ~\ref{tab:combined_images}. The channel with a red cross-mark represents channel corresponding to the velocity centre. The values printed in black boldface on each channel subplot represents the corresponding barycentric velocity (in \kms) corresponding to the \HI-line in for that channel.}
    \label{fig:combined_chanplots}
    \centering
        \includegraphics[width=0.90\linewidth]{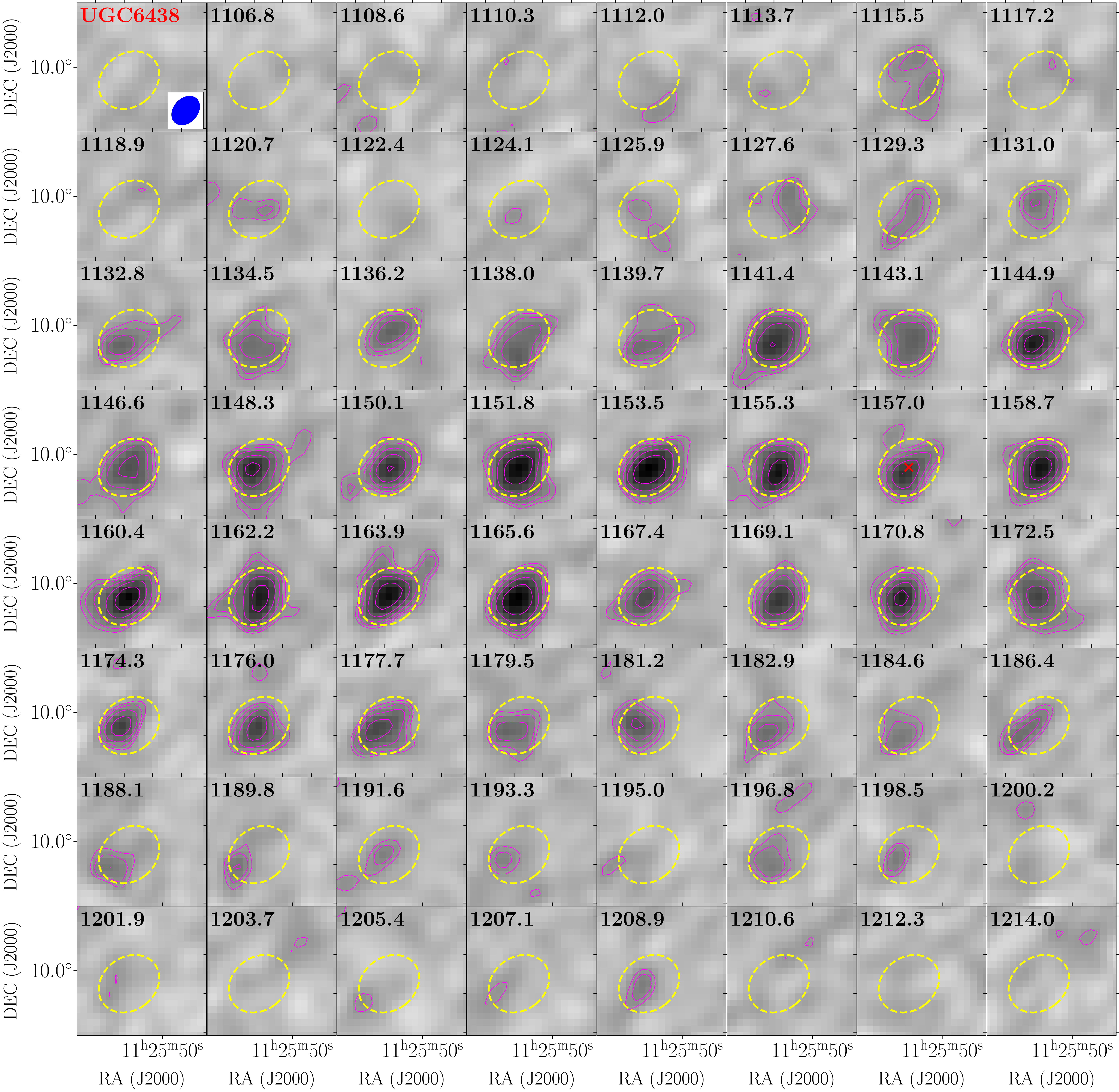}
\end{figure*}

\begin{figure*}
    \ContinuedFloat
    \caption{Continued}
    \centering
        \includegraphics[width=0.90\linewidth]{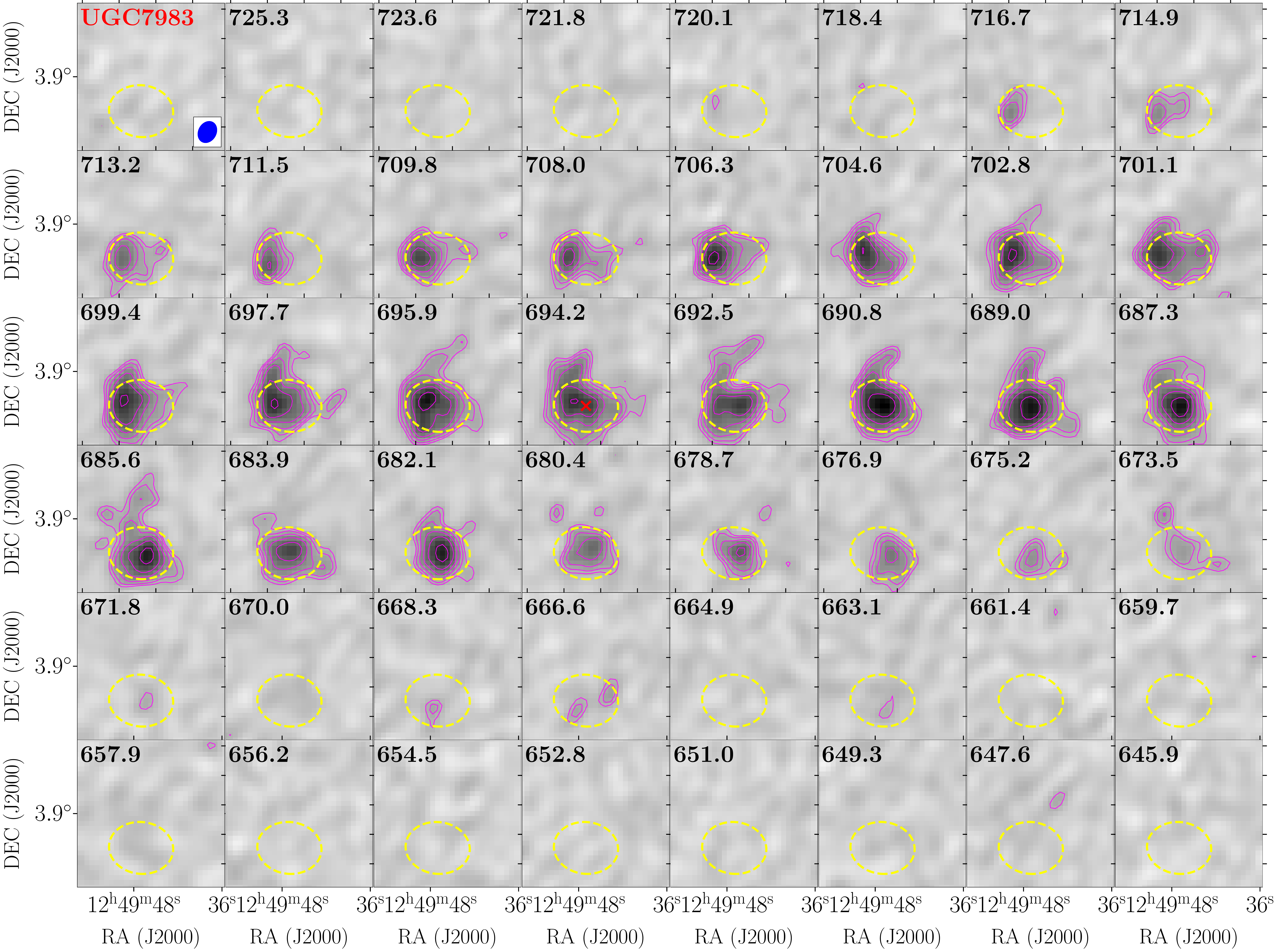} 
        \includegraphics[width=0.90\linewidth]{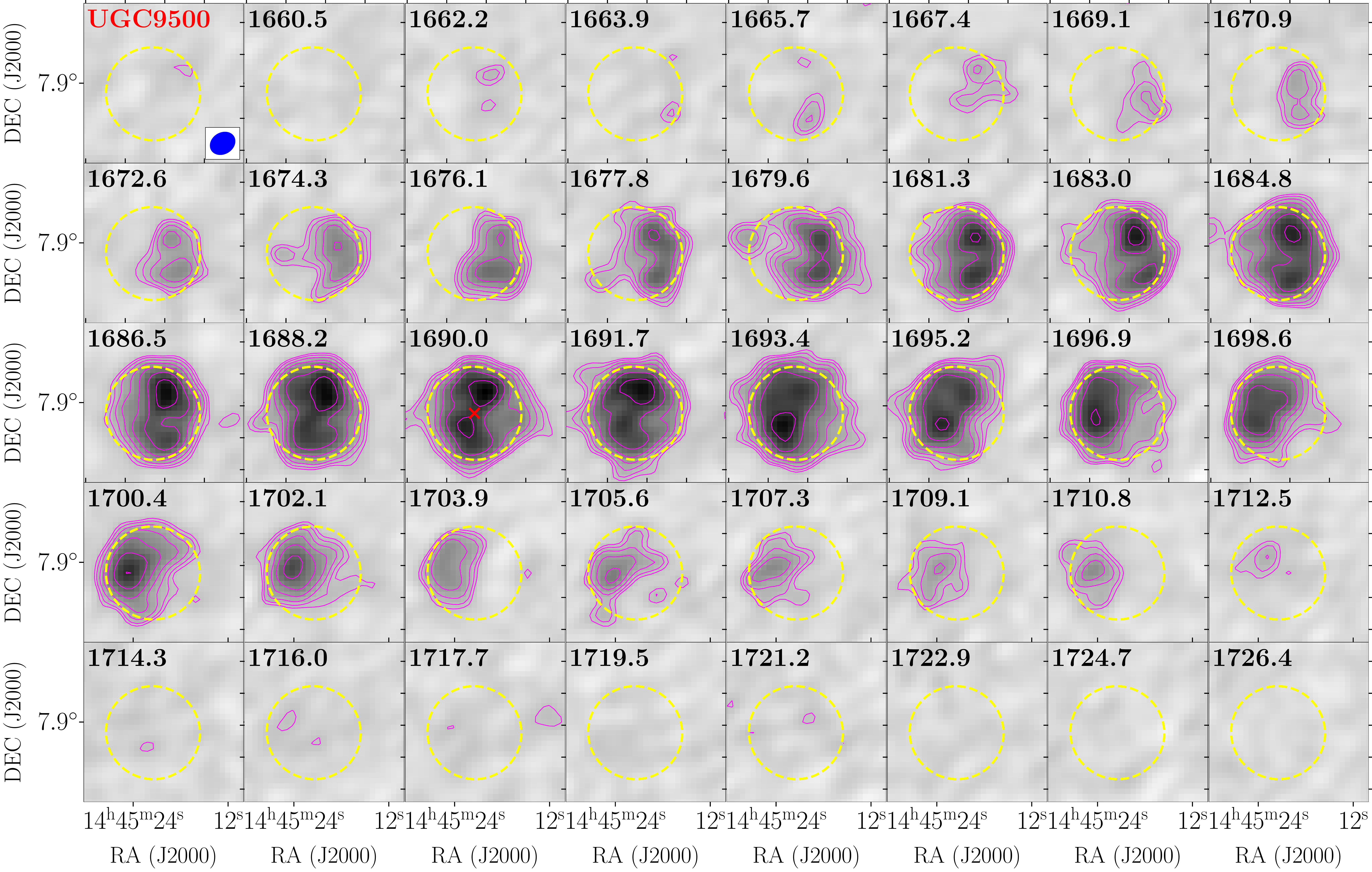}
\end{figure*}

\begin{figure*}
    \ContinuedFloat
    \caption{Continued}
    \centering
        \includegraphics[width=0.90\linewidth]{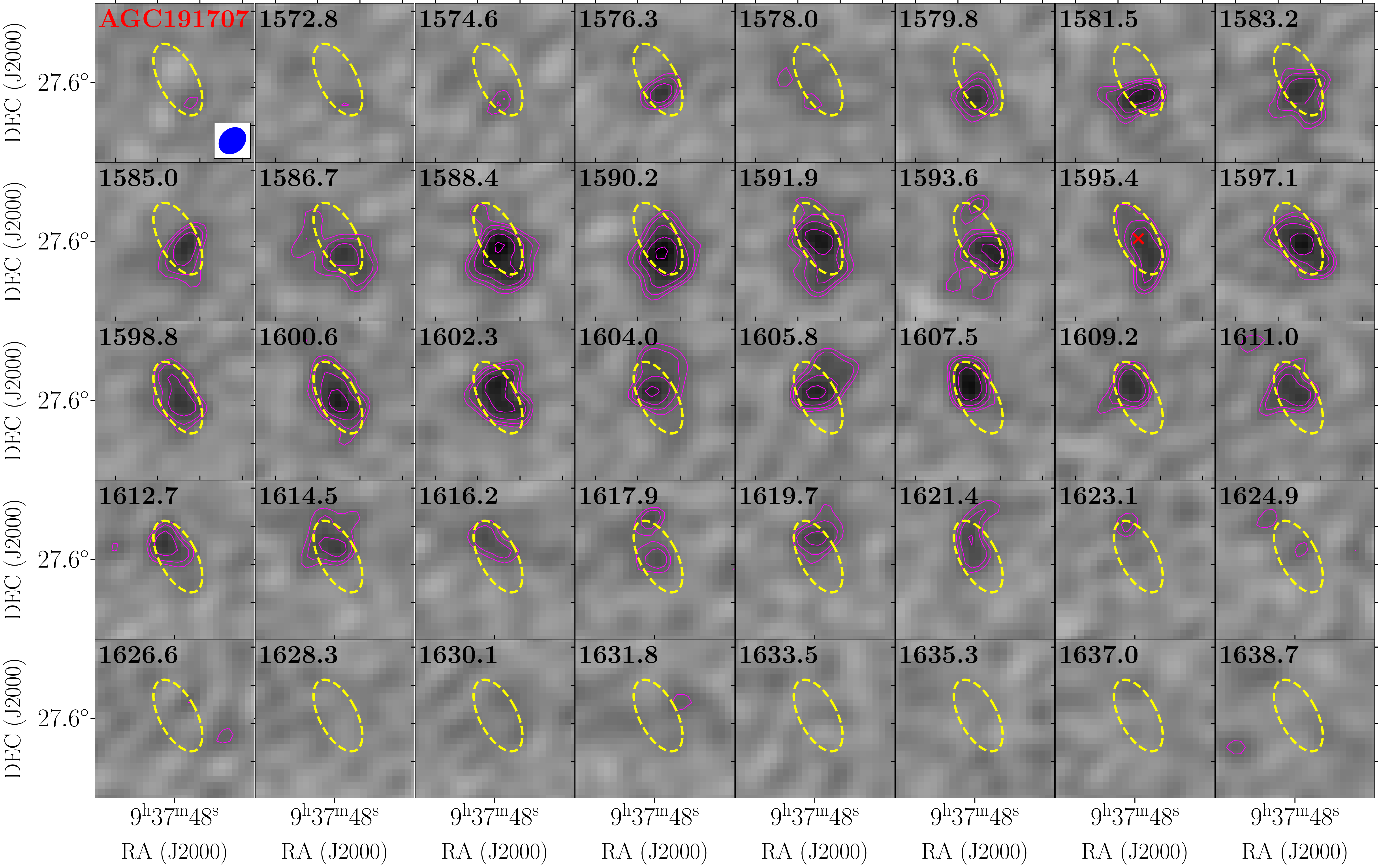} 
        \includegraphics[width=0.90\linewidth]{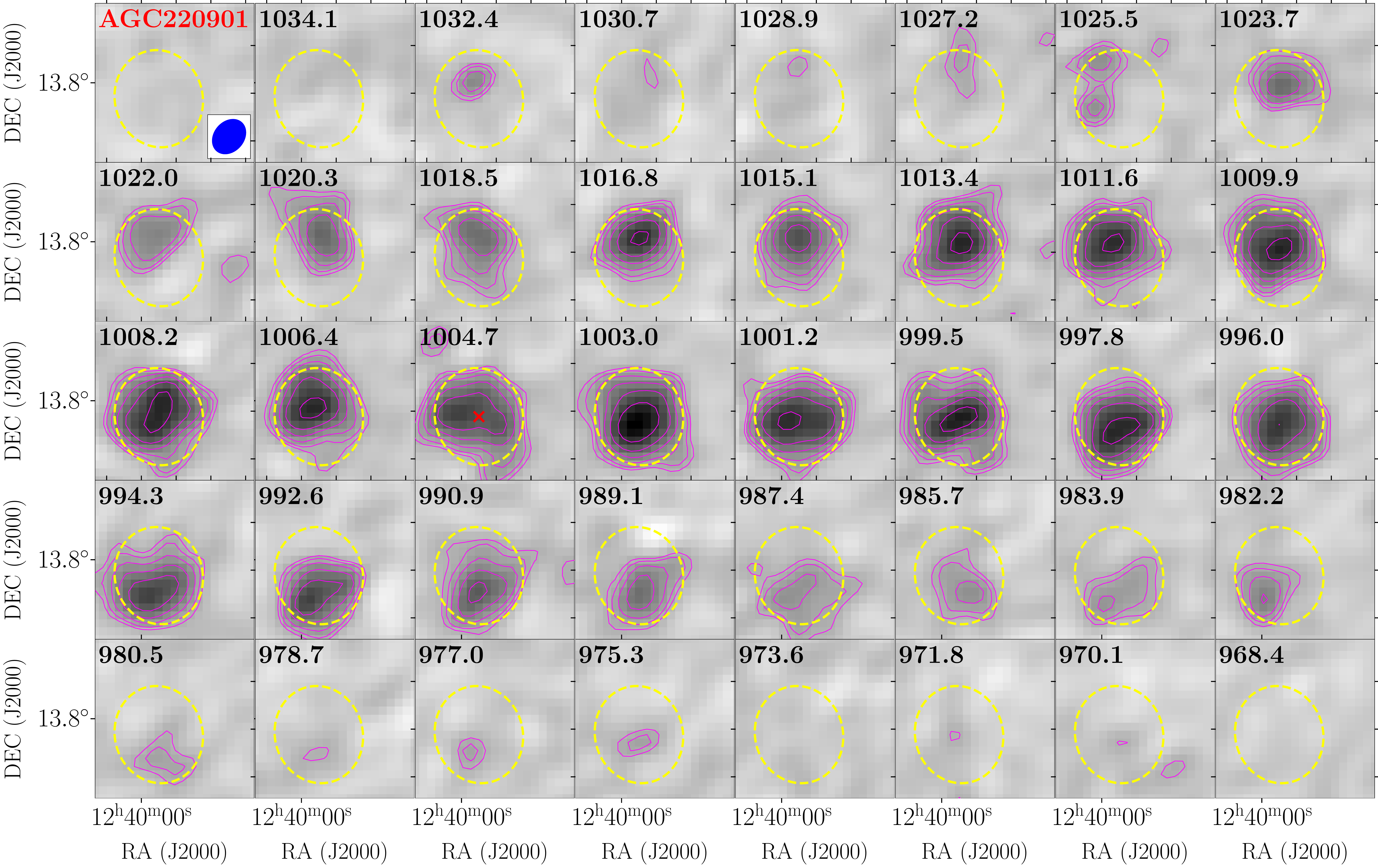} 
\end{figure*}

\begin{figure*}
    \ContinuedFloat
    \caption{Continued}
    \centering
        \includegraphics[width=0.90\linewidth]{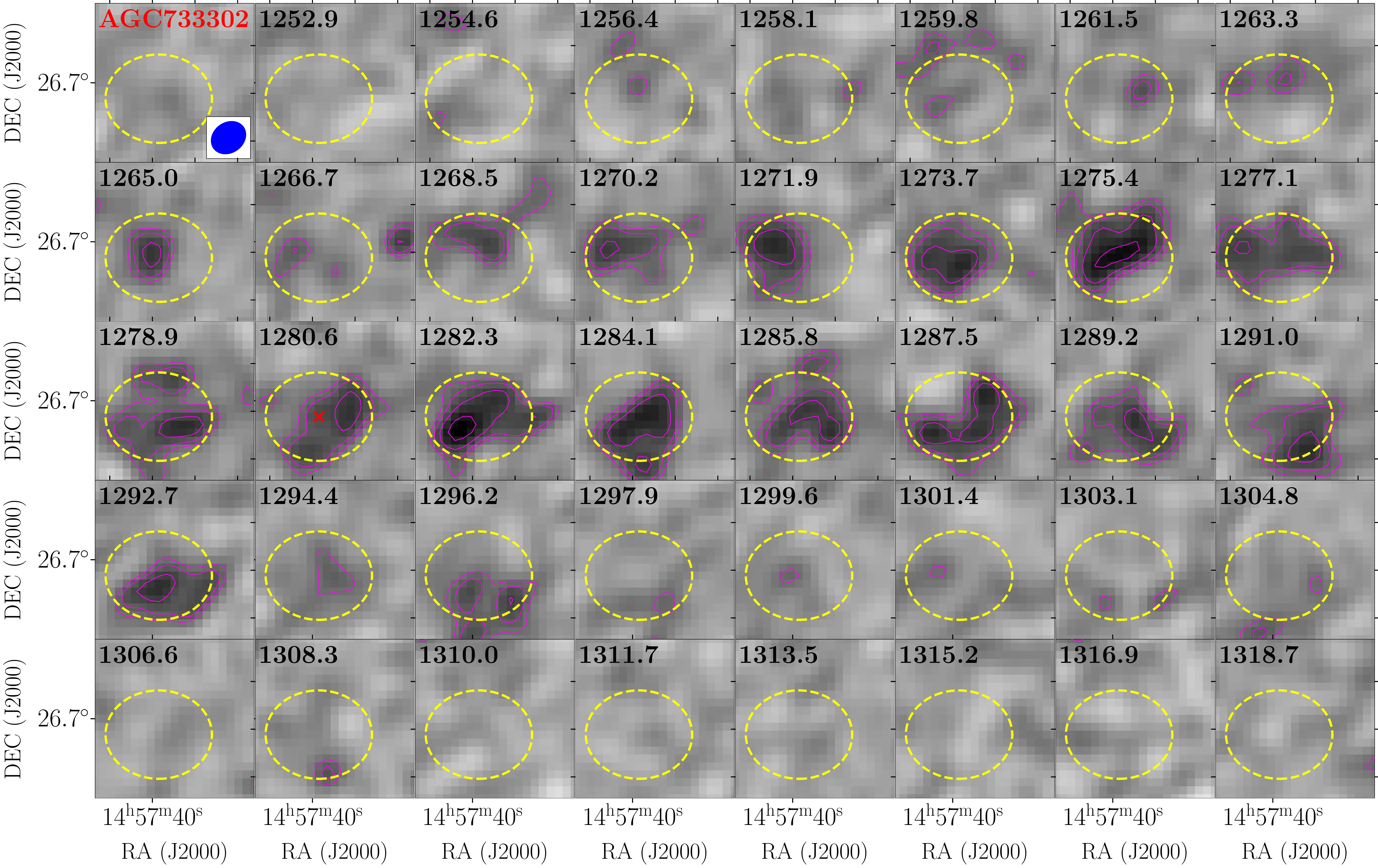}
\end{figure*}

\bsp	
\label{lastpage}
\end{document}